\documentclass{llncs}

\usepackage[utf8x]{inputenc}
\usepackage[T1]{fontenc}
\usepackage{comment}
\usepackage{algorithm}
\usepackage[noend]{algorithmic}
\usepackage{float}

\usepackage{tikz}  
\usetikzlibrary{automata, positioning, arrows}
\usetikzlibrary{shapes}
\usetikzlibrary{shapes.geometric}

\usepackage[a4paper,top=3cm,bottom=2cm,left=3cm,right=3cm,marginparwidth=1.75cm]{geometry}

\usepackage{amsmath}
\usepackage{amsfonts}
\usepackage{graphicx}
\usepackage[colorinlistoftodos]{todonotes}
\usepackage[colorlinks=true, allcolors=blue]{hyperref}

\author{Francisco J. Marmolejo-Coss\'io\thanks{Supported by the Mexican National Council of Science and Technology (CONACyT)}\inst{1} \and Eric Brigham\inst{2} \and Benjamin Sela\inst{3} \and Jonathan Katz\inst{3}}

\institute{University of Oxford\\
\email{francisco.marmolejo@cs.ox.ac.uk}
\and
New College of Florida\\
\email{eric.brigham15@ncf.edu}
\and
University of Maryland\\
\email{jkatz@cs.umd.edu, benj.sela@gmail.com}
}

\title{Competing (Semi)-Selfish Miners in Bitcoin}

\begin{document}

\maketitle

\begin{abstract}
The Bitcoin protocol prescribes certain behavior by the miners who are responsible for maintaining and extending the underlying blockchain; in particular, miners who successfully solve a puzzle, and hence can extend the chain by a block, are supposed to release that block immediately. Eyal and Sirer showed, however, that a selfish miner is incentivized to deviate from the protocol and withhold its blocks under certain conditions. 

The analysis by Eyal and Sirer, as well as in followup work, considers a \emph{single} deviating miner (who may control a large fraction of the hashing power in the network) interacting with a remaining pool of honest miners. 
Here, we extend this analysis to the case where there are \emph{multiple} (non-colluding) selfish miners. We find that with multiple strategic miners, specific deviations from honest mining by multiple strategic agents can outperform honest mining, even if individually miners would not be incentivised to be dishonest. This previous point effectively renders the Bitcoin protocol to be less secure than previously thought. 
\end{abstract}

\keywords{blockchain, selfish mining}

\section{Introduction}
\label{sec:intro}

One of the key innovations in the Nakamoto protocol behind Bitcoin \cite{nakamoto2008bitcoin} is the assumption that agents involved in the upkeep of the digital ledger, so called miners, are strategic rather than adversarial, which invites a game-theoretic analysis of the underlying protocol. Under this relaxed assumption, Bitcoin enjoys more robust guarantees on its security: the adage being ``it is in a miner's best interest to be honest when there is an honest majority of miners''.

This adage however was famously proven to be incorrect by Eyal and Sirer \cite{eyal2014majority}, when they first described ``Selfish Mining'', a non-honest miner strategy that gives more returns to miners than honest mining, even if a majority of other agents are honest. Subsequently, there has been much work exploring the extensions and limitations of selfish mining, but most of this work is limited to the case in which there is a single selfish miner and the rest of the network acts honestly. In this paper we study scenarios where more than one miner deviates from the honest mining protocol. We show that there are substantial game-theoretic differences when multiple miners can be strategic with implications to Bitcoin's security. First of all, there are hash rates where a miner is incentivised to be honest if mining is treated as a one-shot game, yet where the miner is incentivised to be strategic if he is the leader in sequential (Stackleberg) game. Second of all, we show that with multiple strategic miners, specific deviations from honest mining by multiple strategic agents can outperform honest mining, even if individually miners would not be incentivised to be dishonest. These two previous points effectively render the Bitcoin protocol to be less secure than previously thought. 

\subsection{Our Contributions}

We study miner incentives when multiple miners employ variants of selfish mining strategies. Original selfish mining (SM) consists of secretly withholding mined blocks and judiciously publishing private blocks in an attempt to increase stale block rates of other miners. Though such an attack is not immediately profitable, as the block rate of all miners decreases, it can be profitable in a longer time horizon as block difficulty rates decrease. In SM, miners may keep an arbitrarily long private chain, which makes it difficult to analytically solve for relative revenues when more than one miner employs SM. For this reason, we study a truncation of this strategy, semi-selfish mining (SSM), where miners keep a private chain of length at most 2. 

SSM falls within the family of generalised selfish mining strategies of \cite{SSZ16} and \cite{nayak2016stubborn}, and our paper begins by studying analytic properties of SSM's performance against honest mining. In Section \ref{sec:one-SSM}, we show that although SSM achieves less relative revenue than SM against honest mining, it is always a more profitable strategy for a strategic miner than honest mining if the miner has a hash rate larger than $38\%$ of the total system hash rate, and if the strategic miner is able to propagate blocks to other miners quickly, this threshold lowers to around $26.8\%$. In fact, we show the relative revenue of SSM is an asymptotically tight lower bound to the relative revenue of SM as a strategic miner's hash power tends to 0.

As mentioned before, the benefit of SSM is that it can be represented with a reduced state space, and hence we can explicitly solve for relative revenues in the case where multiple strategic miners employ SSM. In Section \ref{sec:two-SSM} we focus on systems with two strategic miners and describe the Markov chain that governs block publishing dynamics. This allows us to explicitly solve for relative revenues of all miners in the steady state. 

With the steady state solutions in hand, we are able to study the incentives that govern the decision whether a miner uses SSM against another strategic miner. To do so, we define a binary action two-player game amongst both strategic miners which we call the SSM game. In the SSM game both miners are denoted by $m_1$ and $m_2$ and they have corresponding utility functions $U_1$ and $U_2$. In addition, each miner has the action set $\{H,S\}$ representing honest mining and SSM mining. Interestingly, we find multiple scenarios at different hash rates:

\begin{itemize}
\item For all pure strategy profiles, $s \in \{(H,H),(S,H),(H,S),(S,S)\}$, there exist hash rates of both strategic miners such that $s$ is a unique pure Nash equilibrium.
\item When both strategic miners have roughly around 0.2 to 0.27 of the system's hash power, both $(H,H)$ and $(S,S)$ are simultaneously pure Nash equilibria of the SSM game.
\item There exist hash rates where a specific miner is not unilaterally incentivised to employ SSM, yet $(S,S)$ is the only pure Nash equilibrium of the game. This effectively lowers the minimum hash rate required for SSM to be profitable by virtue of the existence of another strategic miner.
\item There exist hash rates where $U_1(S,S) < U_1(H,H) < U_1(S,H)$ (once again, an identical result holds with the roles of miners reversed). This is interesting because although SSM is individually rational for the first strategic miner, the second (larger) strategic miner has the ability to ``penalise'' the first miner were they to retaliate by using SSM.
\end{itemize}

We also consider a richer action space for miners: we allow them to partition their hash power into an honest portion and an SSM portion. The game specified by these utilities is called the partition game, and when treated as a one-shot game, it yields the same pure Nash equilibria as the SSM game. The more interesting result stems from treating this game as a Stackelberg game and understanding optimal commitments a miner may make to elicit a desired behaviour in the other miner. It turns out that in the partition game, there exist hash rates with non-trivial Stackelberg equilibria that can result in large gains for leader miners. In fact, there are even hash rates where a miner is honest in the one-shot SSM game, yet strategic in the sequential partition game's Stackelberg equilibrium. This has important consequences for the security of Bitcoin, as miners with smaller hash rates than what was known before may be incentivised to be strategic in a sequential setting.

In Section \ref{sec:N-SSM} we consider the scenario where $M > 2$ miners are strategic. For $1 \leq M \leq 8$, we compute bounds on the minimal $\alpha \in [0,1]$ such that if the $M$ strategic miners each with hash power $\alpha$ have to decide between employing honest mining and SSM, the strategy profile where all such miners employ SSM Pareto-dominates honest mining. For each $M$, we call $\alpha$ the uniform profitability threshold for SSM, and we show that not only is it a decreasing function in $M$, but that already for $M = 8$, $\alpha$ is as low as 0.11. This is striking, because at such hash rates, miners are far from being individually incentivised to employ SSM, implying that the existence of other strategic miners can effectively hurt the stability of Bitcoin.

As an aside, we also note that in Appendix \ref{sec:multiplayer-game-formalism} we explicitly extend our game-theoretic formalism from Section \ref{sec:two-SSM} and Section \ref{sec:partition-games} to the multi-player setting, and we specify how to compute utilities in these games. Furthermore, in Appendix \ref{sec:3SSM} we extensively map incentives of 3 strategic miners akin to Section \ref{sec:two-SSM} and Section \ref{sec:partition-games}. We find that the game-theoretic observations of the two-player setting generalise appropriately.

\subsection{Related Work}

Selfish mining was originally introduced by Eyal and Sirer in \cite{eyal2014majority}. In this work, the authors describe Selfish Mining (SM), a specific mining strategy that deviates from the prescribed honest mining strategy of the Bitcoin network with the key property that it is more profitable than honest mining for miners with over $1/3$ of the hash power of the entire Bitcoin network. Subsequently, \cite{nayak2016stubborn} and \cite{SSZ16} identify a generalised class of selfish mining strategies to which SM belongs and show that in general there are more aggressive and profitable strategies than SM within this family of strategies. In a similar vein, \cite{kiayias2016blockchain} uses game theory to formalise the decision a single strategic miner may take to employ different strategies from the generalised family of selfish mining strategies. In particular, they define analogous complete information games to real-life mining and show that for these games, if no miner has a large enough hash power, honest mining is a Nash equilibrium.

Perhaps most similar to our work is \cite{liu2018strategy}, where the authors simulate multiple strategic miners employing strategies other than honest mining. Their results are simulation-based, whereas we provide closed-form results for the specific SSM strategy. In fact, our model can be seen as a variant of the model used in \cite{BZWXWK18}, which we developed concurrently to allow for an arbitrary number of strategic agents employing SSM. Furthermore we focus on the game-theoretic considerations miners may take in deciding whether to employ SSM in varying degrees.

Subversive mining strategies can also be combined with network level attacks to exacerbate undue profits. This is discussed in \cite{nayak2016stubborn} where the authors combine selfish mining strategies with eclipse attacks; an eclipse attack is when an entity holds all connections with a subset of the mining swarm and can thus control all communication between them and the rest of the miners. The authors show that no combination of a selfish mining strategy and eclipse attack is optimal at all times. The choice of what selfish mining strategy to adopt as well as how to eclipse a victim is highly dependent on the network parameters in which one is operating. These parameters include computational power, percentage of the network that can be eclipsed, and the percentage of remaining miners that can be influenced.

There are additional attacks miners can wage outside the family of selfish mining. At the pool level, managers can wage withholding attacks as per \cite{courtois2014subversive} \cite{eyal2015miner}, where a malicious pool infiltrates a victim pool, submitting shares and withholding full solutions. Indeed this notion of ``partitioning'' one's pool is similar to our partition games from section \ref{sec:two-SSM}. \cite{eyal2015miner} shows that this can be profitable for a single malicious pool, but when multiple pools engage in block withholding attacks, this results in a situation akin to the prisoner's dilemma, where the equilibrium of all malicious pools is to infiltrate and thus reduce the overall profit of every pool in the network. Withholding attacks are further refined in \cite{KwonKSVK17}, where a malicious pool still withholds full solutions from a victim pool, but may share said full solutions when it hears of a full solution being found by a miner outside of the malicious and victim pool. The intent of this strategy is to incentivise the victim pool manager to cause a fork, and this behaviour does away with the prisoner's dilemma of \cite{eyal2015miner}, as there are equilibria where larger pools are strictly better off than honest mining. Furthermore, there is some evidence showing that this family of pool-level attacks can be difficult to detect for victim pools \cite{courtois2014subversive}. 

Along with work covering subversive mining attacks and which strategies miners should adopt based on network parameters, there have also been efforts to defeat these attacks. In \cite{eyal2016bitcoin} the authors outline a new blockchain protocol, Bitcoin Next Generation, which decouples leader election and transaction serialization for better scalability. In addition to this they also modify which chain honest miners adopt as the one they will mine on. Currently, when honest miners are presented with two chains of the same length, they will opt to accept the older one. This fact gives selfish miners an advantage in that they become more powerful the more connected they are to the rest of the network and can lower the necessary computational power needed to selfish mine successfully. In their new protocol, they propose that when an honest miner is presented with two chains of the same length, they  choose which one to mine on uniformly at random. With this change, the lower bound on computational power needed to selfish mine increases, thus making it harder to act subversively. While this was conjectured to be true and showed to be so with simulation, there are contradictory results. In \cite{sapirshtein2016optimal} the authors show that while this change does limit the strength of large selfish miners, it enhances the strength of medium sized selfish miners and that selfish miners with computational power less than 25\% can still gain from acting subversively.

\section{Model Assumptions and Notation}

The decentralised design of Bitcoin consists of clients: users of Bitcoin, who own accounts designated by addresses. A client can send Bitcoin from an address he owns to an arbitrary address by broadcasting a transaction to the Bitcoin P2P network. This transaction will eventually be appended to a a global ledger called the Blockchain. The upkeep of the Blockchain is performed by miners, who collect transactions in blocks and append these blocks to the chain. For this task, miners are rewarded with Bitcoin, either in the form of a block reward or transaction fees.

We model the Blockchain system as a set of $M$ strategic miners, $m_1,...,m_M$, and an implicit honest miner $m_{M+1}$. Each strategic miner $m_i$, controls an $\alpha_i \in (0,0.5]$ portion of the system hash power (we don't consider strategic miners strong enough to perform a 51 percent attack), and the honest miner $m_{M+1}$ controls a $\beta = 1 - \sum_{i=1}^M \alpha_i > 0$ portion of the system hash power. The implicit honest miner is without loss of generality for if any number of miners (beyond the strategic miners $m_1,...,m_M$) employ honest mining, this is equivalent to one miner of their combined hash power employing honest mining. For convenience we denote the set of valid strategic miner hash rates by $\mathcal{H}^M = \{\alpha \in (0,0.5]^M \ | \ \sum_{i=1}^M \alpha_i < 1 \}$.

Given strategic miner hash rates $\alpha \in \mathcal{H}^M$, any found block has an $\alpha_i$ probability of being found by the $i$-th strategic miner $m_i$, and a $\beta$ probability of being found by $m_{M+1}$. We also assume that the system overall finds blocks at a rate of $\lambda$ according to a Poisson process. In terms of the actual implementation of the Bitcoin protocol, $\lambda$ is roughly one block every 10 minutes, which is ensured by dynamically adjusting the difficulty of the block hash target.

The append-only nature of the block renders the Blockchain into a tree with a root at the genesis block. Since the longest path of the tree is the agreed-upon transaction history, a miner's revenue consists of his block rewards and transaction fees arising from blocks that eventually become a part of the longest path in the Blockchain. In this paper we focus on block rewards and normalise such rewards to unit value, hence the revenue of a miner is the number of his blocks that are accepted in the longest path of the blockchain. 

Indeed it could be the case that a longest path in the blockchain is eventually surpassed by a competing path: this is a key aspect to selfish mining strategies. This of course makes it difficult to ascertain revenues when miners are arbitrary agents. In our paper however we pit specific mining strategies against each other and hence obtain a well-defined block creation rates for all agents involved. Furthermore, we assume that agents are rational and that the utility they wish to maximise is their {\em relative revenue}: which is for a miner $m_i$ is the expected number of blocks $m_i$ publishes in the blockchain normalised by the expected number of blocks produced by all miners $m_1,...,m_{M+1}$. The justification behind this utility function comes from the fact that Bitcoin dynamically adjusts its difficulty, hence relative revenue in the long-term corresponds to overall revenue.

\section{Miner Strategies}

Mining strategies are often defined with an implicit assumption that a miner following the strategy will be pitted against miners employing a specific strategy (i.e. honest mining). Since our paper focuses on miner incentives when multiple miners deviate from honest mining, we find ourselves in need of rigorously defining miner strategies with respect to all possible changes in the blockchain, not just those changes that can occur against a specific kind of miner.

In this vein, we formally describe three specific mining strategies: honest mining, selfish mining, and semi-selfish mining. We describe the strategies for an arbitrary miner denoted by $m$. 

To execute these strategies, $m$ must keep track of their private chains, the public chain, a block upon which to mine and an internal state $\ell \in \{0,0'\} \cup \mathbb{N}$. As for additional notation, $priv$ denotes the private chain of $m$, $pub$ denotes the public chain and $F$ (frontier) denotes the set of blocks at the ends of the longest paths of the public chain. Arbitrary blocks are usually denoted by $B$. We also let $len(priv)$ and $len(pub)$ denote the length of the longest path in the miner's private chain and the length of the longest path of the public chain respectively. For a given set of a blocks $S$, we let $oldest(S)$ denote the oldest block in $S$ of which $m$ was aware. Finally, we let $end(priv)$ be the block at the end of the miner's private chain and $p(m)$ be the block upon which $m$ is mining.

The integer of the internal state, $\ell$ represents a miner's ``lead'': how much longer the miner's private chain is than the public chain. For all three mining strategies states 0 and 0' will not only mean that the miner has no lead with respect to the public chain, but that the miner's private chain is in fact the public chain (a fact which follows from the rules governing the strategies). Finally, the difference between 0 and 0' is that the latter state occurs when there is a tie on the public chain, i.e. $|F|>1$. The choices available to miners are where to mine, $p(m)$, and whether to reveal parts of their private chain.

\subsection{Honest Mining}

Honest miners are those who follow the prescribed Bitcoin mining protocol faithfully. We describe the strategy in terms of what actions $m$ takes when in states $\ell \in \{0, 0'\}$:

\begin{itemize}
    \item Case 1: $m$ finds a block, $B$.
        \begin{itemize}
            \item $m$ publishes $B$.
            \item $\ell \leftarrow 0$.
            \item $p(m) \leftarrow B$.
        \end{itemize}
    \item Case 2: $pub$ changes to $pub'$ with frontier $F'$.
        \begin{itemize}
            \item If $|F'| = 1$, then $\ell \leftarrow 0$.
            \item If $|F'| > 1$, then $\ell \leftarrow 0'$.
            \item $p(m) \leftarrow oldest(F')$.
        \end{itemize}
\end{itemize}
\noindent
It is straightforward to check that if all miners mine honestly, their expected relative revenue is precisely their hash rate:
\begin{lemma}
For any $\alpha \in \mathcal{H}^M$, if all strategic miners are honest, the expected (block) reward of any strategic miner $m_i$ is $\alpha_i$ (and $\beta$ for the extra honest miner $m_M$).
\end{lemma}

\subsection{Selfish Mining}
Eyal and Sirer introduced Selfish Mining (SM) in \cite{eyal2015miner} as a specific strategy that outperforms honest mining when a rational agent has sufficient computational resources. SM can be described by the actions $m$ takes in the following states:

\subsubsection*{$\ell = 0$ and $p(m) = oldest(F)$}

\begin{itemize}
    \item Case 1: $m$ finds a block: $B$.
    \begin{itemize}
        \item $m$ keeps $B$ private.
        \item $\ell \leftarrow 1$.
        \item $p(m) \leftarrow B$.
    \end{itemize}
    \item Case 2: $pub$ changes to $pub'$ with frontier $F'$
    \begin{itemize}
        \item If $|F'| = 1$, then $\ell \leftarrow 0$.
        \item If $|F'| > 1$, then $\ell \leftarrow 0'$.
        \item $p(m) \leftarrow oldest(F')$
    \end{itemize}
\end{itemize}

\subsubsection*{$\ell = 0'$ and $p(m) = oldest(F)$}

\begin{itemize}
    \item Case 1: $m$ finds a block: $B$.
        \begin{itemize}
            \item $m$ publishes $B$.
            \item $\ell \leftarrow 0$.
            \item $p(m) \leftarrow B$
        \end{itemize}
    \item Case 2: $pub$ changes to $pub'$ with frontier $F'$.
        \begin{itemize}
            \item If $|F'| = 1$, then $\ell \leftarrow 0$.
            \item If $|F'| > 1$, then $\ell \leftarrow 0'$.
            \item $p(m) \leftarrow oldest(F')$
        \end{itemize}
\end{itemize}

\subsubsection*{$\ell \geq 1$ and $p(m) = end(priv)$}

\begin{itemize}
    \item Case 1: $m$ finds a block: $B$.
    \begin{itemize}
        \item $m$ keeps $B$ private.
        \item $\ell \leftarrow \ell + 1$.
        \item $p(m) \leftarrow B$.
    \end{itemize}
    \item Case 2: $pub$ changes to $pub'$ with frontier $F'$ and $k = len(pub') - len(pub) \leq max(\ell-2, 0)$.
    \begin{itemize}
        \item $m$ publishes $k$-prefix of $priv$.
        \item $\ell \leftarrow \ell - k$.
        \item $p(m) \leftarrow p(m)$.
    \end{itemize}
    \item Case 3: $pub$ changes to $pub'$ with frontier $F'$ and $k = len(pub') - len(pub) > max(\ell-2, 0)$.
    \begin{itemize}
        \item $m$ publishes $priv$, resulting in $pub''$ with frontier $F''$
        \item If $|F''| = 1$, then $\ell \leftarrow 0$.
        \item If $|F''| > 1$, then $\ell \leftarrow 0'$.
        \item $p(m) \leftarrow oldest(F'')$.
    \end{itemize}
\end{itemize}

At a glance, this characterisation of SM may look different to how it is usually described. Upon closer inspection however, one can see that this is equivalent to what was presented in \cite{eyal2015miner}. In particular, the fact that in state $\ell = 0'$, $p(m) = oldest(F)$, means that when a tie involves a block mined by $m$ (as would be the case if they had published a previously private block), they will indeed continue mining upon it, as they will have necessarily seen it first amongst blocks in $F$.

\begin{figure}[h]
    \centering
    
\begin{tikzpicture}[square/.style={regular polygon,regular polygon sides=4},scale=0.5, every node/.style={scale=0.5}]

    \node at (-15.5,-1.5) [circle, draw] (m) {$m$};
    
    \node at (-18,-3) [square, draw] (G) {*};
    \node at (-15.5,-3) [square, draw] (A) {*};
    \draw  (G) -- (A);
    \draw [->] (m) -- (A);

    \draw[->, line width=1.5mm] (-13.5,-2) -- (-12,-0.5);

    \node at (-5,3.5) [circle, draw] (m) {$m$};    

    \node at (-10,0) [square, draw] (G) {*};
    \node at (-7.5,0) [square, draw] (A) {*};
    \node at (-5,2) [square, draw] (1) {$M$};
    \draw  (G) -- (A);
    \draw [dashed] (A) -- (1);
    \draw [->] (m) -- (1);

    \draw[->, line width=1.5mm] (-4,0) -- (-2,0);
    
    \node at (5,3.5) [circle, draw] (m) {$m$};
    
    \node at (0,0) [square, draw] (G) {*};
    \node at (2.5,0) [square, draw] (A) {*};
    \node at (5,2) [square, draw] (1) {$M$};
    \node at (5,-2) [square, draw] (H) {*};
    \draw  (G) -- (A);
    \draw [dashed] (A) -- (1);
    \draw (A) -- (H);
    \draw [->] (m) -- (1);
    
    \draw[->, line width=1.5mm] (6,-0.5) -- (7.5,-2);

    \node at (10,-1.5) [circle, draw] (m) {$m$};
    
    \node at (5,-5) [square, draw] (G) {*};
    \node at (7.5,-5) [square, draw] (A) {*};
    \node at (10,-3) [square, draw] (1) {$M$};
    \node at (10,-7) [square, draw] (H) {*};
    \draw (G) -- (A);
    \draw (A) -- (1);
    \draw (A) -- (H);
    \draw [->] (m) -- (1);

\end{tikzpicture}

\vspace{2mm}

\begin{tikzpicture}[square/.style={regular polygon,regular polygon sides=4},scale=0.465, every node/.style={scale=0.5}]

    \node at (-5,3.5) [circle, draw] (pm) {$m$};    

    \node at (-17.5,0) [square, draw] (1) {*};
    \node at (-15,0) [square, draw] (2) {*};
    \node at (-12.5,2) [square, draw] (3) {$M$};
    \node at (-10,2) [square, draw] (4) {$M$};
    \node at (-7.5,2) [square, draw] (5) {$M$};
    
    \node at (-12.5,-2) [square, draw] (l1) {*};
    \node at (-10,-2) [square, draw] (l2) {*};

    \node at (-5,2) [square, draw] (m) {$M$};
    \draw (1) -- (2);
    \draw [dashed] (2) -- (3);
    \draw [dashed] (3) -- (4);
    \draw [dashed] (4) -- (5);
    \draw [dashed] (5) -- (m);
    \draw [->] (pm) -- (m);

    \draw  (2) -- (l1);
    \draw  (l1) -- (l2);

    \draw[->, line width=1.5mm] (-3.5,0) -- (-1.5,0);
    
    \node at (12.5,3.5) [circle, draw] (pm) {$m$};    

    \node at (0,0) [square, draw] (1) {*};
    \node at (2.5,0) [square, draw] (2) {*};
    \node at (5.5,2) [square, draw] (3) {$M$};
    \node at (7.5,2) [square, draw] (4) {$M$};
    \node at (10,2) [square, draw] (5) {$M$};
    \node at (12.5,2) [square, draw] (m) {$M$};
    
    \node at (5.5,-2) [square, draw] (l1) {*};
    \node at (7.5,-2) [square, draw] (l2) {*};
    
    \draw (1) -- (2);
    \draw (2) -- (3);
    \draw (3) -- (4);
    \draw [dashed] (4) -- (5);
    \draw [dashed] (5) -- (m);
    \draw [->] (pm) -- (m);    

    \draw  (2) -- (l1);
    \draw  (l1) -- (l2);

\end{tikzpicture}

\vspace{2mm}

\begin{tikzpicture}[square/.style={regular polygon,regular polygon sides=4},scale=0.5, every node/.style={scale=0.5}]

    \node at (-7.5,3.5) [circle, draw] (pm) {$m$};    

    \node at (-17.5,0) [square, draw] (1) {*};
    \node at (-15,0) [square, draw] (2) {*};
    \node at (-12.5,2) [square, draw] (3) {$M$};
    \node at (-10,2) [square, draw] (4) {$M$};
    \node at (-7.5,2) [square, draw] (m) {$M$};
    
    \node at (-12.5,-2) [square, draw] (l1) {*};
    \node at (-10,-2) [square, draw] (l2) {*};

    \draw (1) -- (2);
    \draw [dashed] (2) -- (3);
    \draw [dashed] (3) -- (4);
    \draw [dashed] (4) -- (m);
    \draw [->] (pm) -- (m);

    \draw  (2) -- (l1);
    \draw  (l1) -- (l2);

    \draw[->, line width=1.5mm] (-4.5,0) -- (-2.5,0);
    
    \node at (10,3.5) [circle, draw] (pm) {$m$};    

    \node at (0,0) [square, draw] (1) {*};
    \node at (2.5,0) [square, draw] (2) {*};
    \node at (5,2) [square, draw] (3) {$M$};
    \node at (7.5,2) [square, draw] (4) {$M$};
    \node at (10,2) [square, draw] (m) {$M$};
    
    \node at (5,-2) [square, draw] (l1) {*};
    \node at (7.5,-2) [square, draw] (l2) {*};
    
    \draw (1) -- (2);
    \draw (2) -- (3);
    \draw (3) -- (4);
    \draw (4) -- (m);
    \draw [->] (pm) -- (m);    

    \draw  (2) -- (l1);
    \draw  (l1) -- (l2);

    \end{tikzpicture}

    \caption{SM Dynamics. A square with $M$ is a block mined by $m$. The circle with $m$ represents the value of $p(m)$. A solid line means that portion of the chain is public, and a dashed line means that portion of the chain is private. If a miner employing SM has a lead of $\ell = 1$ that is diminished, he publishes his private chain and hopes to win the tie (Top). If the miner has a larger lead that is partially encroached, he publishes a prefix of his private chain to push other miners into a race (Middle). If a miner of lead $\ell > 1$ sees his lead encroached to $\ell = 1$, he publishes all blocks to overtake (Bottom).}
    \label{fig:SM_drawing3}
\end{figure}
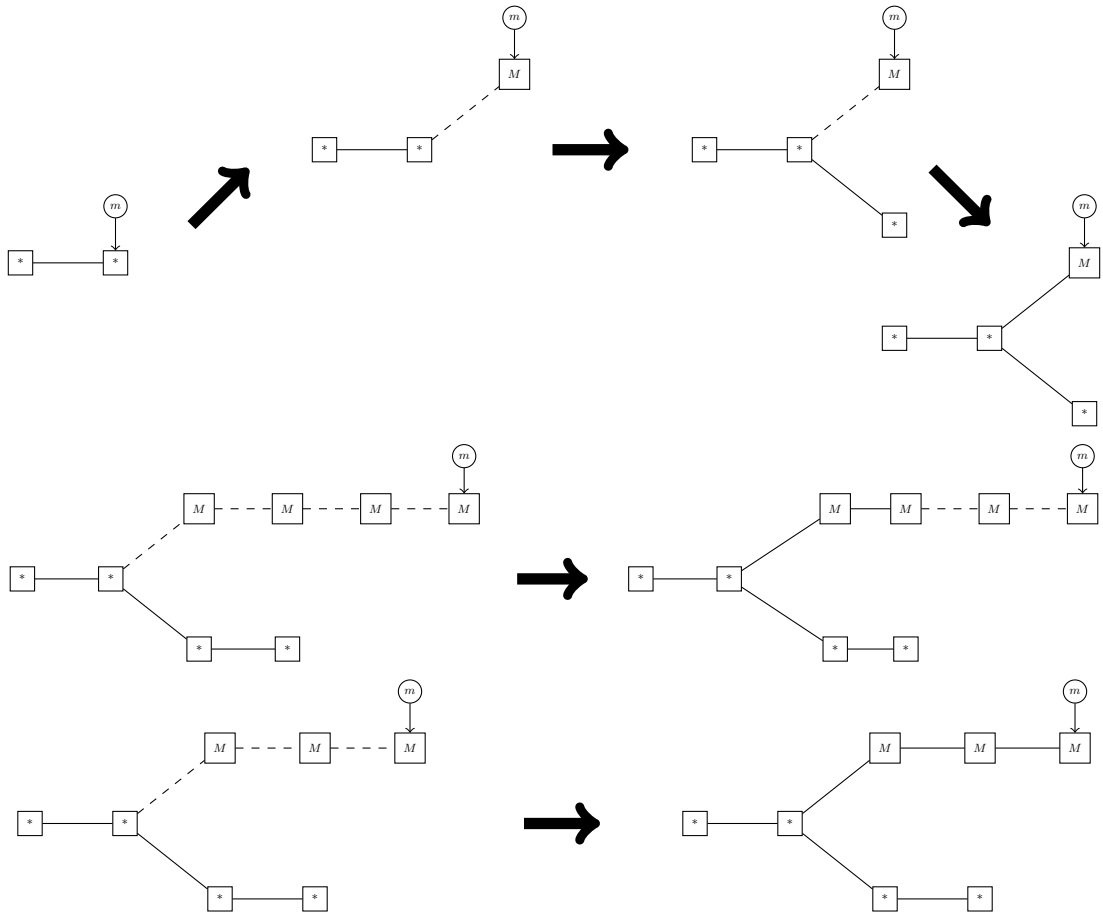

\subsection{Semi-Selfish Mining}

SM can be generalised to a class of strategies where a miner maintains a private chain and has the following actions at hand: publishing a portion of his private chain, mining upon his private chain, and foregoing his private chain to mine upon the public chain. Indeed, this general class of selfish mining strategies is studied in \cite{NKMS16} and \cite{SSZ16}.

We focus on the simplest selfish mining strategies from this family by looking at strategies where the selfish miner never maintains a private chain of length greater than 2. Notice that this is necessary if the selfish miner is to gain any benefit from selfish mining, for if the miner only maintains at most one private block, he can only hurt his chances of having this block (and hence any block) published when facing honest miners. On the other hand, for private chains of length 2, we exhibit a specific strategy Semi-Selfish Mining (SSM) that much like the original SM strategy, leads to increased revenue ratios for the selfish miner if they have sufficient hash power. 

The reason we study such a simple strategy from the rich space of selfish mining strategies is that it still obtains higher relative revenues than honest mining in certain parameter regimes, yet it has a much simpler state space than most selfish mining strategies. This reduced state space will eventually allow us to explicitly solve for expected relative revenues when two selfish miners play against each other. SSM can be described by the actions $m$ takes in the following states:

\subsubsection*{$\ell = 0$ and $p(m) = oldest(F)$}

\begin{itemize}
    \item Case 1: $m$ finds a block: $B$.
    \begin{itemize}
        \item $m$ keeps $B$ private.
        \item $\ell \leftarrow 1$.
        \item $p(m) \leftarrow B$.
    \end{itemize}
    \item Case 2: $pub$ changes to $pub'$ with frontier $F'$
    \begin{itemize}
        \item If $|F'| = 1$, then $\ell \leftarrow 0$.
        \item If $|F'| > 1$, then $\ell \leftarrow 0'$.
        \item $p(m) \leftarrow oldest(F')$
    \end{itemize}
\end{itemize}

\subsubsection*{$\ell = 0'$ and $p(m) = oldest(F)$}

\begin{itemize}
    \item Case 1: $m$ finds a block: $B$.
        \begin{itemize}
            \item $m$ publishes $B$.
            \item $\ell \leftarrow 0$.
            \item $p(m) \leftarrow B$
        \end{itemize}
    \item Case 2: $pub$ changes to $pub'$ with frontier $F'$.
        \begin{itemize}
            \item If $|F'| = 1$, then $\ell \leftarrow 0$.
            \item If $|F'| > 1$, then $\ell \leftarrow 0'$.
            \item $p(m) \leftarrow oldest(F')$
        \end{itemize}
\end{itemize}

\subsubsection*{$\ell = 1$ and $p(m) = end(priv)$}

\begin{itemize}
    \item Case 1: $m$ finds a block: $B$.
    \begin{itemize}
        \item $m$ keeps $B$ private.
        \item $\ell \leftarrow 2$.
        \item $p(m) \leftarrow B$.
    \end{itemize}
    \item Case 2: $pub$ changes to $pub'$ with frontier $F'$ and $k = len(pub') - len(pub) = 0$.
    \begin{itemize}
        \item $m$ does nothing.
    \end{itemize}
    \item Case 3: $pub$ changes to $pub'$ with frontier $F'$ and $k = len(pub') - len(pub) > 0$.
    \begin{itemize}
        \item $m$ publishes $priv$, resulting in $pub''$ with frontier $F''$
        \item If $|F''| = 1$, then $\ell \leftarrow 0$.
        \item If $|F''| > 1$, then $\ell \leftarrow 0'$.
        \item $p(m) \leftarrow oldest(F'')$.
    \end{itemize}
\end{itemize}

\subsubsection*{$\ell = 2$ and $p(m) = end(priv)$}

\begin{itemize}
    \item Case 1: $m$ finds a block: $B$.
    \begin{itemize}
        \item $m$ publishes $oldest(priv \setminus pub)$.
        \item $\ell \leftarrow 2$.
        \item $p(m) \leftarrow B$.
    \end{itemize}
    \item Case 2: $pub$ changes to $pub'$ with frontier $F'$ and $k = len(pub') - len(pub) = 0$.
    \begin{itemize}
        \item $m$ does nothing.
    \end{itemize}
    \item Case 3: $pub$ changes to $pub'$ with frontier $F'$ and $k = len(pub') - len(pub) > 0$.
    \begin{itemize}
        \item $m$ publishes $priv$, resulting in $pub''$ with frontier $F''$
        \item If $|F''| = 1$, then $\ell \leftarrow 0$.
        \item If $|F''| > 1$, then $\ell \leftarrow 0'$.
        \item $p(m) \leftarrow oldest(F'')$.
    \end{itemize}
\end{itemize}

\begin{figure}[h]
    \centering
    
\begin{tikzpicture}[square/.style={regular polygon,regular polygon sides=4},scale=0.365, every node/.style={scale=0.5}]

    \node at (-20,3.5) [circle, draw] (pm) {$m$};    

    \node at (-27.5,0) [square, draw] (1) {*};
    \node at (-25,0) [square, draw] (2) {*};
    \node at (-22.5,2) [square, draw] (3) {$M$};
    \node at (-20,2) [square, draw] (m) {$M$};
    

    \draw (1) -- (2);
    \draw [dashed] (2) -- (3);
    \draw [dashed] (3) -- (m);
    \draw [->] (pm) -- (m);


    \draw[->, line width=1.5mm] (-18,2) -- (-16,4);

    \node at (-5,9.5) [circle, draw] (pm) {$m$};    

    \node at (-15,6) [square, draw] (1) {*};
    \node at (-12.5,6) [square, draw] (2) {*};
    \node at (-10,8) [square, draw] (3) {$M$};
    \node at (-7.5,8) [square, draw] (4) {$M$};
    \node at (-5,8) [square, draw] (m) {$M$};
    

    \draw (1) -- (2);
    \draw [dashed] (2) -- (3);
    \draw [dashed] (3) -- (4);
    \draw [dashed] (4) -- (m);
    \draw [->] (pm) -- (m);


    \draw[->, line width=1.5mm] (-3,4) -- (-1,2);
    
    \node at (10,3.5) [circle, draw] (pm) {$m$};    

    \node at (0,0) [square, draw] (1) {*};
    \node at (2.5,0) [square, draw] (2) {*};
    \node at (5,2) [square, draw] (3) {$M$};
    \node at (7.5,2) [square, draw] (4) {$M$};
    \node at (10,2) [square, draw] (m) {$M$};
    
    
    \draw (1) -- (2);
    \draw (2) -- (3);
    \draw [dashed] (3) -- (4);
    \draw [dashed] (4) -- (m);
    \draw [->] (pm) -- (m);    


    \end{tikzpicture}

    \caption{SSM as a truncation of SM. Once again, a square with $M$ is a block mined by $m$. The circle with $m$ represents the value of $p(m)$. A solid line means that portion of the chain is public, and a dashed line means that portion of the chain is private. Here $m$ has a lead of $\ell = 2$ and upon mining a block, publishes his oldest private block.}
    \label{fig:SSM_drawing1}
\end{figure}
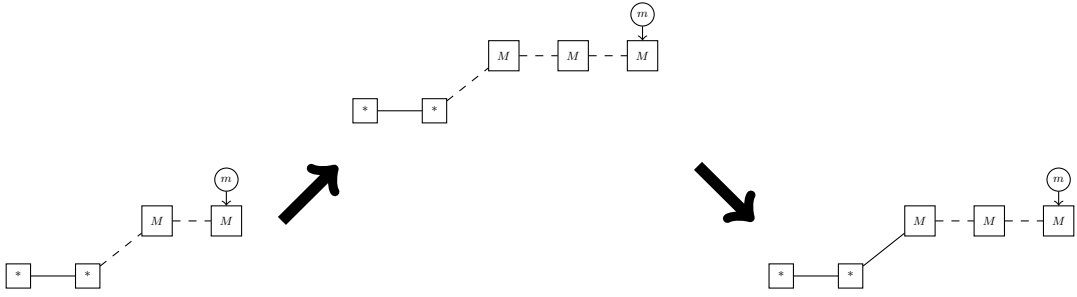

\section{One Strategic Miner}
\label{sec:one-SSM}

We begin by studying how one strategic miner of hash power $\alpha \in \mathcal{H}^1 = (0,0.5]$ performs against honest miners of hash power $\beta = 1 - \alpha$ in terms of relative revenue. As in \cite{eyal2015miner}, we let $\gamma$ be the proportion of honest miners who mine upon an SSM chain in the case of a tie, a parameter which we call the propagation of the strategic miner. In what follows, we let $r_{SM}$ and $r_{SSM}$ be the expected block creation rate of a single miner using SM and SSM respectively against honest miners. Consequently, we let $r_{others}$ be the block creation of other honest miners in the system (this is dependant upon whether SM or SSM is used, but we use the same term for the sake of simplicity). Finally, we let $R_{SM}$ and $R_{SSM}$ denote the relative revenues of a single miner using SM and SSM respectively against honest miners.

\begin{theorem}[Selfish Mining Relative Revenue \cite{eyal2015miner}]
A single strategic miner of hash power $\alpha$ and propagation $\gamma$, attains the following revenue ratio using SM against honest miners:
$$
R_{SM} = \frac{r_{SM}}{r_{SM} + r_{others} } =
\frac{\alpha(1-\alpha)^2(4\alpha + \gamma(1-2\alpha)) - \alpha^3}{1 - \alpha(1 + (2-\alpha)\alpha)}
$$
Asymptotically around $\alpha = 0$ the expression is the following:
$$
R_{SM} = \alpha\gamma + \alpha^2(4-3\gamma) + \alpha^3(4\gamma - 5) + \alpha^4 (7 - 5\gamma) + \alpha^5(6\gamma - 7) + O(\alpha^6)
$$
\end{theorem}

We can use similar Markov chain methods to derive the revenue ratio of SSM against honest miners. The details of the analysis can be found in Appendix \ref{appendix:ssm-vs-honest}. 

\begin{theorem}
A strategic miner of hash power $\alpha$ and propagation $\gamma$, attains the following revenue ratio when using SSM against honest miners: 
$$
R_{SSM} = \frac{r_{SSM}}{r_{SSM} + r_{others}} = 
\frac{\alpha(\alpha(\alpha(2\alpha-5) + 4) - (\alpha-1)^3\gamma)}{(\alpha-1)\alpha^2 + 1}
$$
Asymptotically around $\alpha = 0$, the expression is the following:
$$
R_{SSM} = \alpha\gamma + \alpha^2(4 - 3\gamma) + \alpha^3(4\gamma - 5) - \alpha^4(6 - 5\gamma) + \alpha^5(7\gamma - 9) + O(\alpha^6)
$$
\end{theorem}

\subsection{Comparing Performance of SM and SSM}

\begin{figure}

\begin{center}
\includegraphics[width=.3\textwidth]{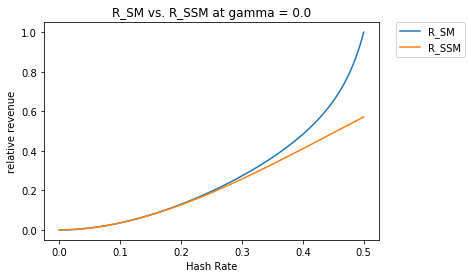}
\includegraphics[width=.3\textwidth]{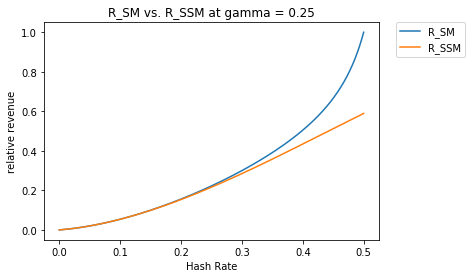}
\includegraphics[width=.3\textwidth]{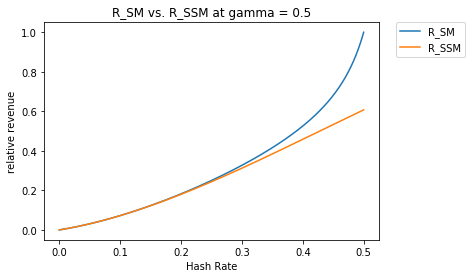}
\end{center}
\caption{$R_{SM}$ and $R_{SSM}$ against honest miners at $\gamma = 0$, $0.25$ and $0.5$}
\label{fig:SSM-vs-honest}
\end{figure}

Asymptotically SM and SSM have the same performance as $\alpha \rightarrow 0$. In fact $R_{SM} - R_{SSM} = O(\alpha^4)$. For all parameter settings SM outperforms SSM, as evidenced in the graphs in Figure \ref{fig:SSM-vs-honest}. At $\gamma = 0$ SM becomes profitable at $\alpha = 1/3$ and SSM becomes profitable at $\alpha = 0.38$. At $\gamma = 0.25$ SM becomes profitable at $\alpha = 0.3$ and SSM becomes profitable at $\alpha = 1/3$. Finally, at $\gamma = 0.5$ SM becomes profitable at $\alpha = 1/4$ and SSM becomes profitable at $\alpha = 0.26795$

\section{Two Strategic Miners}
\label{sec:two-SSM}

The benefit of SSM lies in the fact that it can be a rational strategy distinct from honest mining and more importantly,  describing it in terms of a Markov chain does not require many states. The simplicity of the state space allows us to explore the scenario where two agents of different hash rates employ SSM and analytically solve for relative revenues.

\subsection{Markov Chain Analysis}

Suppose that $\alpha = (\alpha_1,\alpha_2) \in \mathcal{H}^2$ is the strategic hash rate of the system. Since we have two strategic miners, our state space, $S$, consists of nine states of the form $S_{i,j}$ where $0 \leq i,j \leq 2$. These represent the relative lead SSM miners 1 and 2 have with respect to the public chain. Given our description of SSM we can describe the state transitions in the same way as we did for the single SSM case. Both of these can be found in Appendix \ref{appendix:M-SSM}.

\subsection{Transition Matrix and Steady State}

The above state space gives rise to an ergodic Markov chain, so there is a unique stationary distribution we can solve for. In order to do so, we define the following transition matrix, $P$, on $\mathbb{R}^9$, where the coordinate axes of $\mathbb{R}^9$ (in ascending order) represent probability mass in states $S_{0,0}, S_{0,1}, S_{1,0}, S_{0,2}, S_{1,1}, S_{2,0}, S_{1,2}, S_{2,1}$, and $S_{2,2}$ respectively. Each $P_{x,y}$ is the probability of transitioning to state $x$ from state $y$ in the Markov chain. 
\[
P=
\begin{bmatrix}
\beta & \beta & \beta & \beta & \beta & \beta & \alpha_2(1-\alpha_2) + \beta \ & \alpha_1(1-\alpha_1) + \beta \ & 1 \\
\alpha_2 & 0 & 0 & 0 & 0 & 0 & 0 & 0 & 0 \\
\alpha_1 & 0 & 0 & 0 & 0 & 0 & 0 & 0 & 0 \\
0 & \alpha_2 & 0 & \alpha_2 & 0 & 0 & \alpha_2^2 & 0 & 0 \\
0 & \alpha_1 & \alpha_2 & 0 & 0 & 0 & 0 & 0 & 0 \\
0 & 0 & \alpha_1 & 0 & 0 & \alpha_1 & 0 & \alpha_1^2 & 0 \\
0 & 0 & 0 & \alpha_1 & \alpha_2 & 0 & 0 & 0 & 0 \\
0 & 0 & 0 & 0 & \alpha_1 & \alpha_2 & 0 & 0 & 0 \\
0 & 0 & 0 & 0 & 0 & 0 & \alpha_1 & \alpha_2 & 0 
\end{bmatrix}
\]

Since this is an ergodic chain, there is a unique steady state distribution, $\pi$, such that $P \pi = \pi$, which we can solve for with Gaussian elimination. 

\subsection{Propagation and Revenues}

In the original selfish mining paper, much attention was given to the propagation parameter $\gamma$. Indeed block dissemination is important because it allows an attacker to persuade other miners to work on their chain in the case of a tie. We also note from the previous section that the steady state distribution $\pi$ is independent of the propagation of the system. The expected number of blocks published per state however, crucially depends on the propagation of the system, and these two objects specify the relative revenue of agents. 

In our work, when there is a single strategic miner employing SSM, the ability to propogate blocks is parametrised by $\gamma$ as in the original analysis of SM. When there are strategic miners employing SSM however, how propagation is modelled becomes more complicated, since different strategic miners may have a different influence on the P2P network topology. For the rest of the paper we assume that propagation is uniform. In other words, whenever there is a tie in the public chain (of arbitrary size), all miners not involved in the tie are assumed to have a uniformly random chance of contributing their hash power to any element of the tie. Under the assumption of uniform propagation, we can compute the expected block rate per state of the Markov chain for both strategic miners and honest miners. The following matrix $R$ encodes this information: the first and second row are expected block rates per state for the first and second strategic miners respectively, the third column is the block creation rate for honest miners. If $\pi$ is a steady state vector for $M$ above, then $R^T \pi \in \mathbb{R}^3$ gives steady state expected block creation rates for all miners.

\[
R=
\begin{bmatrix}
0 & 0 & \beta  \\
\beta \alpha_1 & 2\beta \alpha_2 + \frac{1}{2}\beta (1- \alpha_2) & \frac{1}{2} \beta \alpha_1 + \frac{3}{2}\beta^2  \\
2\beta \alpha_1 + \frac{1}{2}\beta (1- \alpha_1) \ & \beta \alpha_2 & \frac{1}{2} \beta \alpha_2 + \frac{3}{2}\beta^2 \\
0 & \alpha_2 + 2\beta & 0 \\
2\beta \alpha_1 + \frac{1}{3}\beta^2 & 2\beta \alpha_2 + \frac{1}{3}\beta^2 & \frac{4}{3}\beta^2 \\
\alpha_1 + 2\beta & 0 & 0 \\
0 & 2\beta + 2\alpha_2^2 + 3\alpha_2(1-\alpha_2) & 0 \\
2\beta + 2\alpha_1^2 + 3\alpha_1(1-\alpha_1) & 0 & 0 \\
3\alpha_1 + 3\beta \alpha_1 + \frac{1}{2}\beta^2 & 3\alpha_2 + 3\beta \alpha_2 + \frac{1}{2}\beta^2 & \beta^2  
\end{bmatrix}
\]

For the sake of completeness, in Appendix \ref{appendix:M-SSM} we include a model for different propagation rates when two strategic miners are involved as well as their effects on relative revenues of all miners.

\subsection{To SSM or not to SSM? A Revenue Analysis}

Although our Markov chain analysis gives us a closed-form solution for the relative revenue of both strategic miners when using SSM, the expression is unwieldy. We can however explicitly solve the expression for specific hash values, $\alpha_1$ and $\alpha_2$ and use these values to describe a two-player, binary action game governing the decision as to whether a player employs SSM or not. 

Suppose that $\alpha = (\alpha_1, \alpha_2) \in \mathcal{H}^2$ describes the hash rates of both strategic miners. We let $R_{SSM}(\alpha) = R_{SSM}((\alpha_1,\alpha_2)) \in [0,1]^3$ be the revenue ratios of all miners (including the honest miner $m_3$) when both strategic miners employ SSM. Specifically, $R_{SSM}(\alpha)_i$ is the revenue ratio of $m_i$ for $i = 1,2,3$. With this in place we can define a two-player binary action game governing the incentives behind employing SSM or not for $m_1$ and $m_2$. 

\begin{definition}[Two-player SSM Games]
Suppose that $\alpha = (\alpha_1,\alpha_2) \in \mathcal{H}^2$ is a strategic hash distribution. We define the {\em SSM Game}, $G_\alpha$ as a two-player binary action game. In $G_\alpha$ each strategic miner has a binary action set $\{H,S\} \cong \{0,1\}$, where $H \cong 0$ represents mining honestly and $S \cong 1$ represents employing SSM. We define the utilities of all pure strategy profiles as follows:
\begin{itemize}
\item $U_1(H,H) = \alpha_1$, $U_2(H,H) = \alpha_2$
\item $U_1(H,S) = \frac{\alpha_1}{1- \alpha_2}R_{SSM}((0,\alpha_2))_3$, $U_2(H,S) = R_{SSM}((0,\alpha_2))_2$
\item $U_1(S,H) = R_{SSM}((\alpha_1,0))_1$, $U_2(S,H) = \frac{\alpha_2}{1 - \alpha_1}R_{SSM}((\alpha_1,0))_3$
\item $U_1(S,S) = R_{SSM}(\alpha)_1$, $U_2(S,S) = R_{SSM}(\alpha)_2$
\end{itemize}
For notational convenience, we interchangeably denote a pure strategy profile of all players by either a tuple, as in $(H,H)$ for both miners employing honest mining, or a string, as in $HH$
\end{definition}

\begin{figure}
\centering
\includegraphics[width = .42\textwidth]{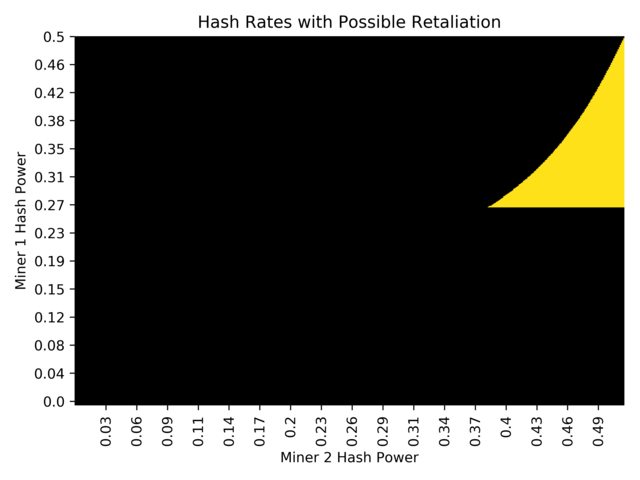}
\includegraphics[width = 0.42\textwidth]{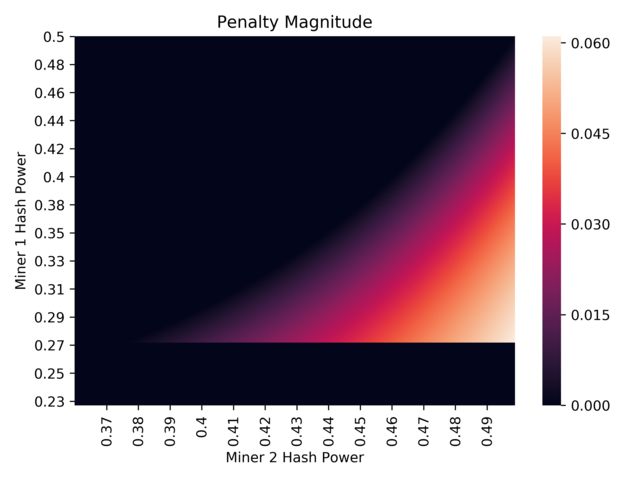}
\caption{Hash rates where $U_1(S,S) < U_1(H,H) < U_1(S,H)$ and subsequent penalty values given by $U_1(S,S) - U_1(H,H)$.}
\label{fig:2SSM-incentive-types}
\end{figure}

As a first region of interest, in Figure \ref{fig:2SSM-incentive-types} we display hash rates where $U_1(S,S) < U_1(H,H) < U_1(S,H)$. For such $\alpha$, although SSM may be unilaterally rational for the first strategic miner, a larger miner can penalise the first strategic miner for deviating from the honest protocol by retaliating with SSM. As a specific example of this phenomenon, let us consider the hash distribution $\alpha = (0.33, 0.48)$ which leads to $G_\alpha$ with utilities summarised in Table \ref{tab:2miner-retaliation}. The second, larger, strategic miner $m_2$ can retaliate from $SH$ by deviating to $SS$, in which case $m_1$ is worse off by approximately $0.04$ in utility than if he had mined honestly at the outset.

\begin{table}[H]
\begin{center}
\caption {Example of $G_\alpha$ where $m_2$ can retaliate against SH} \label{tab:2miner-retaliation} 
\begin{tabular}{||c | c | c | c | c||} 
\hline
\ $\alpha = (0.33, 0.48)$ \ &  HH  &  HS  &  SH  &  SS  \\ [0.5ex] 
\hline\hline
$U_1$  & \ 0.33 \  & \ 0.26794954 \ & \ 0.35517387 \ & \ 0.29387121 \ \\ 
\hline
$U_2$ & 0.48 & 0.57777649 & 0.46196499 & 0.61890781 \\ [1ex] 
\hline
\end{tabular}
\end{center}
\end{table}

\begin{figure}
\centering
\includegraphics[width = .46\textwidth]{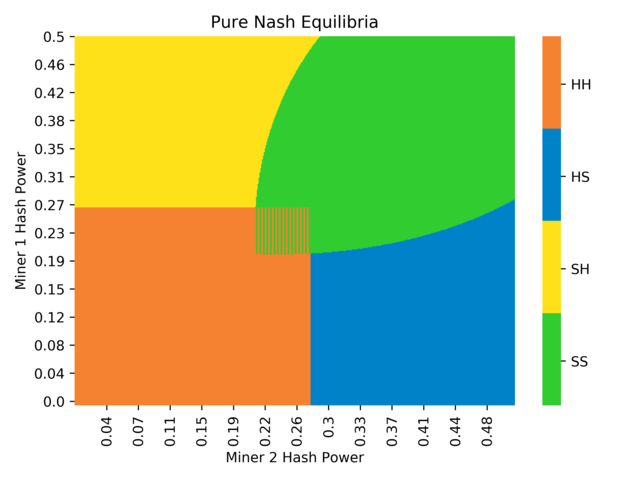}
\includegraphics[width = .46\textwidth]{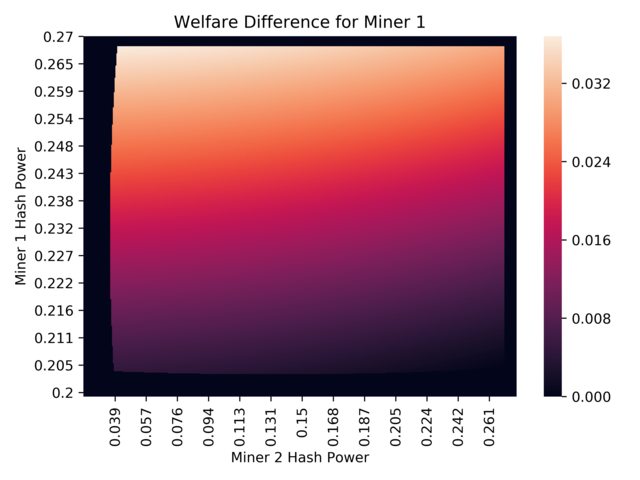}
\caption{PNE types and the welfare surplus of SS over HH for $m_1$ when both are PNE.}
\label{fig:2SSM-PNE-types}
\end{figure}

Now that we have defined the game $G_\alpha$, it is natural to ask about what equilibria it has. Our results suggest that for all values of $\alpha \in \mathcal{H}^2$, $G_\alpha$ has at least one pure Nash equilibrium (PNE), so that if we let PNE$(G)$ denote the PNE of a given game $G$, PNE$(G_\alpha) \neq \emptyset$ for $\alpha \in \mathcal{H}^2$. In the first image of Figure \ref{fig:2SSM-PNE-types} we show which regions of $\mathcal{H}^2$ demonstrate different combinations of PNE. For the most part, hash rates lead to a single PNE in $G_\alpha$, with distinct regions where each pure strategy profile $(HH, HS, SH$, and $SS)$ occurs as a sole equilibrium. The most interesting observation however, is that for $\alpha$ roughly in the region $[0.2,0.27]^2$, $\text{PNE}(G_\alpha) = \{ HH \text{ and } SS \}$. For all of these hash rates, $SS$  Pareto dominates $HH$ as it results in more utility for both agents involved. As a concrete example, consider $G_\alpha$ for $\alpha = (0.24, 0.24)$ with utilities in Table \ref{tab:2miner-PNE}. Clearly $HH$ and $SS$ are PNE in $G_\alpha$, and the utility surplus between $SS$ and $HH$ is approximately $0.02$ for $m_1$ and $m_2$.

\begin{table}[H]
\begin{center}
\caption {Example of $G_\alpha$ with HH and SS as PNE} \label{tab:2miner-PNE}
\begin{tabular}{||c | c | c | c | c||} 
\hline
\ $\alpha = (0.24, 0.24)$ \ & HH & HS & SH & SS \\ [0.5ex] 
\hline\hline
$U_1$ & \ 0.24 \ & \ 0.24293956 \ & \ 0.23069139 \ & \ 0.25911617 \ \\ 
\hline
$U_2$ & 0.24 & 0.23069139 & 0.24293956 & 0.25911617 \\ [1ex] 
\hline
\end{tabular}
\end{center}
\end{table}

\noindent
The second image in Figure \ref{fig:2SSM-PNE-types} focuses on $[0.2,0.27]^2 \subset \mathcal{H}^2$ and visualises the difference in utility between $SS$ and $HH$ for $m_1$. The difference in utility for $m_2$ is symmetric since $G_\alpha$ is an anonymous game, meaning the role $m_1$ and $m_2$ can be interchanged.

\begin{figure}
\centering
\includegraphics[width = .45\textwidth, trim={0 0 7mm 0},clip]{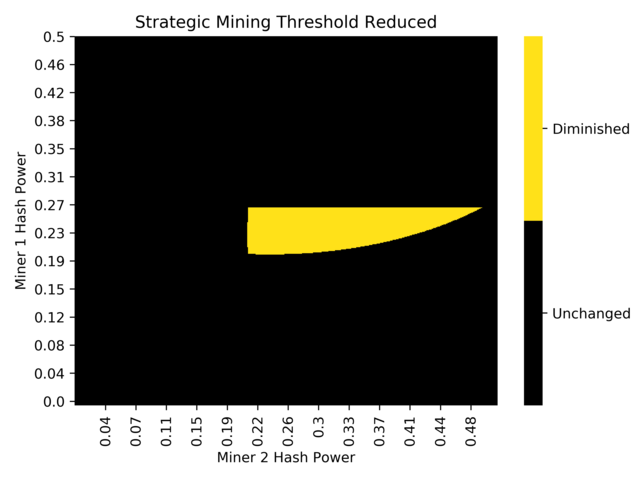}
\caption{Hash rates where the profitability threshold of SSM is reduced.}
\label{fig:2SSM-frontier}
\end{figure}

Interestingly, there are hash rates $\alpha \in \mathcal{H}^2$ where $SS$ is an equilibrium, yet $SH$ is not profitable relative to $HH$ for $m_1$. This means that the existence of another strategic miner can make mining with SSM profitable and stable for $m_1$ whereas this is not the case when $m_1$ with hash power $\alpha_1$ is the only strategic miner in the system. For these $\alpha$ we say the profitability threshold of SSM has decreased. The set of $\alpha \in \mathcal{H}^2$ such that the profitability threshold of SSM decreases is graphed in Figure \ref{fig:2SSM-frontier}. Furthermore, there are hash rates in this region where $SS$ is the only PNE, such as $\alpha = (0.235, 0.345)$ which leads to $G_\alpha$ with utilities in Table \ref{tab:2SSM-threshold}.

\begin{table}[H]
\begin{center}
\caption {Example $G_\alpha$ where SSM Profitability Threshold Decreases} \label{tab:2SSM-threshold}
\begin{tabular}{||c | c | c | c | c||} 
\hline
\ $\alpha = (0.235, 0.345)$ \ & HH & HS & SH & SS \\ [0.5ex] 
\hline\hline
$U_1$ & \ 0.235 \ & \ 0.22352621 \ & \ 0.22418585 \ & \ 0.23160125 \ \\ 
\hline
$U_2$ & 0.345 & 0.37698013 & 0.34987697 & 0.42917647 \\ [1ex] 
\hline
\end{tabular}
\end{center}
\end{table}

The logical next step is to ask about mixed Nash equilibria in $G_\alpha$, however the meaning of mixed strategies is not well-suited for selfish mining attacks. For example, what would the mixed strategy $0.2H + 0.8S$ represent? One interpretation could be a randomised commitment, where with probability $0.2$ a miner commits to $H$ and with probability $0.8$ a miner commits to SSM. This however does not make much sense for selfish mining attacks, since their profitability takes time (due to adjustments in the block difficulty of the system), meaning that an opposing agent would have ample time to perform a best response to the realised commitment over the initial randomisation. 

Another approach is to have $0.2H + 0.8S$ mean that a miner partitions his hash power into honest mining and SSM mining and commits to this partition henceforth. Although utilities of mixed strategies do not directly correspond to convex combinations of utilities, we use this approach to study an extended action space for miners.

\section{Partition Games and Strong Stackelberg Equilibria}
\label{sec:partition-games}

As mentioned at the end of the previous section, we also study incentives when miners are given a richer set of pure strategies beyond that of choosing between honest mining and SSM. In particular, we now allow a given miner with hash power $\alpha_i$ to partition his computational power into a portion following SSM and a portion using honest mining. Before continuing we also clarify notation: for $x,y \in\mathbb{R}^n$, we use $x \circ y$ to denote the Hadamard product of $x$ and $y$.

\begin{definition}[Two-player Partition Games]
Suppose that $\alpha = (\alpha_1,\alpha_2) \in \mathcal{H}^2$ is a strategic hash distribution. We define the {\em Partition Game}, $G^P_\alpha$, as a two-player game, where each player has the same action set $[0,1]$, representing the proportion of their hash power dedicated to employing SSM. For a given pure strategy profile $s = (s_1, s_2) \in [0,1]^2$, we define the utilities of $G^P_\alpha$ as follows:
\begin{itemize}
\item $U_1(s_1,s_2) = s_1 R_{SSM}(s \circ \alpha)_1 + (1 - s_1)\frac{(1-s_1)\alpha_1}{1 - s \cdot \alpha} R_{SSM}(s\circ \alpha)_3$
\item $U_2(s_1,s_2) = s_2 R_{SSM}(s \circ \alpha)_2 + (1 - s_2)\frac{(1-s_2)\alpha_2}{1 - s \cdot \alpha} R_{SSM}(s \circ \alpha)_3$
\end{itemize}
\end{definition}

\begin{figure}
\centering
\includegraphics[width = .46\textwidth]{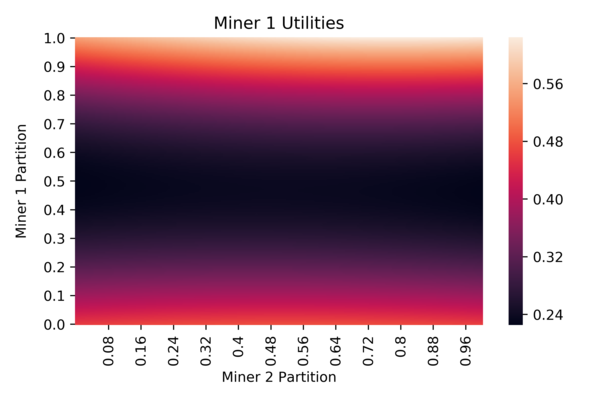}
\includegraphics[width = .46\textwidth]{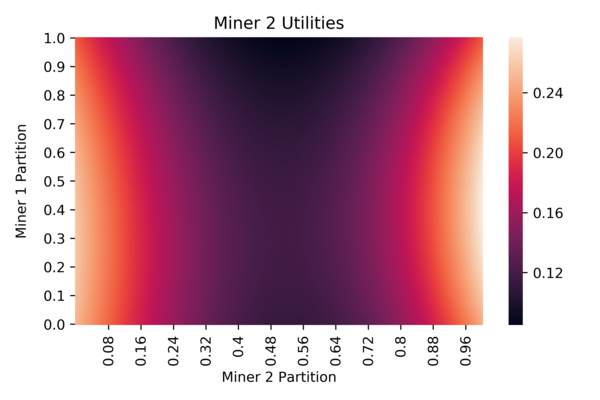}
\caption{Utilities in $G_\alpha^P$ for $\alpha = (0.46,0.25)$.}
\label{fig:utilities_graph}
\end{figure}

In Figure \ref{fig:utilities_graph}, for $\alpha = (0.46,0.25)$ we graph the pure strategy utilities of $m_1$ and $m_2$ as a function of $s \in [0,1]^2$. The most glaring observation is that for fixed $s_{-i}$, $U_i(s_i,s_{-i})$ is a convex function of $s_i$, attaining local maxima at $s_i = 0$ and $s_i = 1$. This is clear from the fact that the blockchain eventually has one common history, so both sides of a miner's partition inherently compete with one another.

Game theoretically, this means best responses for any $m_i$ are always from the set $\{0,1\}$. Immediately, this tells us that the set of pure Nash equilibria of $G^P_\alpha$ are the same as those in $G_\alpha$, since $G^P_\alpha$ restricted to pure strategy profiles in $\{0,1\}^2$ is isomorphic to $G_\alpha$ (recall that $H \cong 0$ and $S \cong 1$ in $G_\alpha$). It may thus seem the augmented strategy space of $G^P_\alpha$ buys us nothing, however if we treat $G^P_\alpha$ as a leadership game, where $m_1$ gets to commit to a pure strategy, $s_1$, to which $m_2$ retaliates, then we get a different story. 

To formally treat $G^P_\alpha$ as a leadership game, we let $m_1$ be the leader and $m_2$ the follower. For a given pure strategy $s_1 \in [0,1]$ of $m_1$, we let $BR(s_1)$ denote the best response $m_2$ has to $s_1$. Since we have observed that best responses for any $m_i$ are always from the set $\{0,1\}$, it follows that $BR(s_1) = \text{argmax}_{x \in \{0,1\}}U_2(s_1,x)$. If $U_2(s_1,0) = U_2(s_1,1)$, then we let $BR(s_1) = \text{argmax}_{x \in \{0,1\}}U_1(s_1,x)$, so that $m_2$ breaks ties in favour of $m_1$. The value of commitment $s_1$ for $m_1$ is denoted by $v_1(s_1) = U_1(s_1, BR(s_1))$ and for the value of commitment $s_1$ for miner 2 is denoted by $v_2(s_1) = U_2(s_1, BR(s_1))$.

In a leadership game, a common solution concept is that of a {\em Strong Stackelberg Equilibrium (SSE)}, which is a strategy pair $(s_1^*,s_2^*)$ such that $s_1^* \in \text{argmax}_{x} v_1(x)$ and $s_2^* = BR(s_1^*)$. This can be seen as a subgame perfect equilibrium of $G^P_\alpha$, or the optimal commitment under $v_1$. Furthermore, we let $\text{SSE}(G)$ denote the SSE of an arbitrary game $G$.

In Figure \ref{fig:optimal-com} we graph (optimal) commitment values for $m_1$ at the SSE of $G^P_\alpha$ for different values of $\alpha \in \mathcal{H}^2$. Furthermore, we graph the value of these optimal commitments when compared to utility players obtain at their respectively optimal PNE of $G_\alpha$ at the given hash rate. 

\subsection{Non-trivial SSE}

Since $G^P_\alpha$ can be seen as an augmented action space to $G_\alpha$, we categorise $\alpha \in \mathcal{H}^2$ depending on how the sets $\text{PNE}(G_\alpha) = \text{PNE}(G^P_\alpha)$ and $\text{SSE}(G^P_\alpha)$ compare.

\begin{definition}[Commitment/SSE Types]
\label{def:com-type}
For every $\alpha \in \mathcal{H}^2$ we associate a {\em commitment type \em} denoted $com(\alpha) \in \{0,1,2,3\}$ defined as follows:

\begin{itemize}
    \item If $\text{SSE}(G^P_\alpha) = \text{PNE}(G_\alpha)$, then $com(\alpha) = 0$. 
    \item If $\text{SSE}(G^P_\alpha) \subset \text{PNE}(G_\alpha)$, then $com(\alpha) = 1$.
    \item If $\text{SSE}(G^P_\alpha) \not \subset \text{PNE}(G_\alpha)$, and $\exists s^* \in \text{SSE}(G^P_\alpha)$ such that $s^*_1 \in \{0,1\}$, then $com(\alpha) = 2$.
    \item If $\text{SSE}(G^P_\alpha) \not \subset \text{PNE}(G_\alpha)$, and $\not \exists s^* \in \text{SSE}(G^P_\alpha)$ such that $s^*_1 \in \{0,1\}$, then $com(\alpha) = 3$.
\end{itemize}
\end{definition}

If $com(\alpha) = 0$ we say $\alpha \in \mathcal{H}^2$ gives rise to a trivial commitment and that the collection of SSE in $G^P_\alpha$ are trivial. Accordingly, if $com(\alpha) \neq 0$, we say $\alpha$ gives rise to a non-trivial commitment and the collection of SSE in $G^P_\alpha$ is non-trivial. Furthermore, we also say that if $\alpha \in \mathcal{H}^2$ is such that $com(\alpha) = i$, then all $s^* \in \text{SSE}(G^P_\alpha)$ are of type $i$ as well. In the two-miner scenario, we make the following observations about $\alpha \in \mathcal{H}^2$ with non-trivial commitment types:

\begin{itemize}
\item $com(\alpha) = 1$ occurs at hash values such that the PNE of $G_\alpha$ are $HH$ and $SS$. $m_1$ commits to $S$ to nudge the system to converge to the $SS$ equilibrium which Pareto-dominates $HH$ in $G_\alpha$.
\item $com(\alpha) = 2$ occurs at hash rates where there is one SSE of $G^P_\alpha$, $s^* = (s_1^*,s_2^*) \in \{0,1\}^2$, yet $s^*$ does not correspond to a PNE of $G_\alpha$. $s^*$ is unstable in $G_\alpha$ from the perspective of $m_1$, who would prefer deviating from $s_1$ when pitted against $s_2$. These SSE make use of the sequentiality of $G^P_\alpha$ but not of the extended action space given by partitioning.
\item $com(\alpha) = 3$ occurs at hash rates such that $HS$ is the only PNE of $G_\alpha$, but where $\alpha$ is close to the region in $\mathcal{H}^2$ where $SS$ arises as the sole PNE of $G_\alpha$. At these values, $m_2$ only slightly prefers $SH$ to $SS$, hence $m_1$ can bait $m_2$ into playing $S$ by reserving a small portion of hash power to mine honestly.
\end{itemize}

\begin{figure}
\centering
\includegraphics[width = 0.35\textwidth]{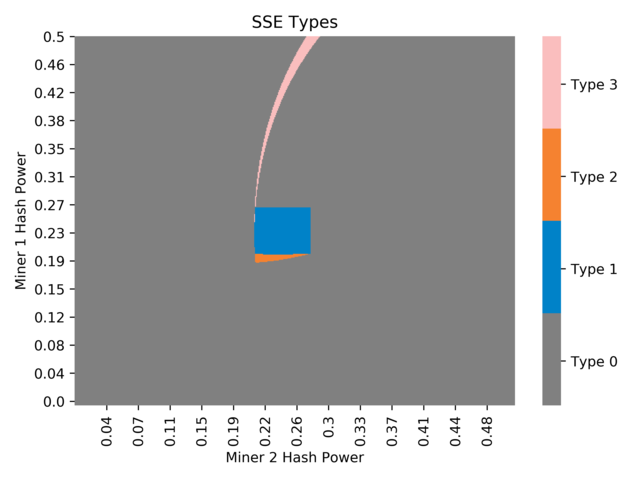}
\includegraphics[width = 0.35\textwidth]{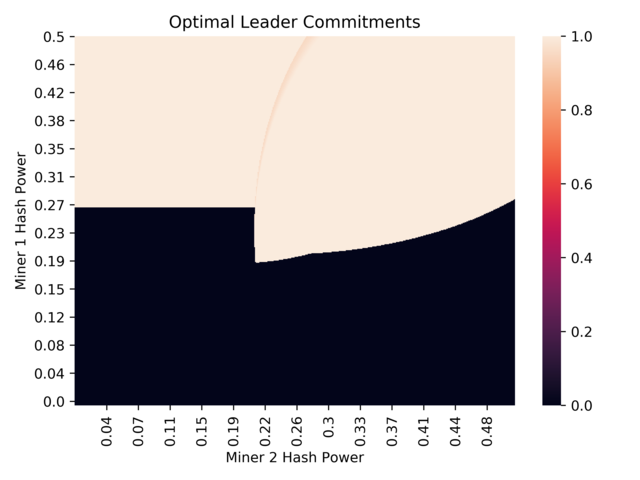}
\includegraphics[width = 0.35\textwidth]{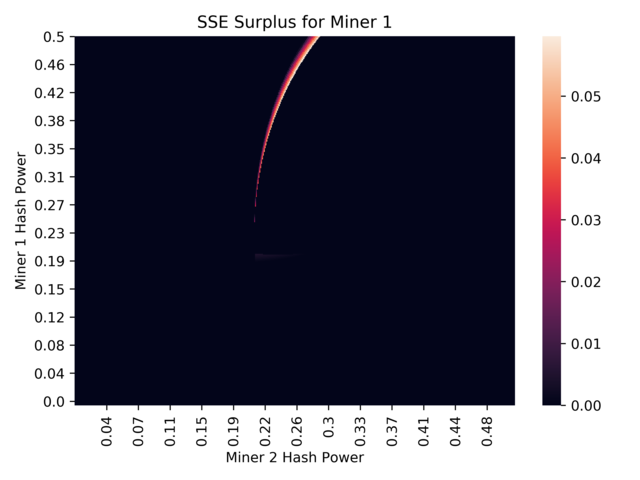}
\includegraphics[width = 0.35\textwidth]{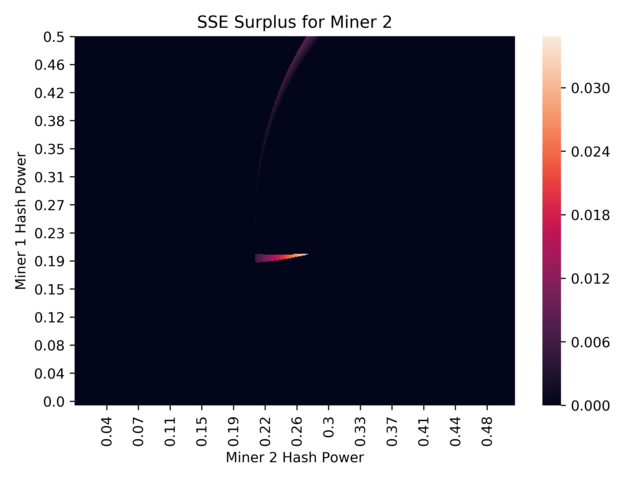}
\caption{SSE types for $m_1$, optimal commitments for $m_1$, and relative surplus of SSE against best PNE for $m_1$ and for $m_2$ respectively.}
\label{fig:optimal-com}
\end{figure}

For any non-trivial SSE, $v_1(s^*_1)$ is lower bounded by the lowest-utility $m_1$ obtains amongst PNE in $G_\alpha$. On the other hand, if $com(\alpha) = 1,3$, the SSE of $G^P_\alpha$ are such that $v_1(s^*_1)$ is strictly greater than the highest utility $m_1$ obtains amongst PNE in $G_\alpha$. This strict  surplus in utility is visible in the latter graphs of Figure \ref{fig:optimal-com}, and we can see that these non-trivial commitments also benefit $m_2$ in spite of being the follower.

\subsection{Plots of Non-Trivial SSE by Type}

We now focus on plotting optimal  $\mathcal{H}^2$ that exhibit SSE of types 1, 2 and 3. For SSE of type 1, it suffices to look at Figure \ref{fig:2SSM-PNE-types} and Table \ref{tab:2miner-PNE} for visualisation of the benefit of SSE over PNE (Since it is just the difference in welfare between PNE in this case).

\begin{figure}
\centering
\includegraphics[width = 0.35\textwidth]{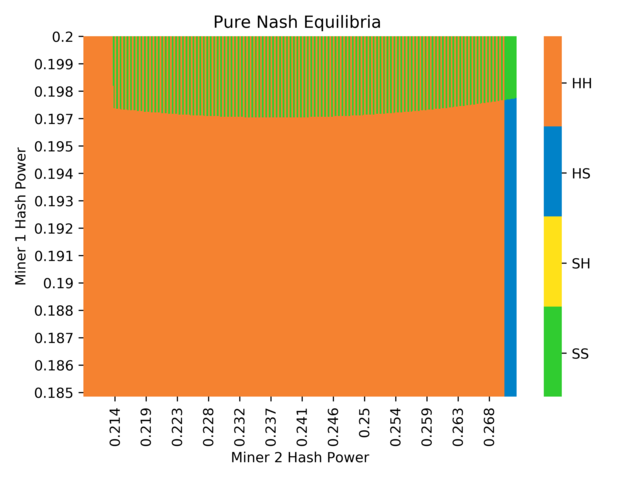}
\includegraphics[width = 0.35\textwidth]{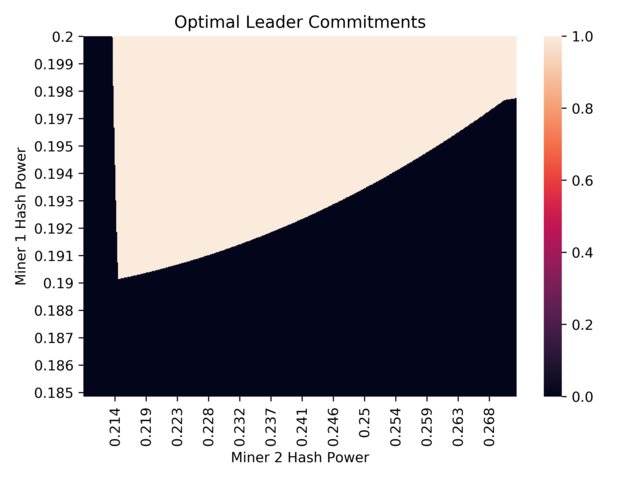}
\includegraphics[width = 0.35\textwidth]{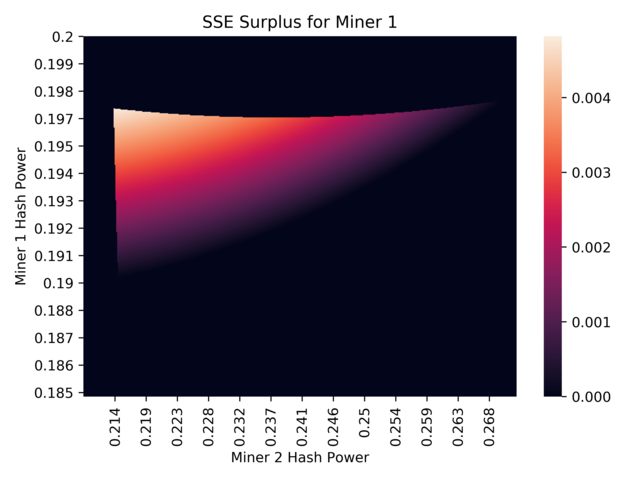}
\includegraphics[width = 0.35\textwidth]{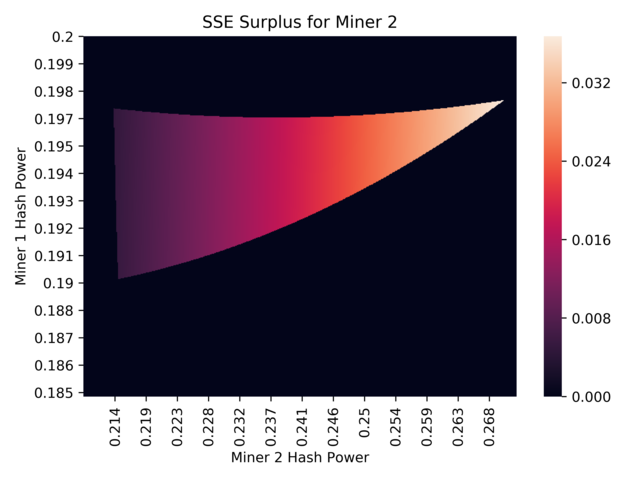}
\caption{Optimal Commitments for $m_1$, as well as SSE surplus against best PNE for $m_1$, and for $m_2$ respectively in the region $[0.185, 0.2] \times [0.21, 0.27]$. This region exhibits SSE of type 2.}
\label{fig:optimal-com-zoom-center}
\end{figure}

As for SSE of type 2, these are plotted in more detail in Figure \ref{fig:optimal-com-zoom-center}. For these values of $\alpha$, we can see that $HH$ is the only PNE in $G_\alpha$, but $SS$ is the SSE of $G^P_\alpha$, which is forcibly unstable in the one shot game, $G_\alpha$, as $m_1$ prefers $HS$ to $SS$. Table \ref{tab:2SSM-type2} shows the utilities for $G_\alpha$ at a specific value of $\alpha$ exhibiting this behaviour. Note that in this example, the leader, $m_1$, has a hash rate of $\alpha_1 = 0.2$, at which normally they would not be incentivised to unilaterally employ SSM in the one-shot SSM game. The power to commit makes SSM viable at smaller hash rates than in the one-shot game.   

\begin{table}[H]
\begin{center}
\caption {Example $G_\alpha$ where the SSE in $G^P_\alpha$ is of  type 2} \label{tab:2SSM-type2}
\begin{tabular}{||c | c | c | c | c||} 
\hline
\ $\alpha = (0.2,0.225)$ \ & HH & HS & SH & SS \\ [0.5ex] 
\hline\hline
$U_1$ & \ 0.2 \ & \ 0.20352746 \ & \ 0.18016529 \ & \ 0.20179681 \ \\ 
\hline
$U_2$ & \ 0.225 \ & 0.21133109 & 0.23057851 & 0.23905979 \\ [1ex] 
\hline
\end{tabular}
\end{center}
\end{table}

\begin{figure}
\centering
\includegraphics[width = 0.35\textwidth]{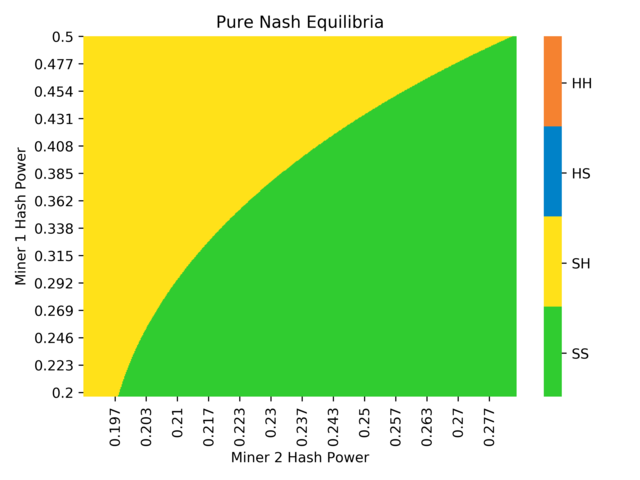}
\includegraphics[width = 0.35\textwidth]{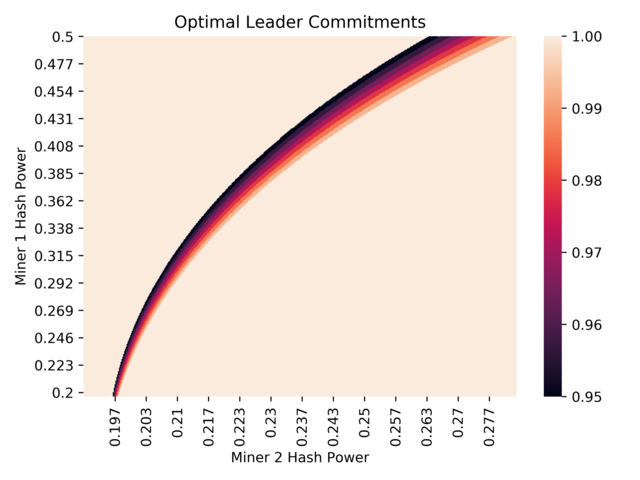}
\includegraphics[width = 0.35\textwidth]{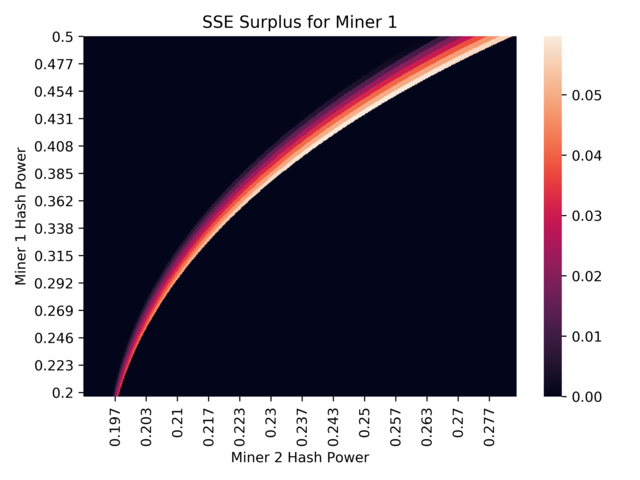}
\includegraphics[width = 0.35\textwidth]{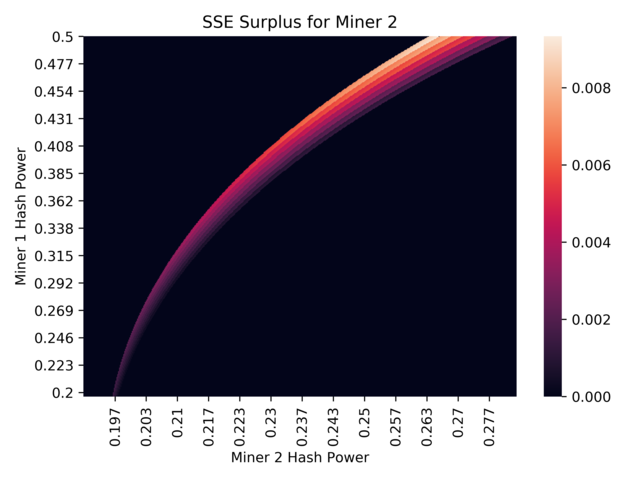}
\caption{Optimal Commitments for $m_1$, as well as SSE surplus against best PNE for $m_1$, and for $m_2$ respectively in the region  $[0.2, 0.5] \times [0.19, 0.28]$. This region exhibits SSE of type 2.}
\label{fig:optimal-com-zoom-upper-center}
\end{figure}

\begin{figure}
\centering
\includegraphics[width = 0.35\textwidth]{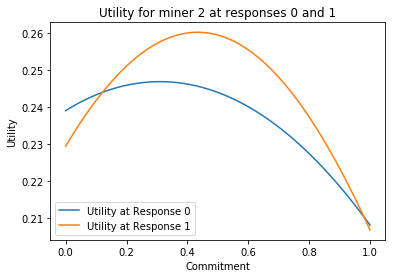}
\includegraphics[width = 0.35\textwidth]{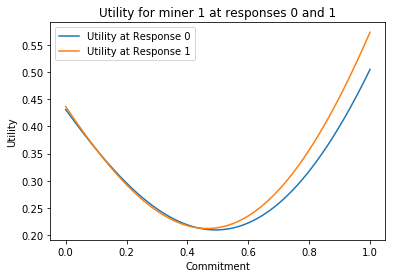}
\includegraphics[width = 0.35\textwidth]{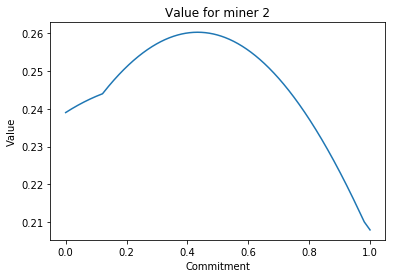}
\includegraphics[width = 0.35\textwidth]{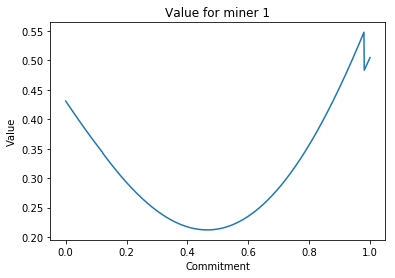}
\caption{Partition Game Analysis for $\alpha = (0.431,0.239)$. The top left image plots follower utilities when playing $H$ or $S$ against a leader commitment partition. The bottom left image plots follower utility when best responding to a leader commitment. The best response at a given commitment dictates which of the two utilities the leader obtains in the top right plot. Putting everything together, the bottom right plot gives the value of a leader commitment (for the leader) as a function of their commitment. Note how this function is maximised at approximately 0.98, where the follower is indifferent between $H$ and $S$.}
\label{fig:leadership-single-hash}
\end{figure}

Figure \ref{fig:optimal-com-zoom-upper-center} focuses on hash rates where SSE are of type 3. Furthermore, Figure \ref{fig:leadership-single-hash} looks specifically at $\alpha = (0.431,0.239)$, which is a hash rate such that $G^P_\alpha$ has an SSE of type 3, and graphs utilities and best responses as a function of the leader commitment in $G^P_\alpha$. This gives a better way of visualising how $s_1 = 0.98$ is an optimal commitment where $m_2$ is rendered indifferent between $S$ and $H$.

\section{$M > 2$ Strategic Miners}
\label{sec:N-SSM}

Our analysis from Section \ref{sec:two-SSM} extends in a straightforward fashion to when there are $M > 2$ strategic miners. Consequently, for any hash distribution $\alpha \in \mathcal{H}^M$, we can compute $R_{SSM}(\alpha) \in [0,1]^{M+1}$, the revenue ratio of all $M$ strategic miners and all other honest miners, when all strategic miners of hash power $\alpha_i$ employ SSM. The full details of the corresponding Markov chain and reward vectors can be found in Appendix \ref{appendix:M-SSM}.

It is also straightforward to extend the game-theoretic formalism of Section \ref{sec:two-SSM} to study incentives when $M > 2$ strategic miners interact. This formalism can also be found in Appendix \ref{sec:multiplayer-game-formalism}. In Appendix \ref{sec:3SSM} we also plot similar graphs to Section \ref{sec:two-SSM} for $M = 3$ at different hash rates to visualise strategic miner behaviour. When $M > 3$ however, it becomes difficult to visualise how aspects of $G_\alpha$ and $G^P_\alpha$ precisely vary with $\alpha$. That being said, we do find very similar structures as in the $M = 2$ and $M = 3$ case, such as: penalising coalitions, existence of PNE, and for some regions multiple PNE, in $G_\alpha$, non trivial commitments in $G^P_\alpha$, and finally, hash rates $\alpha$ where the SSM profitability threshold decreases with the existence of other strategic miners. We expand upon this final point to specifically see how the number of strategic miners $M$ affects the profitability threshold of SSM.

\subsection{Decreasing SSM Profitability Threshold}

To study the effect of the number of miners on the profitability threshold of SSM, we define the following:

\begin{definition}[Uniform Profitability Threshold for SSM]
For $M \geq 1$ miners we say the {\em uniform profitability threshold for SSM} is the smallest $\eta \in [0,1]$ such that $\alpha = \eta \vec{1} \in \mathcal{H}^M$ and $\vec{1} \in \text{PNE}(G_\alpha)$ (all players employing SSM is a PNE in $G_\alpha)$. 
\end{definition}

\begin{figure}
\centering
\includegraphics[width = 0.4\textwidth]{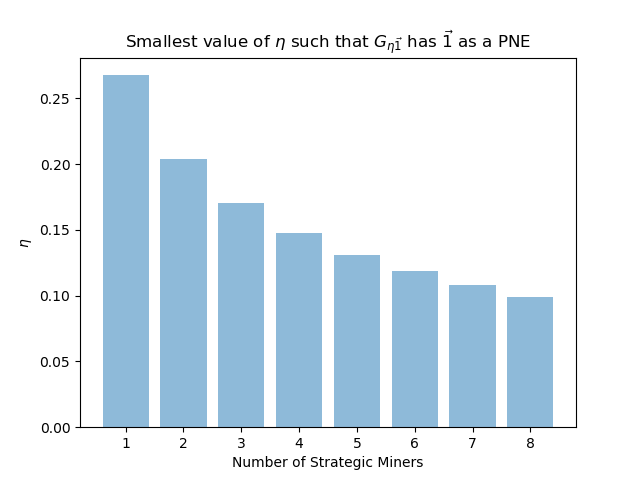}
\includegraphics[width = 0.4\textwidth]{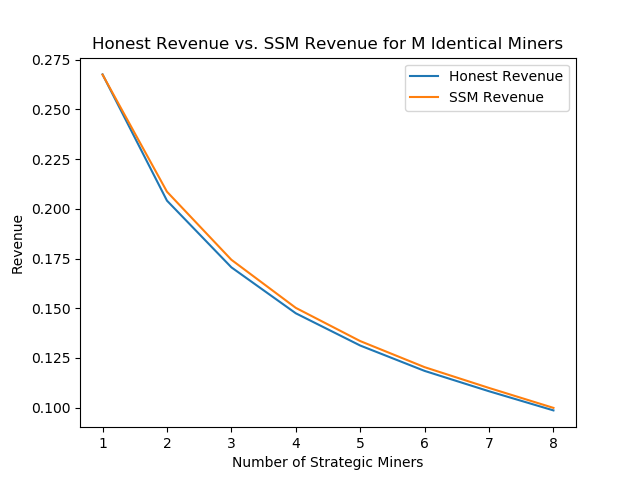}
\caption{Upper bounds on the uniform profitability threshold for SSM as a function of the number of strategic miners. We also plot the welfare of $\vec{1}$ (all SSM) versus $\vec{0}$ (all honest).}
\label{fig:NSSM-allS-PNE}
\end{figure}

With our methods from Appendix \ref{sec:multiplayer-game-formalism}, we can approximate the uniform SSM profitability threshold for various values of $M$. In particular, Figure \ref{fig:NSSM-allS-PNE} shows these threshold values for $M = 1,...,8$. Furthermore, the second plot takes the uniform SSM profitability threshold, $\eta$, and for $\alpha = \eta \vec{1}$, computes the utilities of both $\vec{0}$ and $\vec{1}$ which are both PNE in $G_\alpha$. Interestingly, for $M = 1,..,8$, not only does the uniform profitability threshold decrease as a function of $M$, but all miners employing SSM is a PNE that Pareto dominates all miners being honest. These results thus show that the presence of multiple strategic miners may have more of an impact on the stability of Bitcoin than previously thought.

\section{Conclusion and Further Work}

In this paper we have described a specific miner strategy, semi-selfish mining (SSM) that is a truncated variant of Selfish Mining (SM). SSM has the benefit of being a profitable strategy for large enough miners (in the same way as SM), and also structured enough for us to explicitly solve for relative revenues when more than one strategic miner employs SSM. With this in hand, we have been able to use a game-theoretic lense to glean some information on miner incentives when more than one miner is strategic within the bitocin system. 

In particular, for any $\alpha \in \mathcal{H}^M$, we define the SSM game $G_\alpha$ which governs strategic miner incentives in choosing to employ SSM or mine honestly, and the partition game $G^P_\alpha$, which extends the action space of $G_\alpha$ to allow miners to partition their hash power between honest mining and SSM. For $M > 1$ strategic miners we find the following main takeaways from studying $G_\alpha$ and $G^P_\alpha$:

\begin{itemize}
\item All $\alpha \in \mathcal{H}^M$ seem to lead to $G_\alpha$ with pure Nash equilibria. Furthermore, there are regions in $\mathcal{H}^M$ such that $G_\alpha$ has multiple PNE.
\item A single miner might prefer to use SSM over honest mining in $G_\alpha$, but there can exist a coalition of miners who may retaliate against this action and punish the original SSM miner into receiving less utility than their hash power.
\item Though the set of PNE in $G^P_\alpha$ is identical to those of $G_\alpha$, when treating $G^P_\alpha$ as a sequential game leads to non-trivial commitments, some of which involve a miner employing SSM even though SSM is not rational in the one-shot SSM game.
\item Finally, there exist hash rates, $\alpha \in \mathcal{H}^M$ such that $m_1$ does not unilaterally prefer to employ SSM, but some PNE of $G_\alpha$ includes $m_1$ employing SSM, effectively reducing the profitability threshold of SSM and consequently affecting the stability of Bitcoin.
\end{itemize}

The action spaces in $G_\alpha$ and $G^P_\alpha$ may seem limited due to the fact that they only interpolate between honest mining and SSM, but there is nothing barring a variant $G_\alpha$ and $G^P_\alpha$ from studying the choice of employing other subversive mining strategies over honest mining. In fact, $G_\alpha$ and $G^P_\alpha$ can be defined by using empirical estimates to steady state payoffs instead of closed form solutions, which could glean some information into how mining dynamics change when a larger palette of subversive strategies is available to interdependent strategic miners. In fact, $G^P_\alpha$ could be extended so that the action space of miners is no longer simply partitioning mining power between honest mining and SSM, but any partition of mining power amongst a given list of subversive mining strategies. 

In addition, the fact that penalising coalitions exist hints at the possibility of modelling such structures in a repeated game framework. The issue of course comes in modelling how much utility a penalising coalition gains in maintaining everyone honest, but there could be interesting subgame perfect Nash equilibria in an appropriate model. Finally, along the same vein of penalising coalitions, there is also scope for a more fine-grained cooperative game theoretic analysis of SSM and Partition games.

\bibliographystyle{splncs}
\bibliography{biblio}

\appendix

\section{SSM vs. Honest Mining}
\label{appendix:ssm-vs-honest}

We can use a similar Markov chain analysis to derive the revenue ratio of SSM against honest miners. We recall that the strategic miner, $m_1$,  has hash power $\alpha \in \mathcal{H}^1 = (0,0.5]$ and the honest miner $m_2$ has hash power $\beta = 1 - \alpha$. Let us define the state space $S = \{S_0,S_1,S_2\}$ corresponding to the number of private blocks belonging to the miner employing SSM. We can now describe the transitions and their corresponding revenues (expected block creation rate per state):

\subsubsection{Transitions from state $S_0$}
\begin{itemize}
\item $S_0 \rightarrow S_0$ occurs if $m_2$ find a block. The probability of this transition is $\beta$ and $m_2$ wins a block. 
\item $S_0 \rightarrow S_1$ occurs if $m_1$ finds a block. The probability of this transition is $\alpha$ and no players win a block.
\end{itemize}

\subsubsection{Transitions from state $S_1$}
\begin{itemize}
\item $S_1 \rightarrow S_0$ occurs if $m_2$ finds and publishes a block, which occurs with probability $\beta$. A fork is created when $m_1$ subsequently publishes his hidden block and from here three events can occur: A first scenario occurs when $m_1$ finds another block to resolve the tie in his favour, resulting in two blocks for $m_1$. This occurs with probability $\alpha$. A second scenario occurs when an honest miner finds a block that resolves the tie in favour of $m_1$, resulting in one block for $m_1$ and one block for $m_2$. This occurs with probability $\gamma \beta$. A final scenario occurs when an honest miner finds a block that resolves the tie in favour of $m_2$ which results in two blocks for $m_2$. This final event occurs with probability $(1-\gamma)\beta$. In all aforementioned scenarios the resulting state is $S_0$, thus the probability of the transition to state $S_0$ is $\beta$.
\item $S_1 \rightarrow S_2$ occurs if $m_1$ finds a block and keeps it private as per SSM. This event occurs with probability $\alpha$ and no blocks are awarded to any agent. 
\end{itemize}

\subsubsection{Transitions from state $S_2$}
\begin{itemize}
\item $S_2 \rightarrow S_0$ occurs when $m_2$ finds a block. The probability of this transition is $\beta$ and $m_1$ wins two blocks.
\item $S_2 \rightarrow S_2$ occurs if $m_1$ finds a block. The probability of this transition is $\alpha$ and $m_1$ wins a block.
\end{itemize}

\noindent

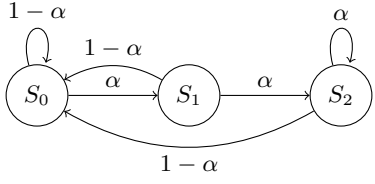
\begin{figure}
\center{
\begin{tikzpicture}
\node[state] (q0) {$S_0$};
\node[state, right of=q0] at (1, 0) (q1) {$S_1$};
\node[state, right of=q1] at (3, 0) (q2) {$S_2$}; 

\path[every node/.style={font=\sffamily\small}]
    (q0) [->] edge[above] node {$\alpha$} (q1)
    (q0) edge[loop above] node {$1-\alpha$} (q0)
    (q1) edge[above] node{$\alpha$} (q2)
    (q1) edge[bend right, above] node{$1-\alpha$} (q0)
    (q2) edge[bend left, below] node{$1-\alpha$} (q0)
    (q2) edge[loop above] node {$\alpha$} (q2);
    
\end{tikzpicture}
\caption{States and Transitions for SSM vs. Honest Miners}\label{fig:transitions-ssm-vs-honest}
}
\end{figure}

The transitions are visualised in Figure \ref{fig:transitions-ssm-vs-honest}. Furthermore, we can fully express the transition matrix of the Markov chain as follows: 
\[
M=
\begin{bmatrix}
1- \alpha & 1-\alpha & 1-\alpha\\
\alpha & 0 & 0\\
0 & \alpha & \alpha\\
\end{bmatrix}
\]
For a given probability distribution $x \in \mathbb{R}^3$ over the state space $S$, $Mx$ gives the resulting probability distribution over $S$ after one transition under the Markov chain above. Since the chain is easily seen to be ergodic, there exists a unique steady state distribution, $\pi$, such that $M \pi = \pi$. Using Gaussian elimination we obtain $ \pi = (1-\alpha, \alpha(1-\alpha), \alpha^2)^T$ as the unique steady state. Furthermore, from the transitions mentioned above we obtain the following expected block creation rates (denoted by $r_{SSM}$ and $r_{others}$) per state:

\begin{table}[H]
\begin{center}
\caption {Expected revenue per state} \label{tab:1miner-revenue} 
\begin{tabular}{|c|c|c|}
\hline
State & $\mathbb{E}(r_{SSM} \mid S_i)$ & $\mathbb{E}(r_{others} \mid S_i)$ \\
\hline
$S_0$ & 0 & $1-\alpha$\\
\hline
\ $S_1$ & $(1-\alpha)(\gamma(1-\alpha)+2\alpha)$ \ & \ $(1-\alpha)(\gamma(1-\alpha)+2(1-\alpha)(1-\gamma))$ \ \\
\hline
$S_2$ & $\alpha + 2(1-\alpha)$ & $0$\\
\hline
\end{tabular}
\end{center}
\end{table}

We let $r_{SSM}$ and $r_{others}$ denote the expected revenue per round at steady state $\pi$ for $m_1$ and $m_2$. We also let $R_{SSM}$ and $R_{others}$ denote the revenue ratios of $m_1$ and $m_2$ at steady state. Given our expected revenues per state, we obtain $r_{SSM} = (2-\gamma)\alpha^4 + (3\gamma - 5)\alpha^3 + (4-3\gamma)\alpha^2 + \gamma \alpha$ and $r_{others} = (1-\alpha)^2 \left( (\gamma - 2)\alpha^2 + (2 - \gamma)\alpha + 1 \right)$.

\begin{theorem}
A strategic miner of hash power $\alpha$ attains the following revenue ratio when playing against honest miners: 
$$
R_{SSM} = \frac{r_{SSM}}{r_{SSM} + r_{others}} = 
\frac{\alpha(\alpha(\alpha(2\alpha-5) + 4) - (\alpha-1)^3\gamma)}{(\alpha-1)\alpha^2 + 1}
$$
Asymptotically around $\alpha = 0$, the expression is the following:
$$
R_{SSM} = \alpha\gamma + \alpha^2(4 - 3\gamma) + \alpha^3(4\gamma - 5) - \alpha^4(6 - 5\gamma) + \alpha^5(7\gamma - 9) + O(\alpha^6)
$$

\end{theorem}

\section{Markov Chain Formalism for $M 
\geq 2$ Strategic Miners and Arbitrary Propagation}
\label{appendix:M-SSM}

In this section we delve into the Markov chain governing revenues (block creation rates) when multiple strategic miners employ SSM. In what follows we assume that $\alpha \in \mathcal{H}^M$. This implies that $m_1,...,m_M$ are strategic miners with hash power $\alpha_1,...,\alpha_m$, and $m_{M+1}$ is an honest miner with hash power $\beta = 1 - \sum_{i=1}^M \alpha_i$. 

As in the one miner case, we let $S = \{0,1,2\}^M$ be the state space of all possible private leads held by $m_1,...,m_M$ employing SSM. For a given $x \in S$, $x_i$ denotes the private lead of $m_i$. In addition, for a given $x \in S$, we let $A_x = \{i \in [M] \ | \ x_i = 1\}$ and $B_x = \{i \in [M] \ | \ x_i = 2\}$. Clearly $A_x \cap B_x = \emptyset$, furthermore, we can completely establish transition probabilities from $x$ by looking at $A_x$ and $B_x$.

\subsection{State Transitions}

Let us suppose $x \in S$ is arbitrary. In what follows we let $e_i \in \{0,1\}^M$ be the unit vector with 1 in the $i$-th coordinate. Furthermore, we let $P_{x \rightarrow y}$ denote the probability of transitioning from $x$ to $y$. To fully describe all transitions for any $x \in S$, we look at four different cases depending on $A_x$ and $B_x$.

\subsubsection{$|A_x| \geq 0$, $|B_x| = 0$}

If any strategic miner $m_i$ obtains a block, they keep it private as per SSM extending their private chain by 1 (which they forcibly have a margin to do so). This results in state $x + e_i$ and occurs with probability $\alpha_i$. If $m_{M+1}$ finds a block, they publish it as per the honest mining protocol, which occurs with probability $\beta$. All $m_i$ such that $x_i \neq 0$ then publish their private chains as per SSM and a race ensues. The conditions of SSM and honest mining dictate that the race is settled in the following turn, and hence we return to state 0. In summary:

\begin{itemize}
    \item $P_{x \rightarrow x + e_i} = \alpha_i$ for all $i \leq M$
    \item $P_{x \rightarrow 0} = \beta$
\end{itemize}

\subsubsection{$|A_x| = 0$, $B_x = \{j\}$}
In this case a single miner has a private lead of 2 and all other miners have no private lead. If any strategic miner $m_i$ such that $i \neq j$ finds a block, SSM dictates that they keep this block private and proceed to having a private chain of length 1. This corresponds to transitioning from $x$ to $x + e_i$, which occurs with probability $\alpha_i$. If the $m_j$ finds a block, an event which happens with probability $\alpha_j$, SSM dictates he publish his oldest private block. Since $A_x = \emptyset$, this block will be the longest public chain, and the resulting state will be $x$ again. Finally, if $m_{M+1}$ finds a block, honest mining dictates he publish it. $m_j$ in turn sees his private lead decrease to 1 and hence publishes his entire private chain. As a consequence state 0 ensues, and this transition occurs with probability $\beta$. In summary we have the following transitions:

\begin{itemize}
    \item $P_{x \rightarrow x + e_i} = \alpha_i$ for all $i \neq j$
    \item $P_{x \rightarrow x} = \alpha_j$
    \item $P_{x \rightarrow 0} = \beta$
\end{itemize}

\subsubsection{$|A_x| \geq 0$, $|B_x| > 1$}
Suppose that $m_i$ such that $i \notin B_x$ finds a block, which occurs with probability $\alpha_i$. As per SSM $m_i$ has a margin to keep this block private, hence state $x + e_i$ ensues. On the other hand, if $m_i$ is such that $i \in B_x$, then by SSM, $m_i$ publishes their oldest private block. As a result, all miners in $A_x$ publish their private leads to start a race, and all miners in $B_x$ publish their private leads to overtake. $m_i$ thus sees his private lead diminish to 1, hence by SSM he publishes his entire private chain. This chain is the longest of all miners, hence we return to state 0. Finally, if $m_{M+1}$ finds a block, which occurs with prability $\beta$, he publishes it as per honest mining, all strategic miners with hidden chains once again publish their hidden chains. There is a multi-way race amongst all miners in $B_x$, but as per SSM and honest mining, this race is decided in the following turn and we return to state 0. In summary we have the following transitions:

\begin{itemize}
    \item $P_{x \rightarrow x+e_i} = \alpha_i$ for $i \notin B_x$
    \item $P_{x \rightarrow 0} = \beta + \sum_{i \in B_x} \alpha_i$
\end{itemize}

\subsubsection{$|A_x| > 1$, $B_x = \{j\}$}

If any $m_i$ such that $i \neq j$ finds a block, an event which occurs with probability $\alpha_i$, then SSM dictates they keep this block private and the resulting state is $x + e_i$. If $m_{M+1}$ finds a block, which occurs with probability $\beta$, they publish it as per honest mining, and $m_j$ sees his lead diminished and by the rules of SSM, publishes his private chain to create the longest public chain. The resulting state is thus 0. Finally, if $m_j$ finds the following block, he publishes his oldest private block as per SSM, and consequently the public tie is amongst a prefix of the chain of $m_j$ the chains of all $m_i$ such that $i \in A_x$, since they also publish their private chains. At this point $m_j$ is mining upon his private chain whereas all other miners, including $m_{M+1}$ mine upon some of the chains partaking in the public tie. From here there are two scenarios. Either $m_j$ also finds the following block, in which case SSM dictates he publish it, and the new public prefix of his chain is the longest public chain and the ensuing state is $2e_j$, or any miner other than $m_j$ finds the next block, in which case $m_j$ sees his lead diminished and publishies his entire private chain resulting in state 0. The overall probability of the first scenario is $\alpha_2^2$ and the overall probability of the second scenario is $\alpha_2(1 - \alpha_2)$. In summary we have the following transitions:

\begin{itemize}
    \item $P_{x \rightarrow x+e_i} = \alpha_i$ for $i \neq j$
    \item $P_{x \rightarrow 2e_j} = \alpha_j^2$
    \item $P_{x \rightarrow 0} = \beta + \alpha_j(1- \alpha_j)$
\end{itemize}


\subsection{Propagation Formalism }
In the original analysis of selfish mining, much attention was given to a data propagation parameter $\gamma$. Propagation is important because it allows an attacker to persuade honest miners to work on their end the public chain when forks occur.

When there are $M \geq 2$ strategic miners however, propagation intricacies cannot be captured by a single parameter, as different strategic agents have different abilities to convinces other miners of their own chains. To encompass this generality, let us suppose that $D \subseteq [M+1]$ is a subset of miners engaged in a tie (we recall that $m_{M+1}$ is the implicit honest miner in the system). For $j \in D$ and $i \in [M+1]$ we let $\gamma^D_{i,j}$ be the probability that $m_i$ mines upon the chain of $m_j$ in the tie composed of all $D$ miners. The only restriction we place on these parameters is that $\gamma^D_{i,i} = 1$ for $i \neq M+1$. The reason for this is that a strategic miner will mine upon their public chain in case of a tie. Finally, we note that in the uniform propagation model we use throughout the paper, we simply let $\gamma^D_{i,j} = \frac{1}{|D|}$ if $i \notin D$ and $i \neq j$ or if $M+1 \in D$ and $i = j = M+1$.

\subsection{Expected Revenue per State with Arbitrary Propagation}

For a given state $x \in S$, we compute the expected revenue per agent under the underlying Markov chain governing SSM dynamics. We denote this quantity by $rev(x) \in \mathbb{R}^{M+1}$, where $rev(x)_i$ denotes the expected revenue of $m_i$ when the system is in state $x \in S$. 

In order to compute these quantities, it will be useful to define the expected revenue all agents obtain when there is an arbitrary tie involving a set $D \subseteq [M]$ of miners. As we have seen in state transitions, for a given $x \in S$, ties can involve either agents with a private lead of 1 or agents with a private lead of 2. We denote the expected revenue all agents receive when a tie of $D \subseteq [M]$ miners with private lead of $i = 1,2$ occurs by $T_i(D) \in \mathbb{R}^{M+1}$: 

\begin{itemize}
\item $T_1(D) = \sum_{i= 1}^{M+1} \alpha_i \left( \sum_{j \in D} \gamma_{i,j}^D (e_i + e_j) \right)$
\item $T_2(D) = \sum_{i= 1}^{M+1} \alpha_i \left( \sum_{j \in D} \gamma_{i,j}^D (e_i + 2e_j) \right)$
\end{itemize}

As with state transitions, for a given $x \in S$, we can characterise $rev(x)$ by looking at $A_x$ and $B_x$, the indices of strategic miners with a private lead of 1 and 2 respectively.

\subsubsection{$|A_x| = |B_x| = 0$}
In this case, only $m_{M+1}$ revceives a block if he finds one, which occurs with probability $\beta$.
\begin{itemize}
\item $rev(x) = \beta e_{M+1}$
\end{itemize}

\subsubsection{$|A_x| > 0$, $|B_x| = 0$}
In this case, blocks are only won in the event of a tie, which in turn only happens if $m_{M+1}$ originally finds a block with probability $\beta$. In such a case, there is a tie amongst the indices $A_x \cup \{M+1\}$.

\begin{itemize}
\item $rev(x) = \beta  \left( (T_1(A_x \cup \{ M+1\}) \right)$
\end{itemize}

\subsubsection{$|A_x| = 0$, $B_x = m_j$}
In this case, if $m_j$ finds a block, he publishes his oldest block as per SSM and thus wins a block in the turn. If $m_{M+1}$ finds and publishes a block as per honest mining, $m_j$ publishes his entire chain as per SSM and wins two blocks in the turn.

\begin{itemize}
\item $rev(x) = \alpha_j (e_j) + \beta (2e_j)$
\end{itemize}

\subsubsection{$|A_x| \geq 0$, $|B_x| > 1$}
If any $m_j$ such that $j \in B_x$ finds a block, by SSM they publish their oldest private block. Other miners with indices in $B_x$ thus publish their entire private chains of length 2, and consequently $m_j$ publishes his entire private chain of length 2 (relative to the original fork so still longer than all other private chains), winning 3 blocks overall. If any $m_i$ such that $i \notin B_x$ finds a block, they simply keep it private as per SSM and no blocks are definitively won. Finally, if $m_{M+1}$ finds a block with probability $\beta$, then all $m_j$ such that $j \in B_x$ publish their private chains and a tie ensues amongst these agents, which results in $T_2(B_x)$ expected revenue for all miners.

\begin{itemize}
\item $rev(x) = \sum_{j \in B} \alpha_j (3 e_j) + \beta T_2(B_x)$
\end{itemize}

\subsubsection{$|A_x| > 1$, $B_x = m_j$}
If $m_{M+1}$ finds a block with probability $\beta$, $m_j$ sees his lead diminished and publishes his entire private chain, thus winning two blocks. If any $m_i$ such that $i \notin B_x$ finds a block, they simply keep it private as per SSM and no one immediately wins blocks. Finally, if $m_j$ finds two blocks in a row he publishes a prefix of his private chain and wins two blocks (transitioning to state $2e_j$ in the process). If $m_j$ finds a block (thus leading him to publish his oldest private block as per SSM), and subsequently any other miner finds the next one, $m_j$ sees his lead diminished and publishes his entire private chain, winning 3 blocks overall.

\begin{itemize}
\item $\beta(2e_j) + \alpha_j^2(2e_j) + \alpha_j(1 - \alpha_j)(3e_j)$
\end{itemize}

\section{Game Theoretic Formalism for $M > 2$ Strategic Miners}
\label{sec:multiplayer-game-formalism}

Our analysis from Section \ref{sec:two-SSM} extends in a straightforward fashion to when there are $M > 2$ strategic miners. Consequently, for any hash distribution $\alpha \in \mathcal{H}^M$, we can compute $R_{SSM}(\alpha) \in [0,1]^{M+1}$, the revenue ratio of all $M$ strategic miners and all other honest miners, when all strategic miners of hash power $\alpha_i$ employ SSM. The full details of the corresponding Markov chain and reward vectors can be found in Appendix \ref{appendix:M-SSM}. In this section, we extend the game-theoretic formalism of Section \ref{sec:two-SSM} to to study incentives when $M > 2$ strategic miners interact. 

\subsection{To SSM or not to SSM in the Multiplayer Setting}

We recall that Section \ref{sec:two-SSM} introduced SSM games, a family of binary action games $G_\alpha$ that governed the incentives behind choosing to employ SSM or honest mining. We extend this game to the multiplayer setting in a natural way. Suppose that $\alpha \in \mathcal{H}^M$ is a hash rate of all strategic miners. Once again, we let $R_{SSM}(\alpha) \in [0,1]^{M+1}$ be the revenue ratios of all miners. Just as before, $R_{SSM}(\alpha)_i$ is the relative revenue of $m_i$.

\begin{definition}[Multi-player SSM Games]
For every $\alpha \in \mathcal{H}^M$, we define the {\em SSM Game}, $G_\alpha$ as a $M$-player binary action game. Each strategic miner has a binary action set $\{H,S\}$, where $H$ represents mining honestly and $S$ represents employing SSM. For convenience, we associate this action space with $\{0,1\}^M$, where action 0 denotes honest mining and action 1 denotes employing SSM. In order to specify utilities, we suppose that $x \in \{0,1\}^M$ is a pure action profile such that $x_i = 1$ and $x_j = 0$: 

\begin{itemize}
\item $U_i(x) = R_{SSM}(\alpha \circ x)_i$
\item $U_j(x) = \frac{\alpha_j}{1 - \alpha \cdot x} R_{SSM}(\alpha \circ x)_{M+1}$
\end{itemize}
\end{definition}

\subsubsection{Penalising Coalitions}

In Section \ref{sec:two-SSM} we explored scenarios where an agent might be unilaterally incentivised to use SSM, but a second larger agent can retaliate by employing SSM to make the original agent worse off than when everyone mines honestly. In the multiplayer setting, any subset of agents can retaliate in a similar fashion, thus we formally define what constitutes a penalising coalition. In what follows, we use the notation $\vec{\chi}_C \in \{0,1\}^M$ to denote an indicator vector for a subset $C \subset [M]$. 

\begin{definition}[Penalising Coalition]\label{def:multiplayer-penalising}
Suppose that $\alpha \in \mathcal{H}^M$ is a distribution of hash power amongst $M$ strategic miners. We say that $C \subset \{2,...,M\}$ is a penalising coalition for miner 1 if the following hold:
\begin{itemize}
\item $U_1(\vec{\chi}_1) > U_1(\vec{0})$ 
\item $U_i(\vec{\chi}_{ 1 \cup C}) > U_i(\vec{\chi}_{1 \cup C \setminus i})$ for all $i \in C$
\item $U_1(\vec{\chi}_{1 \cup C}) < U_1 (\vec{0})$
\end{itemize}
Furthermore, we say that $C$ incurs a penalty of $U_1 (\vec{0}) - U_1(\vec{\chi}_{1 \cup C})$ on miner 1 when retaliating
\end{definition}

The first condition ensures that miner 1 has a unilateral incentive to deviate and employ SSM. The second condition ensures that each miner in the penalising coalition is better off retaliating than defecting from the retaliation (ensuring retaliation is in a loose sense a credible threat), and finally the third condition ensures that miner 1 is worse off when being retaliated against than when everyone is honest.

\subsection{Partition Games in the Multiplayer Setting}
\label{subsec:partition-games-multiplayer}

In Section \ref{sec:partition-games}, we introduced the notion of a partition game, $G^P_\alpha$, which extended the action space of $G_\alpha$ to allow miners to partition their hash power into honest mining and employing SSM. This definition extends naturally to the $M$-player case.

\begin{definition}[Multi-player Partition Games]
Suppose that $\alpha \in \mathcal{H}^M$ is a hash distribution for $M$ strategic miners. We define the {\em Partition Game}, $G^P_\alpha$, as a $M$-player game, where each player has the same action set $[0,1]$, representing the proportion of their hash power dedicated to employing SSM. For a given pure strategy profile $s \in [0,1]^M$, we define the utility of the $i$-th player in $G^P_\alpha$ as follows:
\begin{itemize}
\item $U_i (s) = s_i R_{SSM}(s \circ \alpha)_i + (1 - s_i) \frac{\alpha_i (1 - s_i)}{1 - \sum \alpha_i s_i} R_{SSM}(s \circ \alpha)_{M+1}$
\end{itemize}
\end{definition}

\subsubsection{Optimal Commitments in $G^P_\alpha$}

As in the $M = 2$ miner case, for any miner $i$, every action $s_i \in (0,1)$ is dominated by either $s_i = 0$ or $s_i = 1$ if $G^P_\alpha$ is treated as a one shot game. The reason for this is that partitioning hash power results in unnecessary self competition, hence it will never be a best response to fixed opponent strategies. Consequently, the PNE of $G^P_\alpha$ as a one shot game are identical to the PNE of $G_\alpha$. 

On the other hand, we can once again treat $G^P_\alpha$ as a full information sequential game where $m_1$ commits to a strategy and all other $M-1$ players react. The subgame perfect Nash equilibria (SGPNE) of this game are generalisations of the Stackelberg equilbria of Section \ref{sec:partition-games}. The most subtle issue with generalising SSE however arises in tie-breaking. The assumption in SSE for two player games is that the follower will break ties in favour of the leader. This is a fair assumtion in the two-player setting, because it is often the case that commitments that lead to indifference in responses are of lower measure than those that invoke unique best responses. For this reason a leader can commit to strategies in an arbitrarily small neighbourhood of an SSE to ellicit the desired best response in the case of a tie for the follower. 

In the multi-player setting however, it can be the case that a non-trivial neighbourhood of leader commitments give rise to subgames with multiple PNE. For this reason it may be unfeasible to assume that follower agents converge to a PNE that maximises the welfare of the leader, as there is nothing in the power of the leader to even approximately guarantee this behaviour. For this reason, we take a pessimistic approach to SGPNE of $G^P_\alpha$. In particular, we assume that for a leader commitment, all other agents will settle on a PNE that minimises welfare for the leader. To be precise, for a given pure strategy $s_1 \in [0,1]$, we let $G^P_\alpha(s_1,-)$ denote the $(M-1)$-player subgame for miners $2,...,M$ conditioned on miner 1 committing to $s_1$. Furthermore, we let $WSN(s_1)$ (Worst sub-Nash) be the lowest utility pure Nash equilibrium of $G^P_\alpha(s_1,-)$ for miner 1. The value of commitment $s_1$ in the leadership game $G^P_\alpha$ for miner 1 is $v_1(s_1) = U_1(s_1,WSN(s_1))$, and for any other miner $i = 2,...,M$, $v_i(s_1) = U_i(s_1,WSN(s_1))$. We call the family of all such pure strategy profile the collection of {\em Pessimistic Sub-game Perfect Nash Equilibria}, (P-SGPNE). In particular, we are interested in values of $\alpha$ where the set of P-SGPNE of $G^P_\alpha$ result in strictly larger welfare for $m_1$, implying that either the possiblity of commitment or partitioining strictly benefits $m_1$ in the worst case.  

In a similar fashion to the two-player case, we study how different values of $\alpha \in \mathcal{H}^M$ give rise to different P-SGPNE($G^P_\alpha)$ vs PNE($G_\alpha$) = PNE($G^P_\alpha$). 

\begin{definition}[Multiplayer Commitment/ SGPNE Types]
\label{def:multi-com-type}
Suppose that $\alpha \in \mathcal{H}^M$, we classify its commitment type, $com(\alpha)$, depending on the relationship between the sets P-SGPNE($G^P_\alpha)$ and PNE$(G_\alpha)$
\begin{itemize}

	\item If $\text{P-SGPNE}(G^P_\alpha) = \text{PNE}(G_\alpha)$, then $com(\alpha) = 0$.
	\item If $\text{P-SGPNE}(G^P_\alpha) \subset \text{PNE}(G_\alpha)$, then $com(\alpha) = 1$.
	\item If $\text{P-SGPNE}(G^P_\alpha) \not \subset \text{PNE}(G_\alpha)$ and $\exists s^* \in \text{P-SGPNE}$ such that $s^*_1 \in \{0,1\}$, then $com(\alpha) = 2$
	\item If $\text{P-SGPNE}(G^P_\alpha) \not \subset \text{PNE}(G_\alpha)$ and $\not \exists s^* \in \text{P-SGPNE}$ such that $s^*_1 \in \{0,1\}$, then $com(\alpha) = 3$. 
\end{itemize}
\end{definition}

As in the two-player case, if $com(\alpha) = 0$ we say $\alpha \in \mathcal{H}^M$ gives rise to a trivial commitment and that the collection of P-SGPNE in $G^P_\alpha$ are trivial. Accordingly, if $com(\alpha) \neq 0$, we say $\alpha$ gives rise to a non-trivial commitment and the collection of P-SGPNE in $G^P_\alpha$ is non-trivial. Furthermore, we also say that if $\alpha \in \mathcal{H}^M$ is such that $com(\alpha) = i$, then all $s^* \in \text{P-SGPNE}(G^P_\alpha)$ are of type $i$ as well. 

\section{Results for $M = 3$ Miners}
\label{sec:3SSM}

In order to visualise results for $M = 3$ miners, we fix the hash rate of the third player, $\alpha_3$ and repeat our analysis for Section \ref{sec:two-SSM} when $\alpha_1$ and $\alpha_2$ are allowed to vary. We observe qualitative difference in the family of games $G_\alpha$ for four different regions of $\alpha_3$ values: $R_1 = [0,0.17]$, $R_2 = [0.175,0.203]$, $R_3 = [0.208, 0.27]$ and $R_4 = [0.274, 0.5]$. 



\subsection{Pure Nash Equilibria in $G_\alpha$}

As mentioned in the previous section, our results show four main regimes of results as a function of $\alpha_3$. In terms of PNE, When $\alpha_3 \in R_1$, $H$ strictly dominates $S$ for miner 3, hence the three player games $G_\alpha$ and $G^P_\alpha$ reduce to a two player game conditioned on player 3 playing $H$. For $\alpha_3 \in R_2$, we see the emergence of $SSS$ as a PNE near the centre of the hash space, and the size of this region grows as a function of $\alpha_3$. For $\alpha_3 \in R_3$, $SSS$ is still a PNE for central values of $\alpha$, however we see the emergence of distinct regions where $SHS$ and $HSS$ are PNE. Finally, when $\alpha_3 \in R_4$, $S$ strictly dominates $H$ for player 3, and once again $G_\alpha$ and $G^P_\alpha$ reduce to subgames conditioned on miner 3 playing $S$. 

\begin{figure}
\centering
\includegraphics[width = 0.24\textwidth]{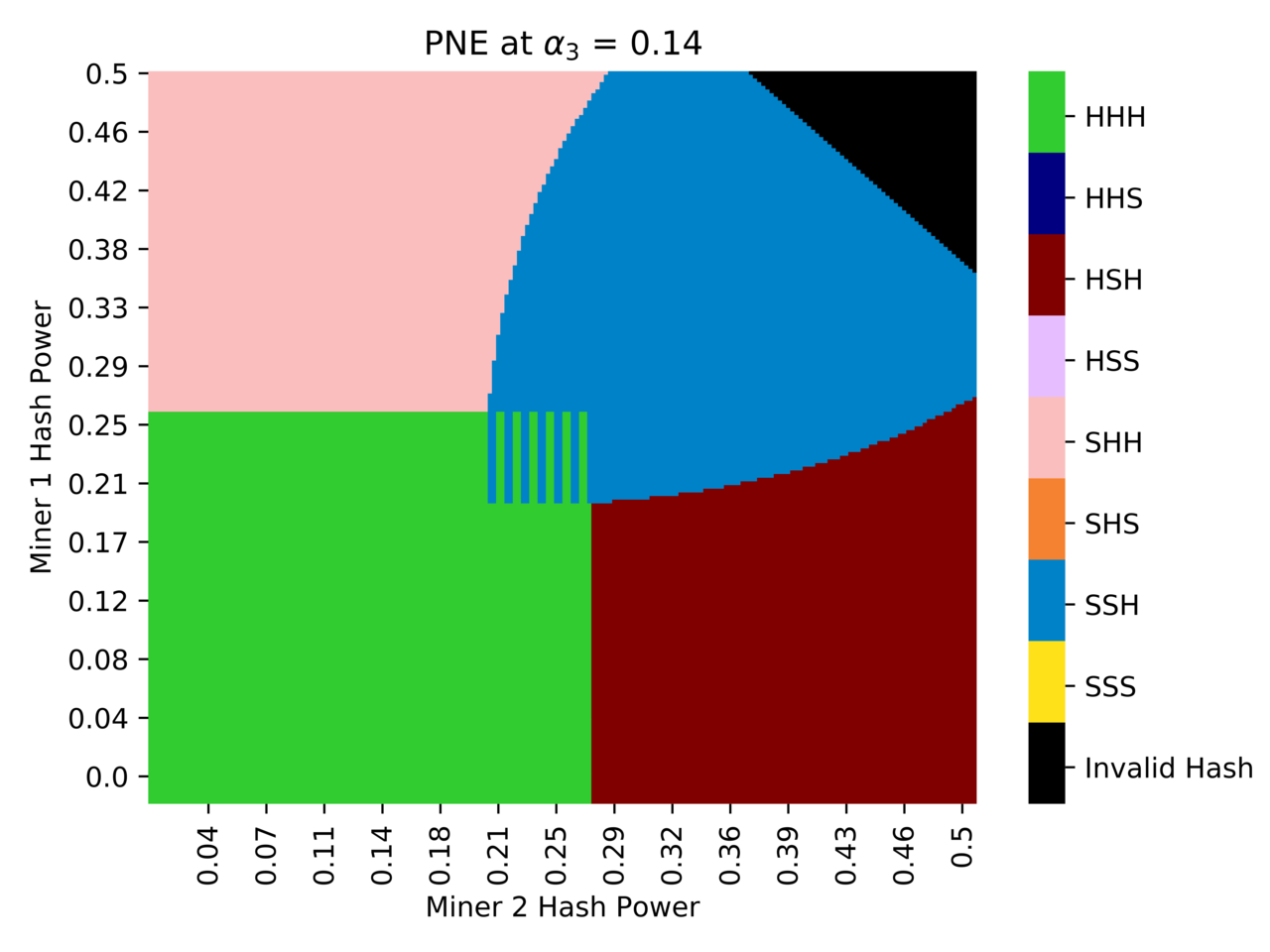}
\includegraphics[width = 0.24\textwidth]{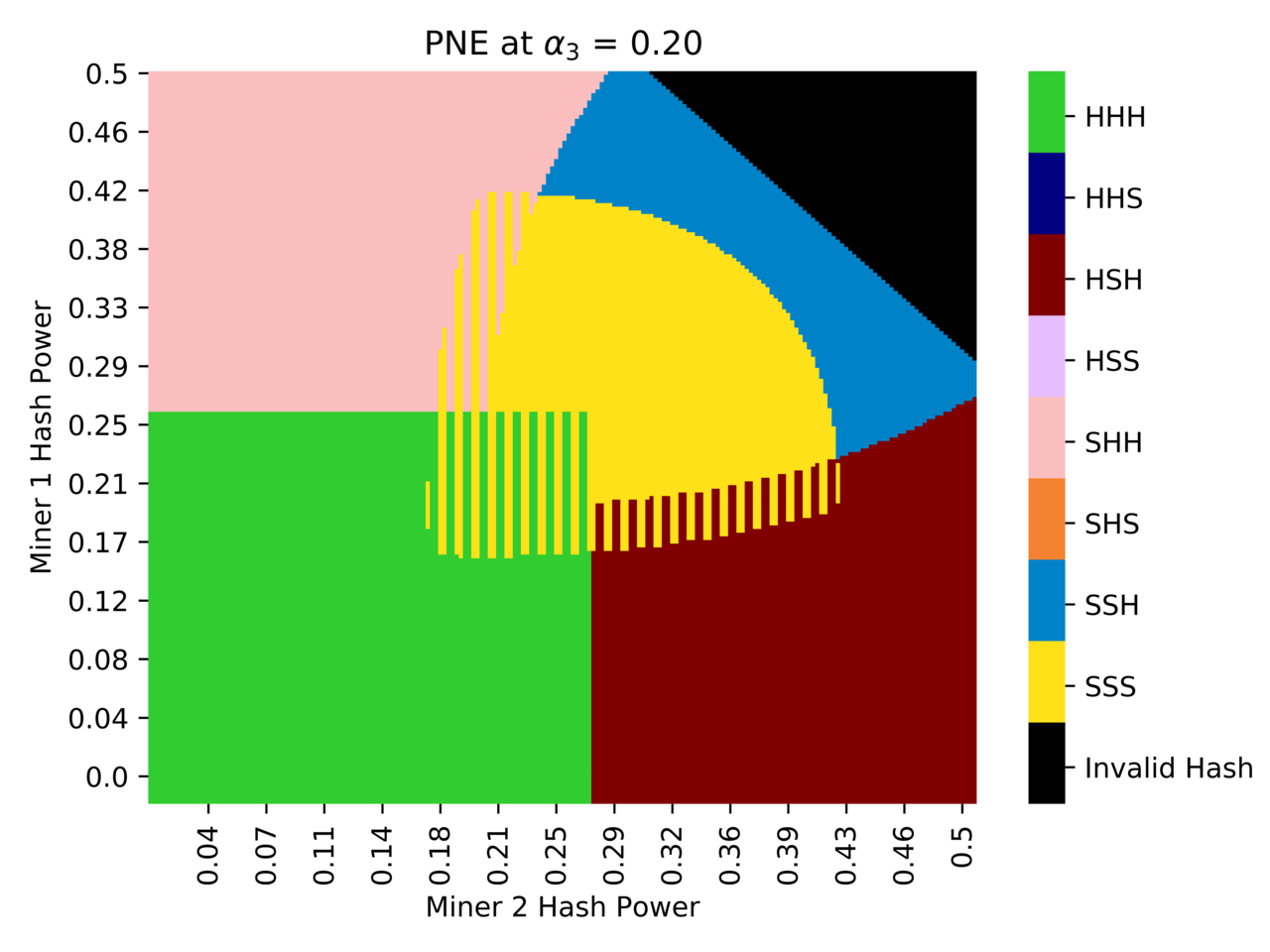}
\includegraphics[width = 0.24\textwidth]{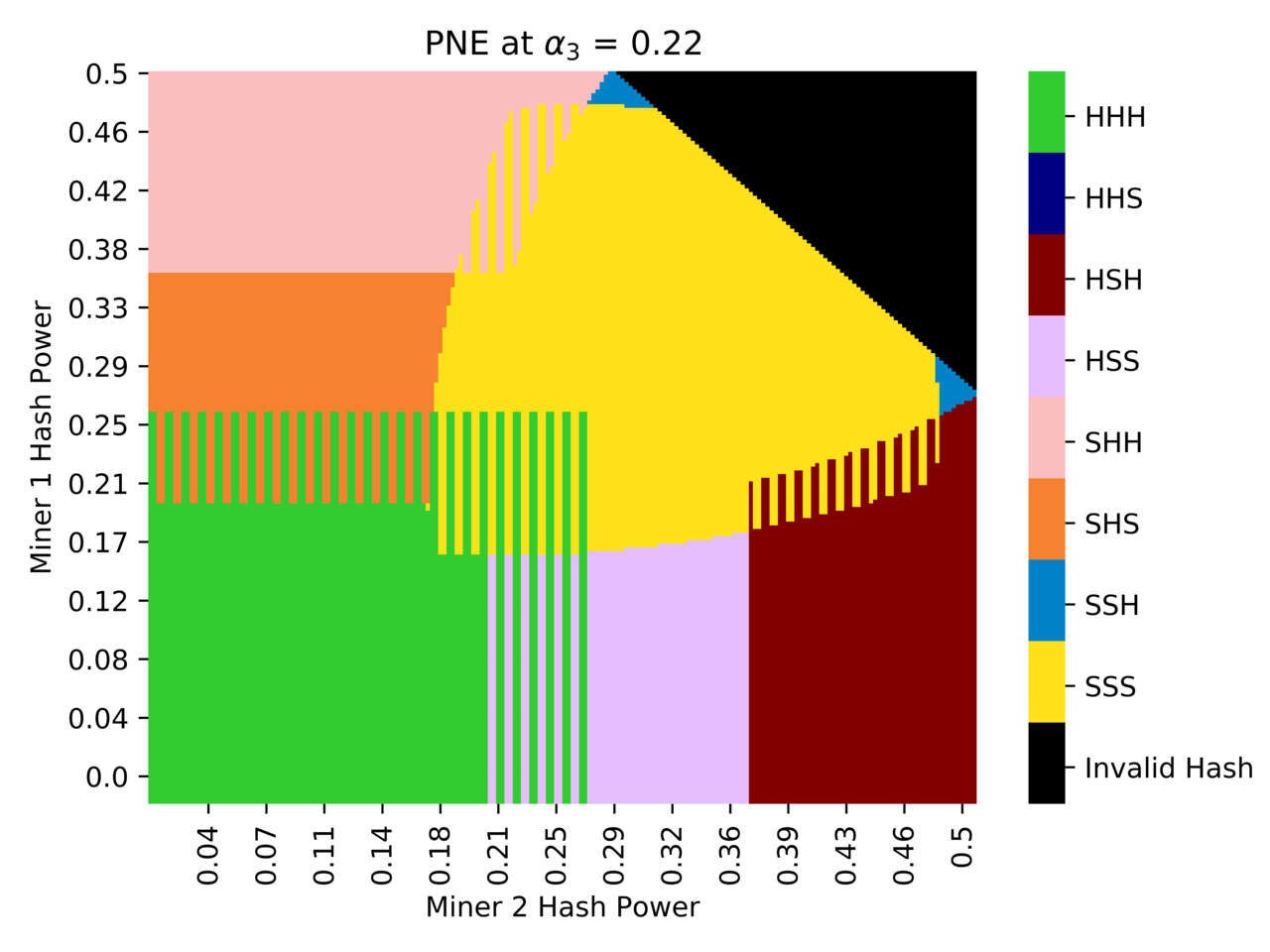}
\includegraphics[width = 0.24\textwidth]{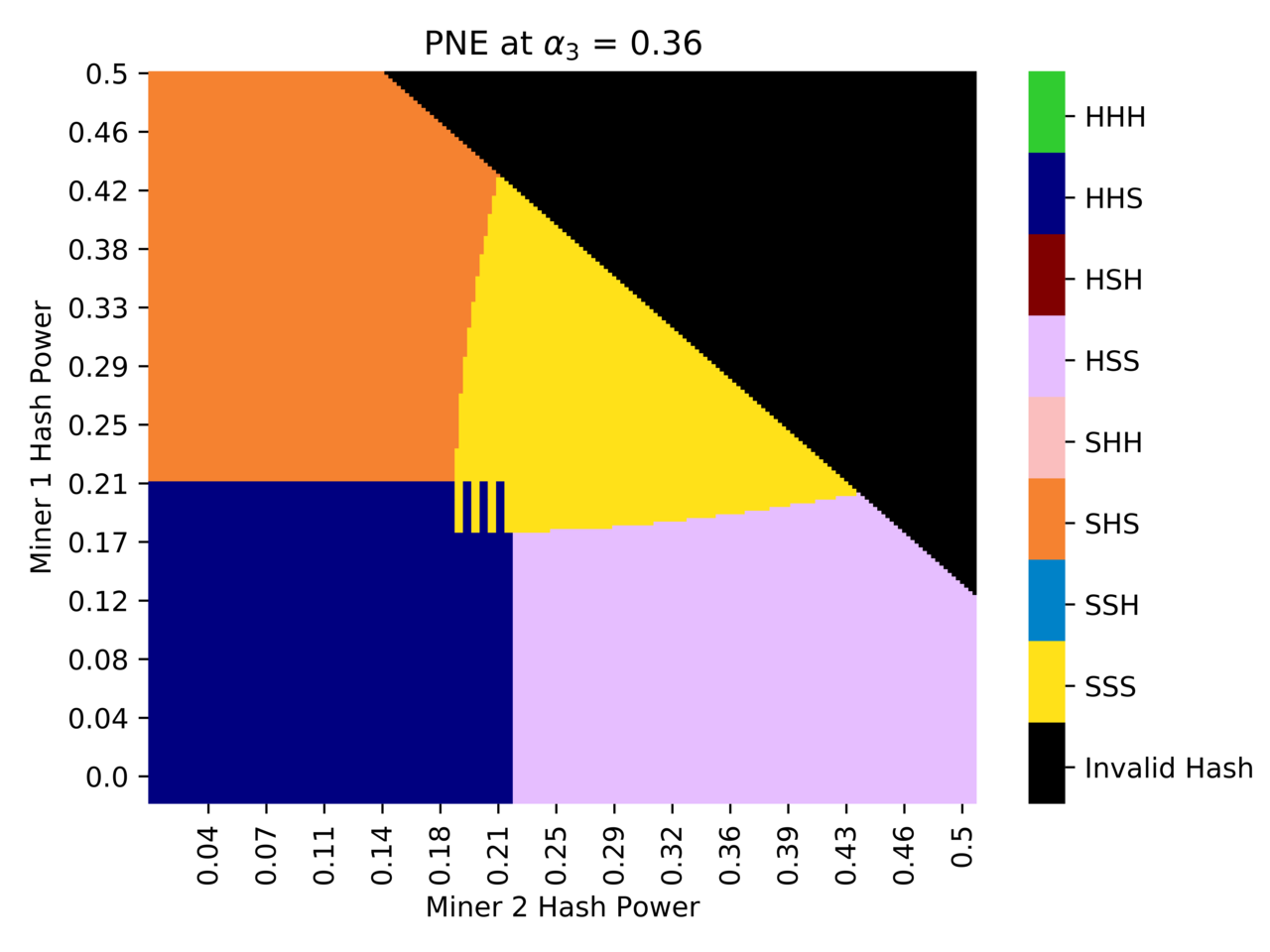}
\includegraphics[width = 0.24\textwidth]{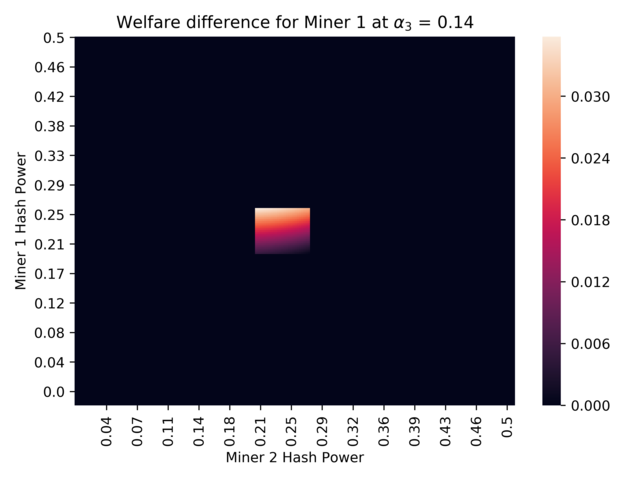}
\includegraphics[width = 0.24\textwidth]{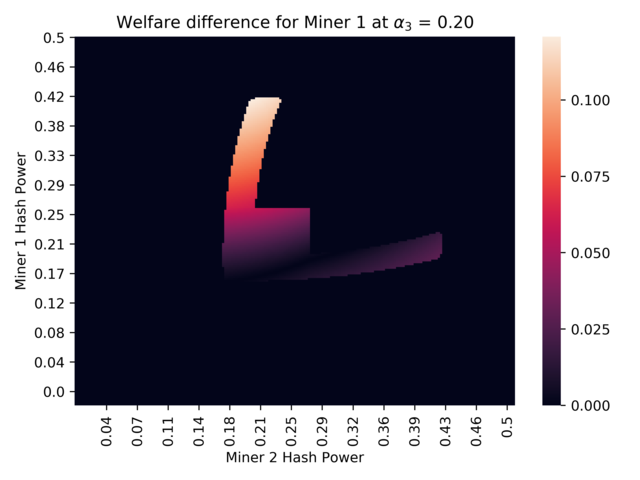}
\includegraphics[width = 0.24\textwidth]{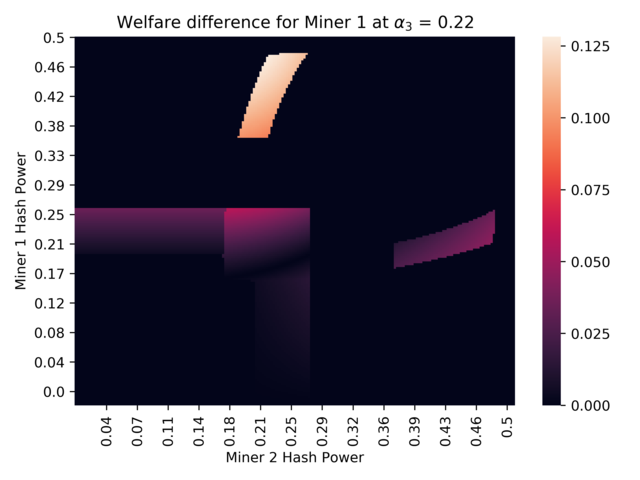}
\includegraphics[width = 0.24\textwidth]{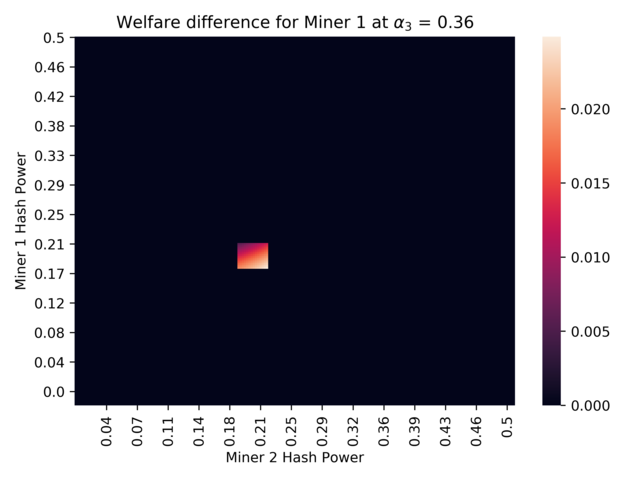}
\includegraphics[width = 0.24\textwidth]{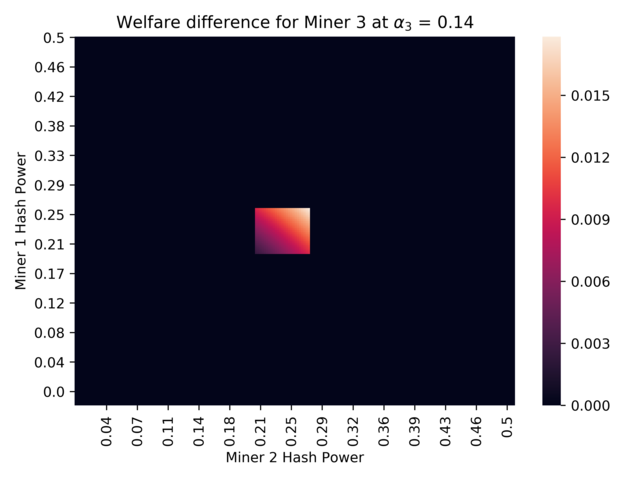}
\includegraphics[width = 0.24\textwidth]{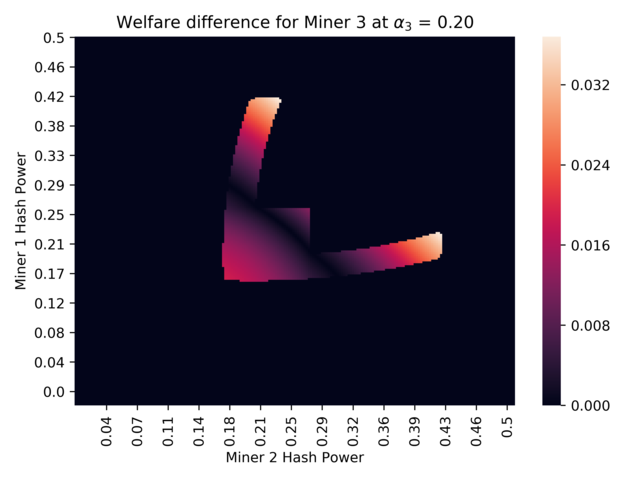}
\includegraphics[width = 0.24\textwidth]{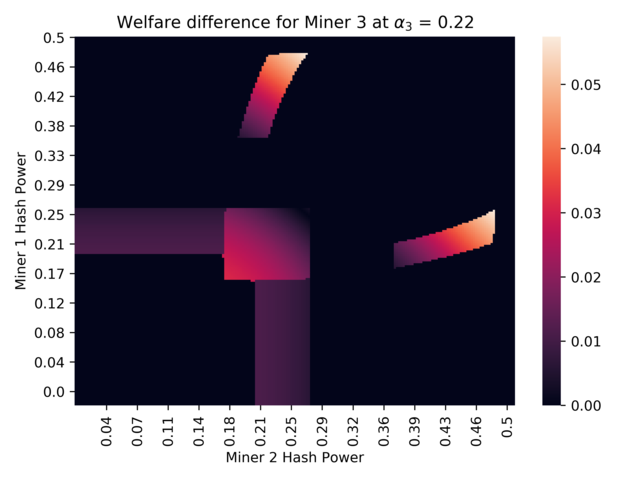}
\includegraphics[width = 0.24\textwidth]{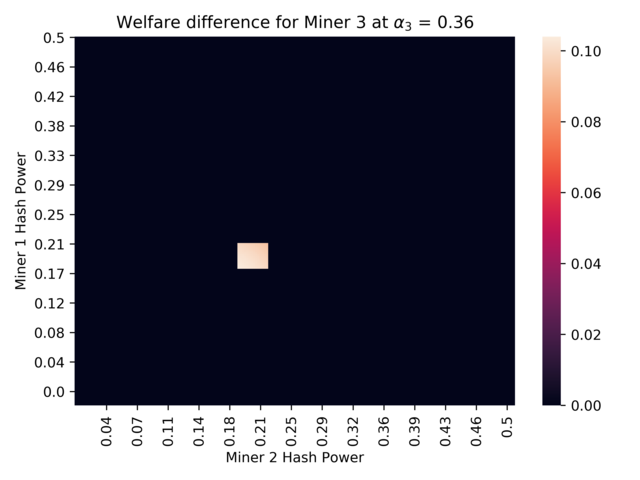}
\caption{PNE for $\alpha_3 \in \{0.14, 0.2, 0.22, 0.36\}$. For the areas that have multiple PNE, the difference in welfare at the best PNE and worst PNE for miner 1 and miner 3 are mapped in the second and third rows respectively. Since $G_\alpha$ is an anonymous game, the difference in welfare for miner 2 is the same as that of miner 1 reflected about the axis $y = x$. 
}
\label{fig:3SSM-PNE-welfare}
\end{figure}

In Figure \ref{fig:3SSM-PNE-welfare}, we visualise this phenomenon by picking representative values of $\alpha_3$ in $R_1,R_2,R_3$ and $R_4$ and graphing regions where distinct PNE occur in $G_\alpha$ as well as the difference in welfare between the best and worst PNE for each player respectively.

\subsection{SSM Profitability Threshold Diminished}

\begin{figure}
\centering
\includegraphics[width = 0.24\textwidth, trim={0 0 9mm 0},clip]{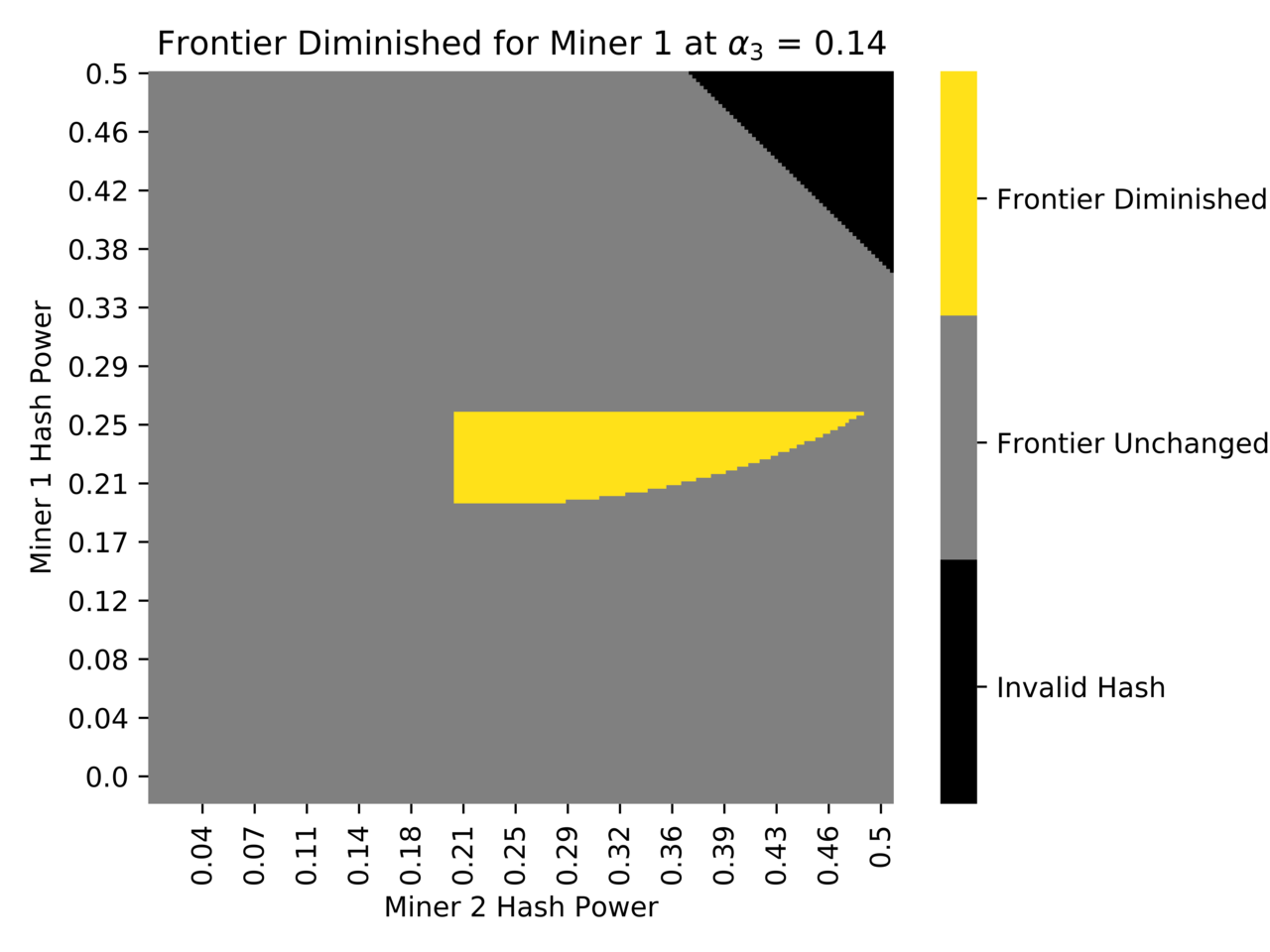}
\includegraphics[width = 0.24\textwidth, trim={0 0 9mm 0},clip]{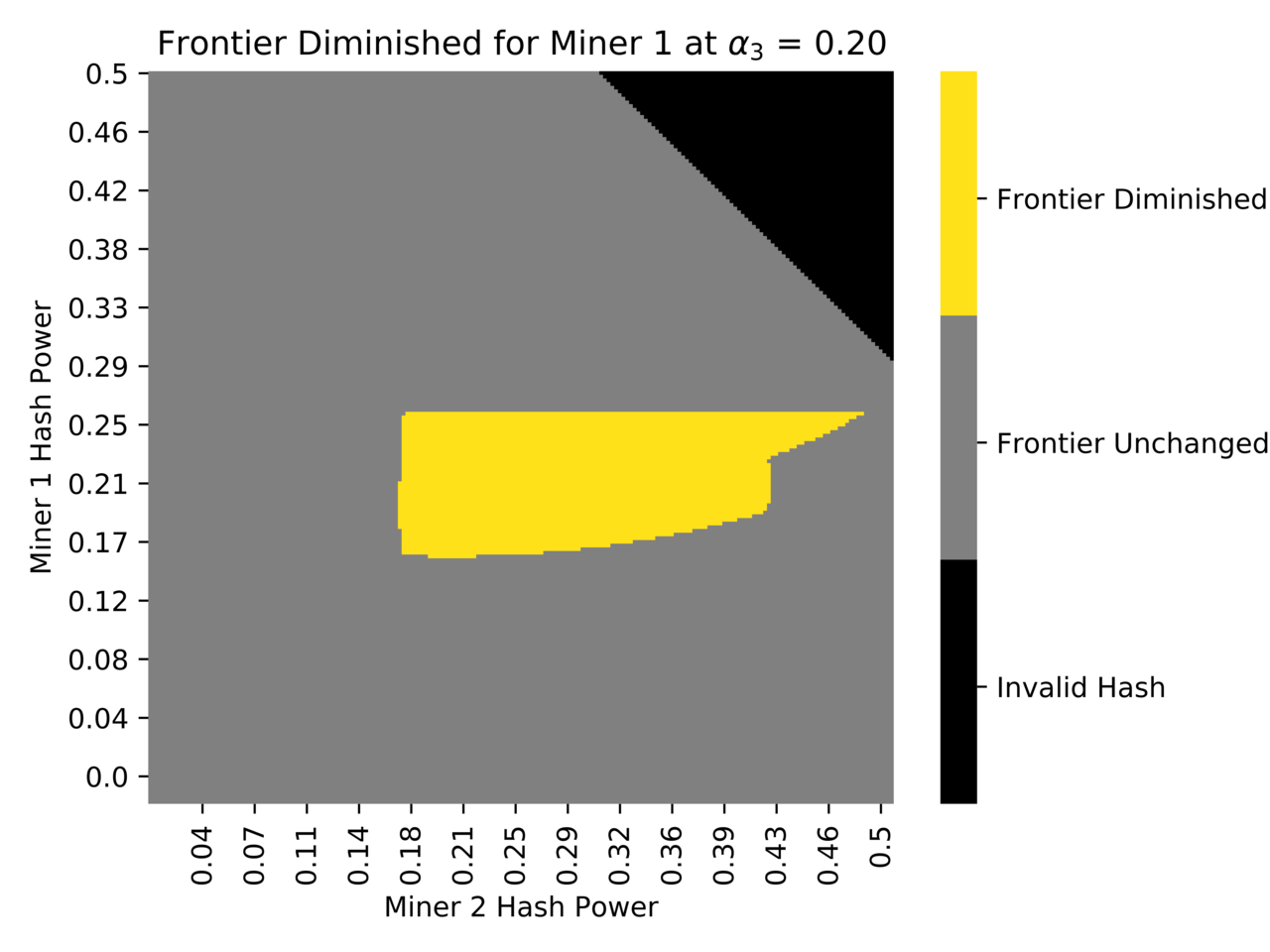}
\includegraphics[width = 0.24\textwidth, trim={0 0 9mm 0},clip]{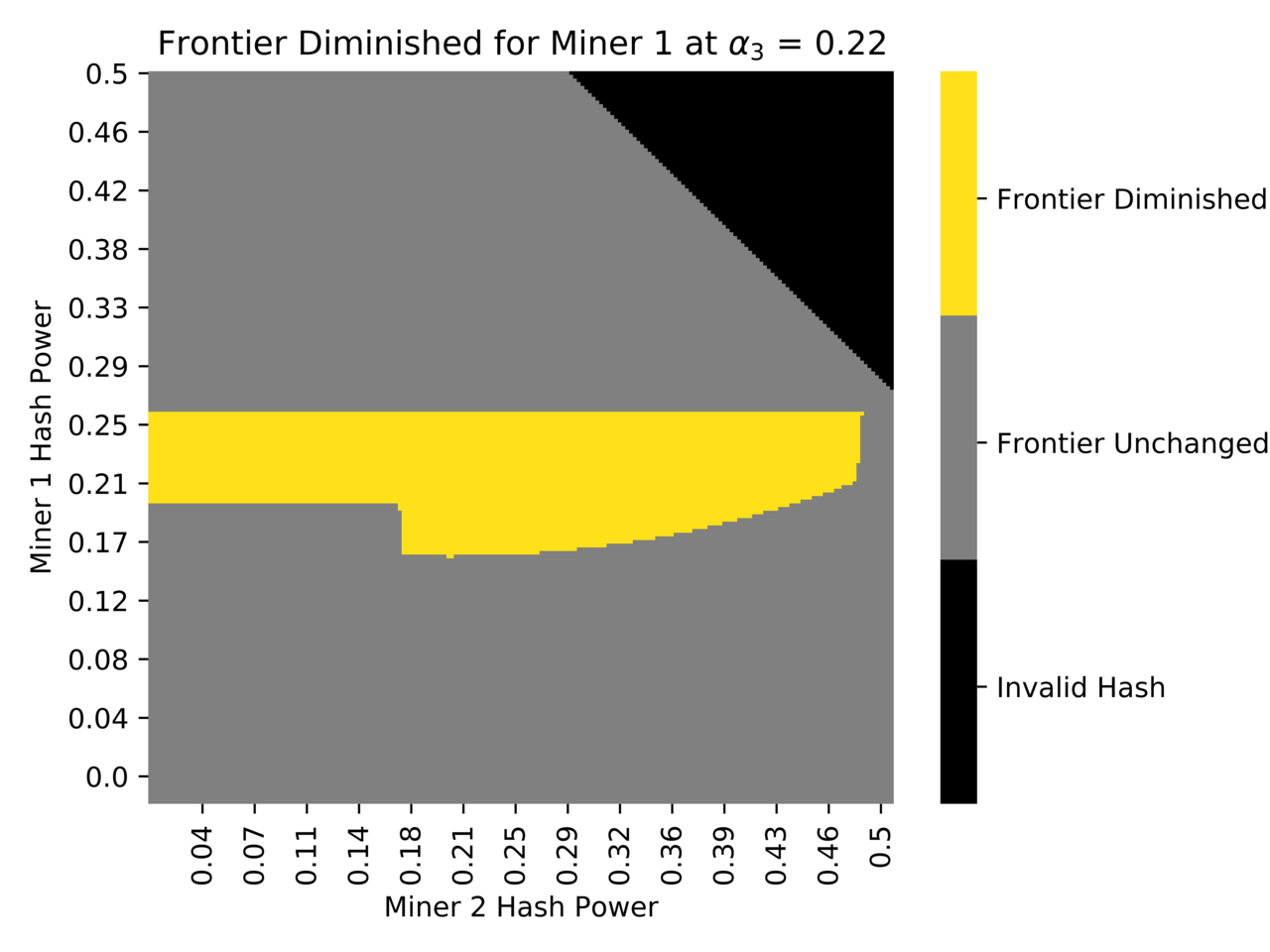}
\includegraphics[width = 0.24\textwidth, trim={0 0 9mm 0},clip]{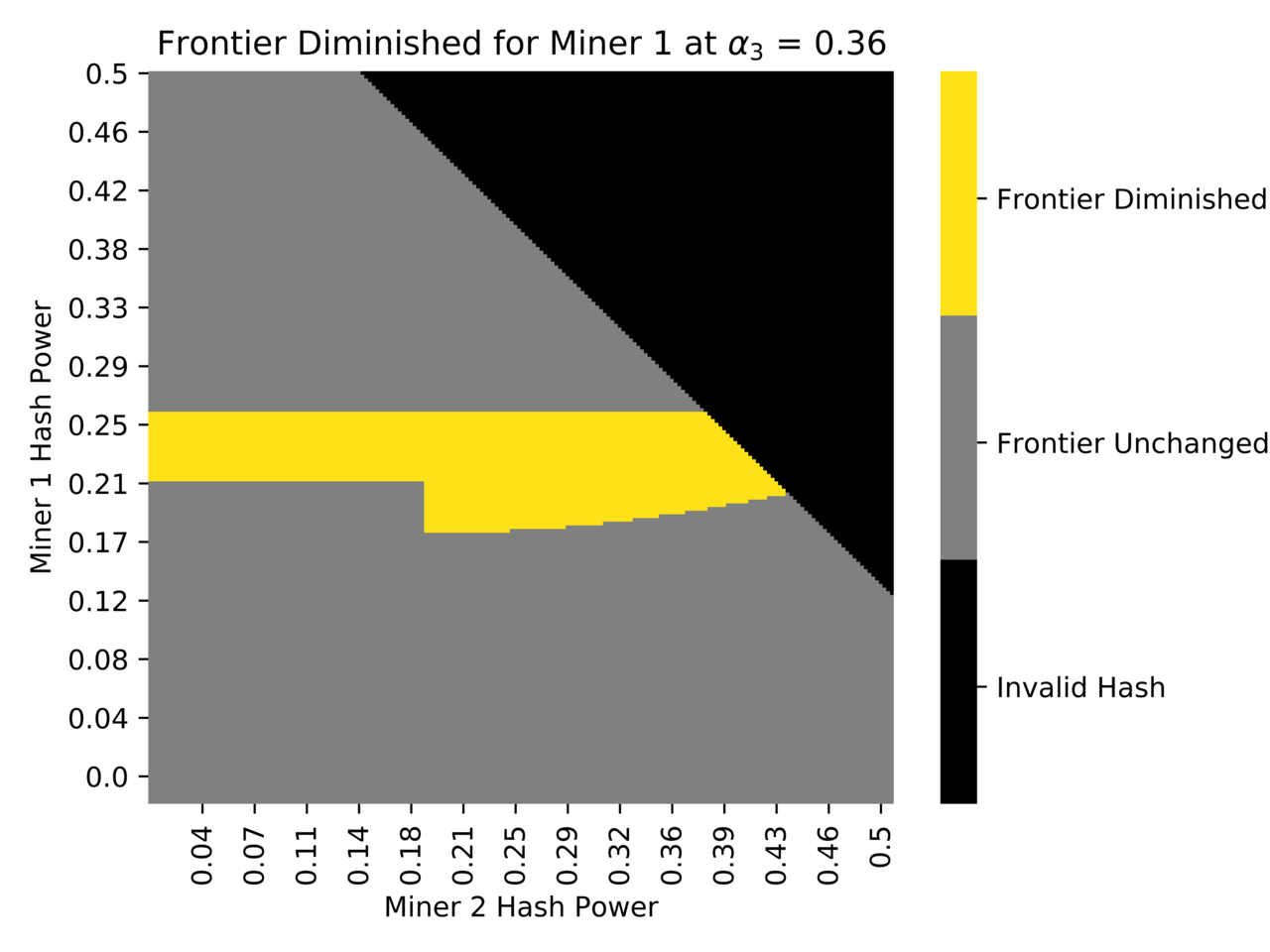}
\caption{Profitability Threshold Diminished for $\alpha_3 \in \{0.14, 0.2, 0.22, 0.36\}$}
\label{fig:3SSM-frontier}
\end{figure}

As in the two-player case, we find that there are hash rates where $m_1$ is not unilaterally incentivised to employ SSM, yet there exist equilibria where $m_1$ employs SSM. In Figure \ref{fig:3SSM-frontier} we visualise the hash rates where this happens for all $R_i$ relevant regions of $\alpha_3$ values. In particular, for $\alpha_3$ values in $R_2, R_3$ and $R_4$, we see that the emergence of SSS as a PNE can occur when $m_1$ has much smaller hash power than the 0.26795 necessary to make SSM profitable unilaterally.

\subsection{Optimal Commitments}

As mentioned in Section \ref{sec:multiplayer-game-formalism}, we treat $G^P_\alpha$ as a full information sequential game where $m_1$ commits to a strategy and all other miners subsequently act. We recall that for a given pure strategy commitment $s_1 \in [0,1]$ for $m_1$, the worst Nash equilibrium of the resulting subgame $G_\alpha(s_1,-)$ for $m_1$ (so in terms of $U_1$) is denoted by $WSN(s_1)$. In addition, $v_i(s_1) = U_i(s_1,WSN(s_1))$ for $i = 1,...,M$, denotes the utility obtained by each miner at $(s_1,WSN(s_1))$. 

\begin{figure}
\centering
\includegraphics[width = 0.24\textwidth]{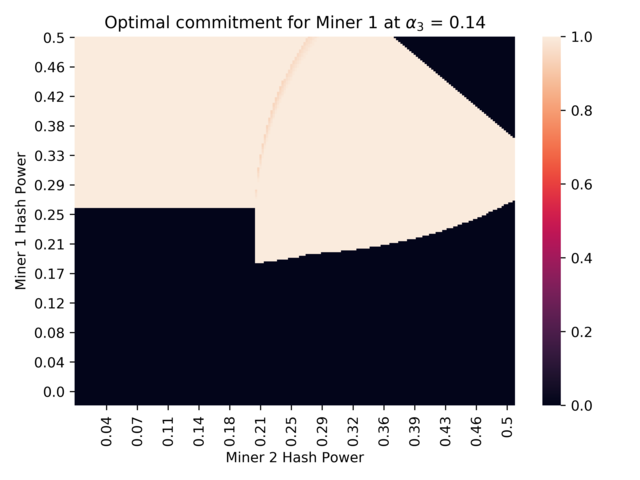}
\includegraphics[width = 0.24\textwidth]{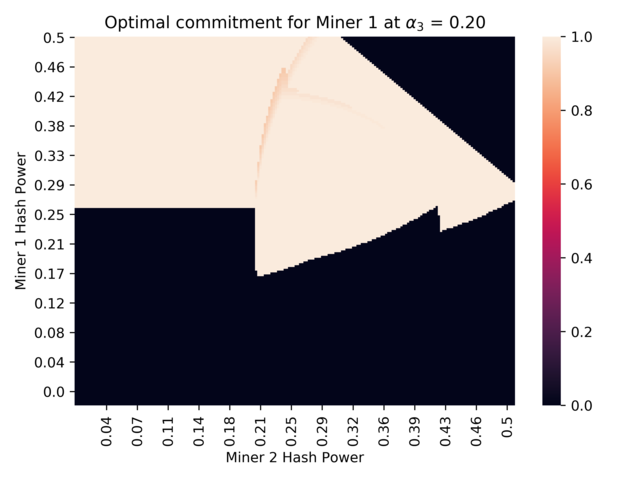}
\includegraphics[width = 0.24\textwidth]{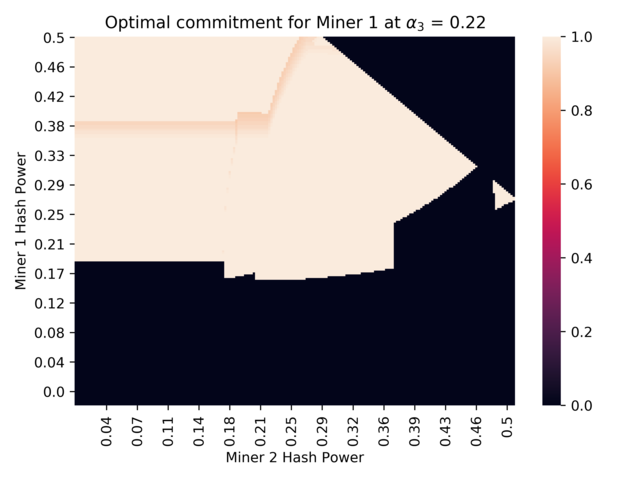}
\includegraphics[width = 0.24\textwidth]{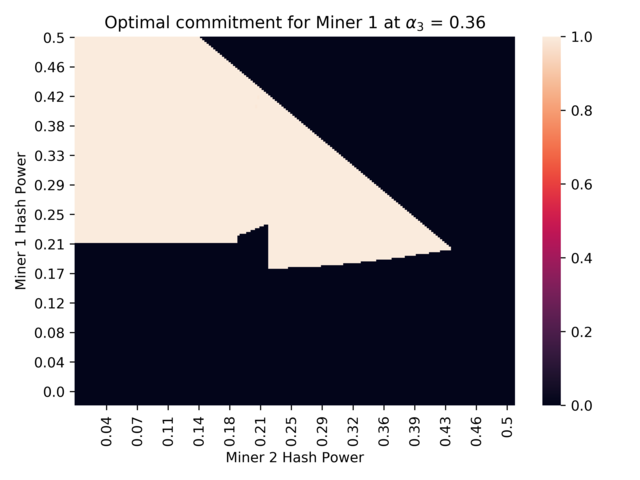}
\includegraphics[width = 0.24\textwidth]{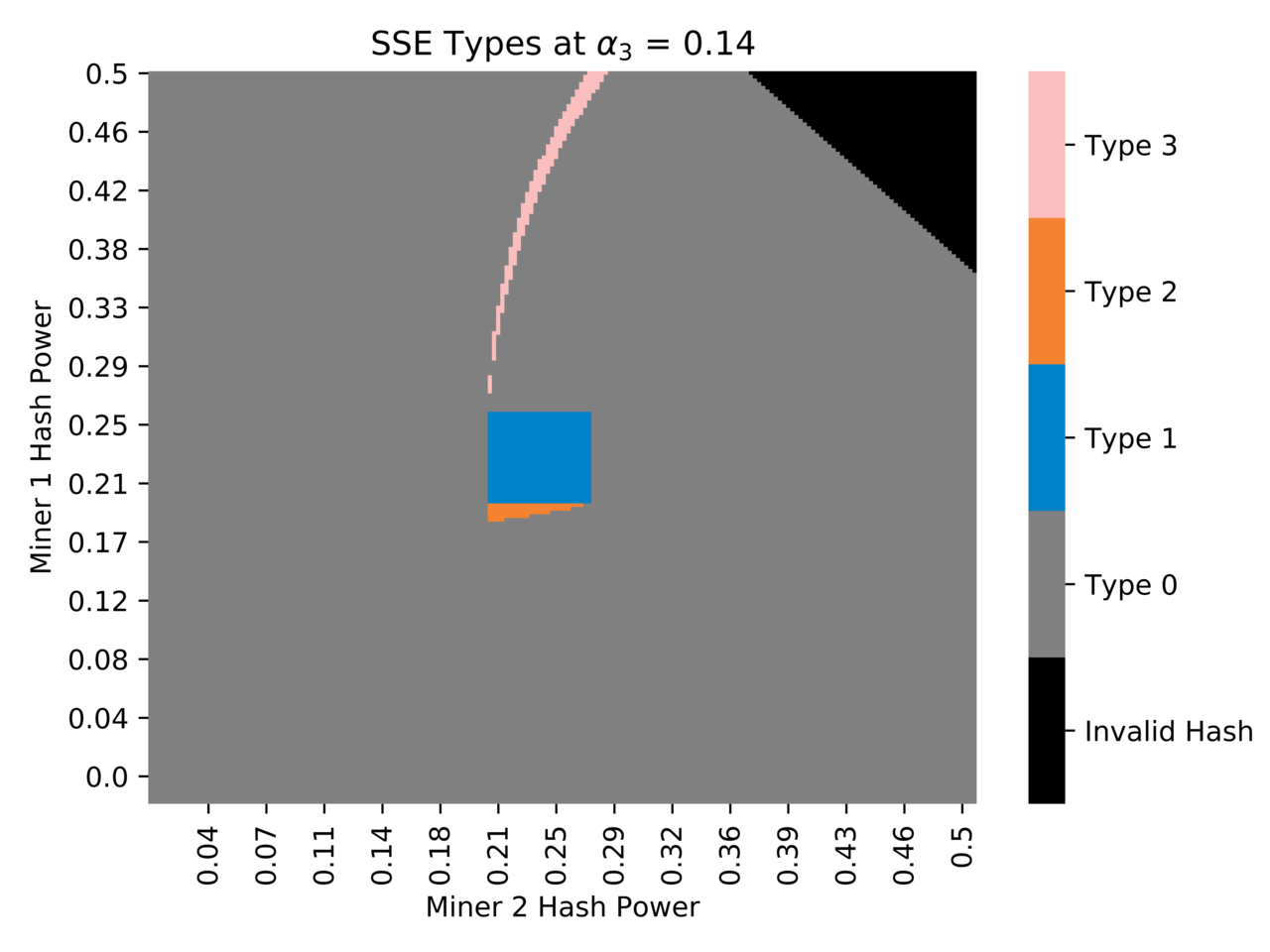}
\includegraphics[width = 0.24\textwidth]{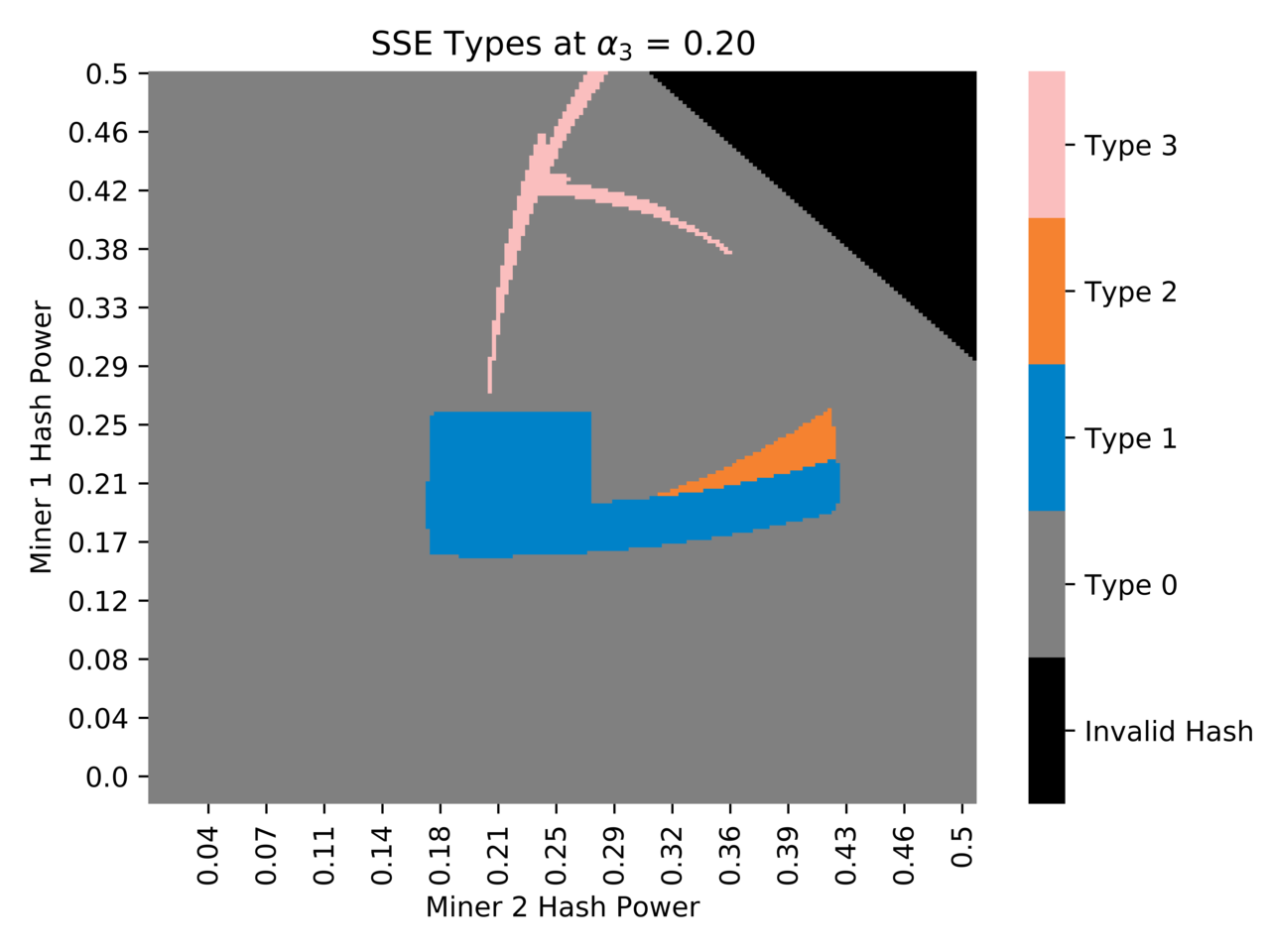}
\includegraphics[width = 0.24\textwidth]{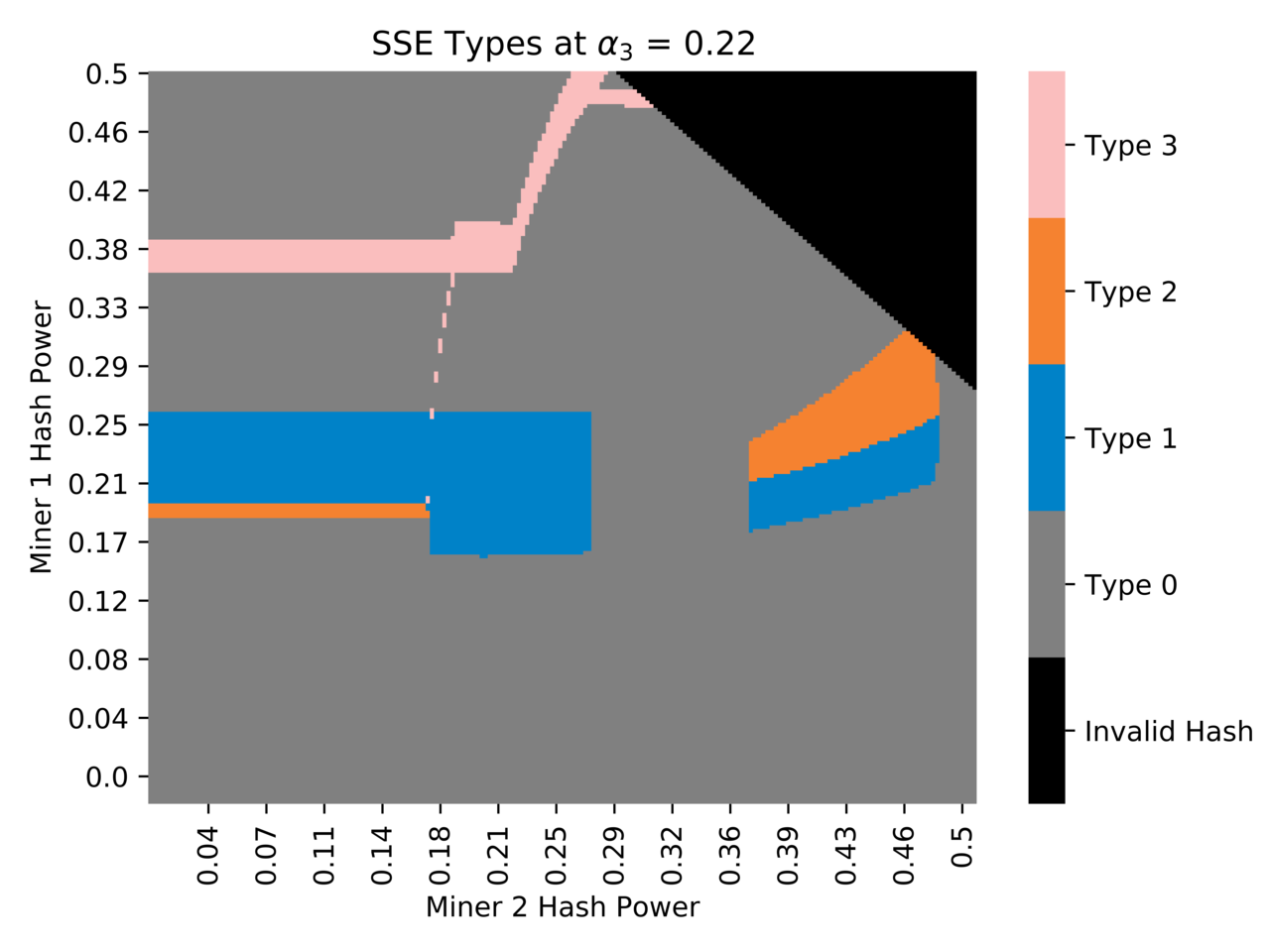}
\includegraphics[width = 0.24\textwidth]{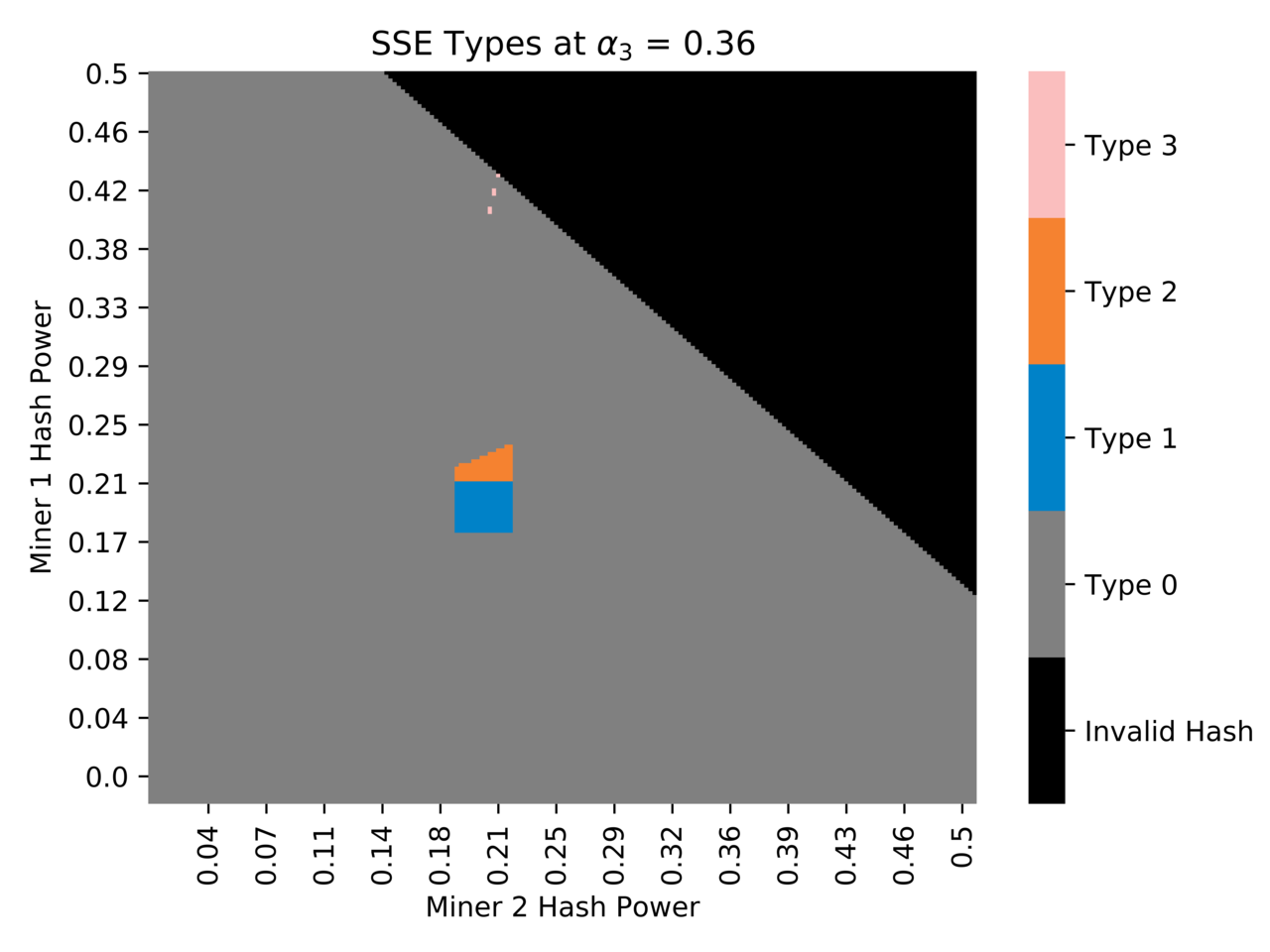}
\includegraphics[width = 0.24\textwidth]{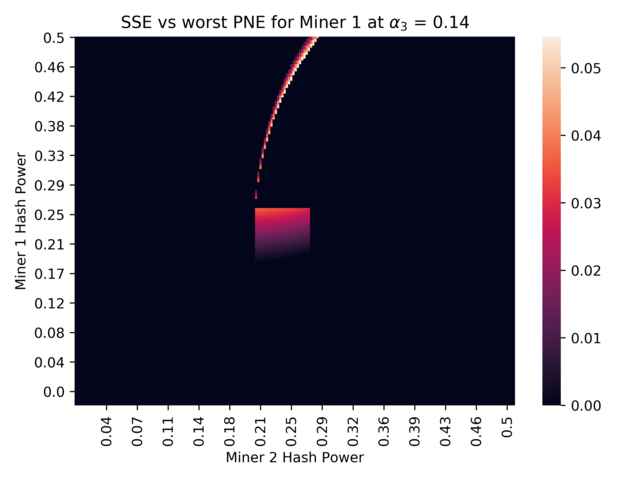}
\includegraphics[width = 0.24\textwidth]{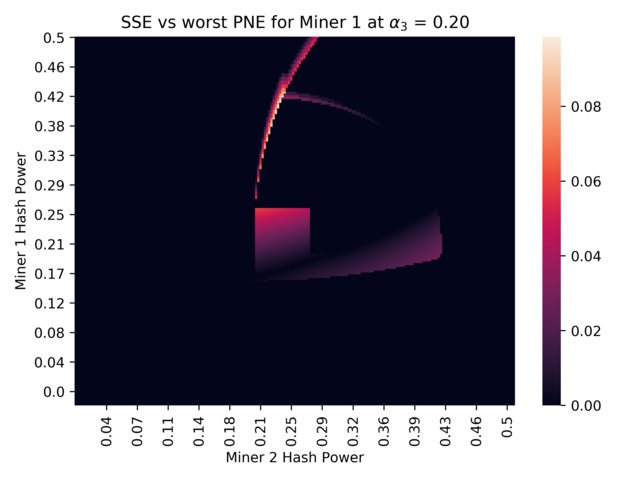}
\includegraphics[width = 0.24\textwidth]{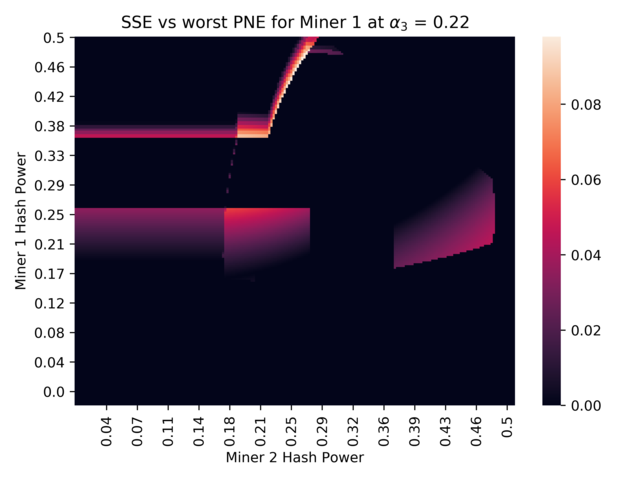}
\includegraphics[width = 0.24\textwidth]{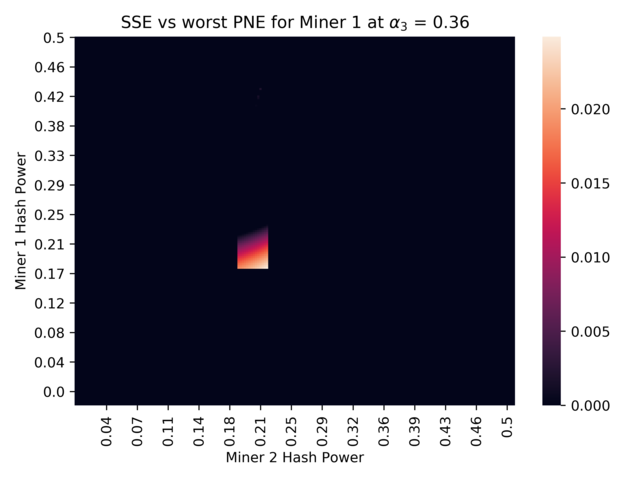}

\caption{Optimal $m_1$ commitment for $\alpha_3 \in \{0.14, 0.2, 0.22, 0.36\}$, commitment types, and utility surplus in P-SGPNE vs. worst PNE for $m_1$.}
\label{fig:3SSM-opt_com}
\end{figure}

For $\alpha_3 \in R_1,R_2,R_3,R_4$, we look at what values of $s_1$ optimise $v_1(s_1)$ as optimal commitments from the leader of $G_\alpha^P$, $m_1$. As mentioned in Section \ref{sec:multiplayer-game-formalism}, any $s^* = (s_1^*, WSN(s_1^*))$ that optimises $v_1$ is necessarily a (pessimistic) subgame perfect Nash equilibrium. In Figure \ref{fig:3SSM-opt_com} we plot these optimal commitments with fixed $\alpha_3$ as a function of $(\alpha_1,\alpha_2)$. Furthermore, we also plot $com(\alpha)$ as per Definition \ref{def:multi-com-type}, and the subsequent surplus between $v_1$ at the aforementioned pessimistic SGPNE and the worst lowest utility PNE for $m_1$. Similar observations can be made as in the two-player case of Section \ref{sec:partition-games}:

\begin{itemize}
    \item When $com(\alpha) = 0$, pessimistic SGPNE of $G^P_\alpha$ are identical to $PNE$ of $G_\alpha$, so the ability to partition and the ability to commit to strategies do not give $m_1$ an undue advantage in the worst case.
    \item When $com(\alpha) = 1$ it is generally the case that $G_\alpha$ has multiple PNE, and the commitment of $m_1$ ``nudges'' other players to a PNE that Pareto-dominates the worst PNE in $G_\alpha$.  
    \item The only exception to the previous observation is the left-most region of $com(\alpha) = 1$ when $\alpha_3 = 0.2$ is fixed (second column of Figure \ref{fig:3SSM-opt_com}). In this area, the optimal commitment for $m_1$ is $s_1 = 0$. In response to this, the subgame $G_\alpha(0,-)$ only has $HH$ as a PNE. As a consquence, the only pessimistic SGPNE at these $\alpha$ values is $(0,0,0)$, yet both $(0,0,0)$ and $(1,1,1)$ are PNE in $G_\alpha$. The reason for this however, is that if we consider the commitment $s_1 = 1$ (i.e. $m_1$ employing SSM), then $G_\alpha(1,-)$ actually has two PNE: $HH$ and $SS$. The worst of these two equilibria however is $HH$, and thus the strategy profile $(1, WSN(1)) = (1,0,0)$, which is strictly worse than $(0,0,0)$ fpr $m_1.$ 
    \item When $com(\alpha) = 2$ there exist pessimistic SGPNE, $s^* = (s_1^*, WSN(s_1^*)) \notin \text{PNE}(G_\alpha)$ such that $s_1 \in \{0,1\}$ and $s^*_1$ is not a best response to $WSN(s_1^*)$ for $m_1$.These non-trivial commitments make use of sequentiality of $G^P_\alpha$ but not of the augmented action space granted by partitioning. 
    \item When $com(\alpha) = 3$, $m_1$ has enough hash power that PNE($G_\alpha$) only has strategy profiles that exemplify $m_2$ and $m_3$ being disincentivised to use SSM. That being said, at these values of $\alpha$, $m_2$ and $m_3$ are almost indifferent between employing SSM and honest mining (hence the reason $com(\alpha) = 3$ occurs along boundaries of where PNE($G_\alpha$) changes values), hence $m_1$ can bait them into employing SSM by judiciously giving away some hash power to honest mining in a partition. 
\end{itemize}

\subsection{Penalising Coalitions}

\begin{figure}
\centering
\includegraphics[width = 0.24\textwidth]{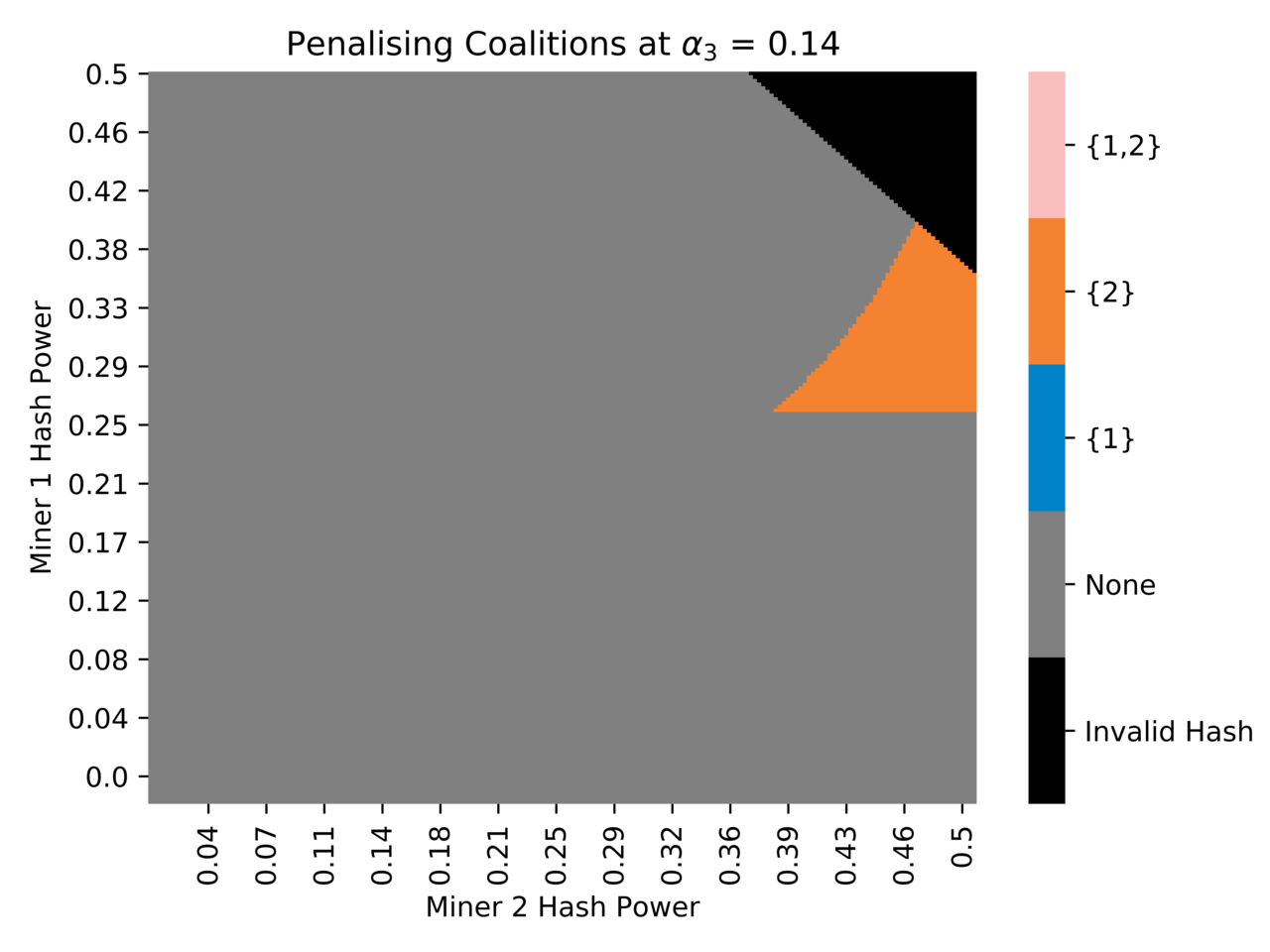}
\includegraphics[width = 0.24\textwidth]{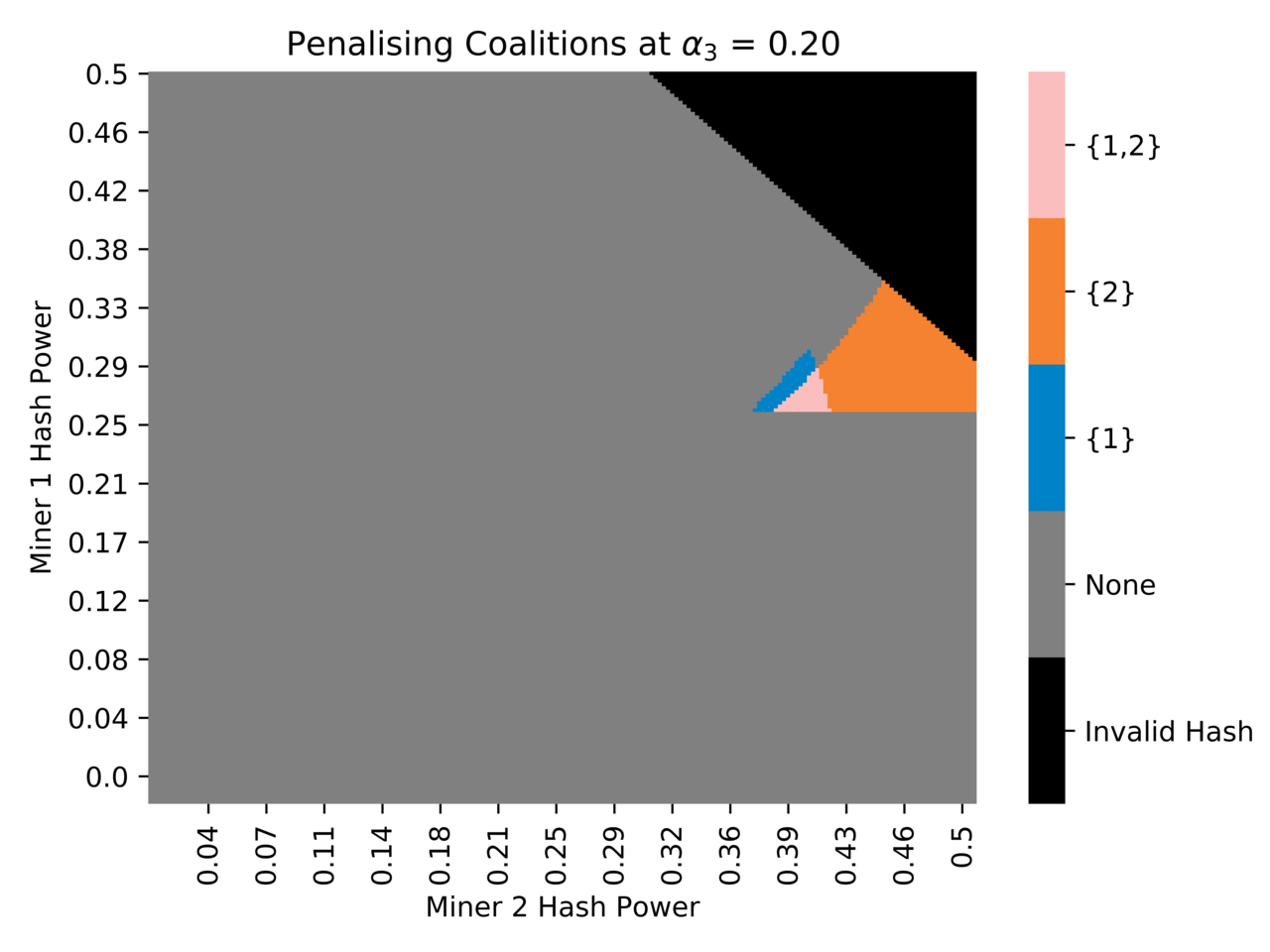}
\includegraphics[width = 0.24\textwidth]{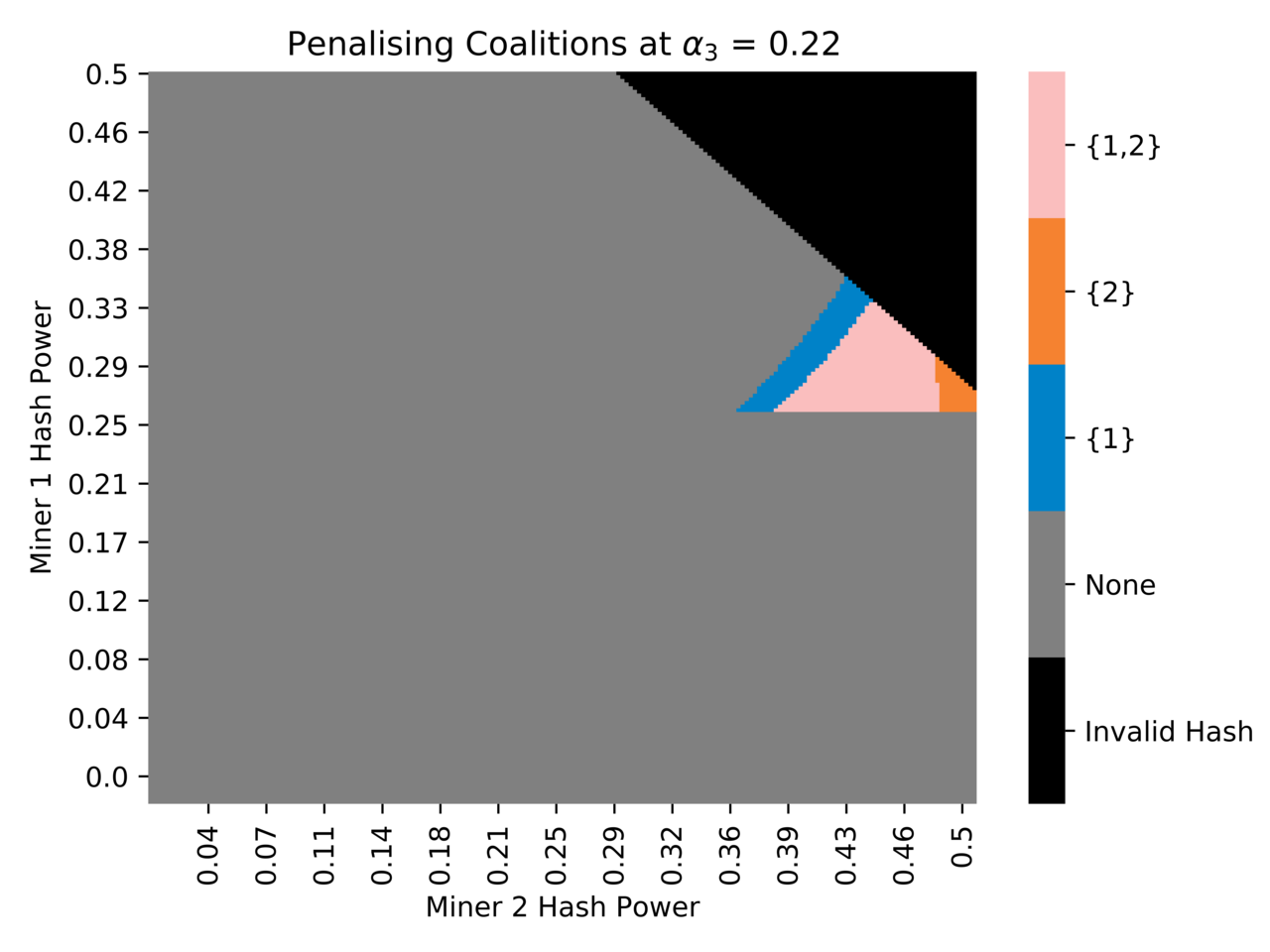}
\includegraphics[width = 0.24\textwidth]{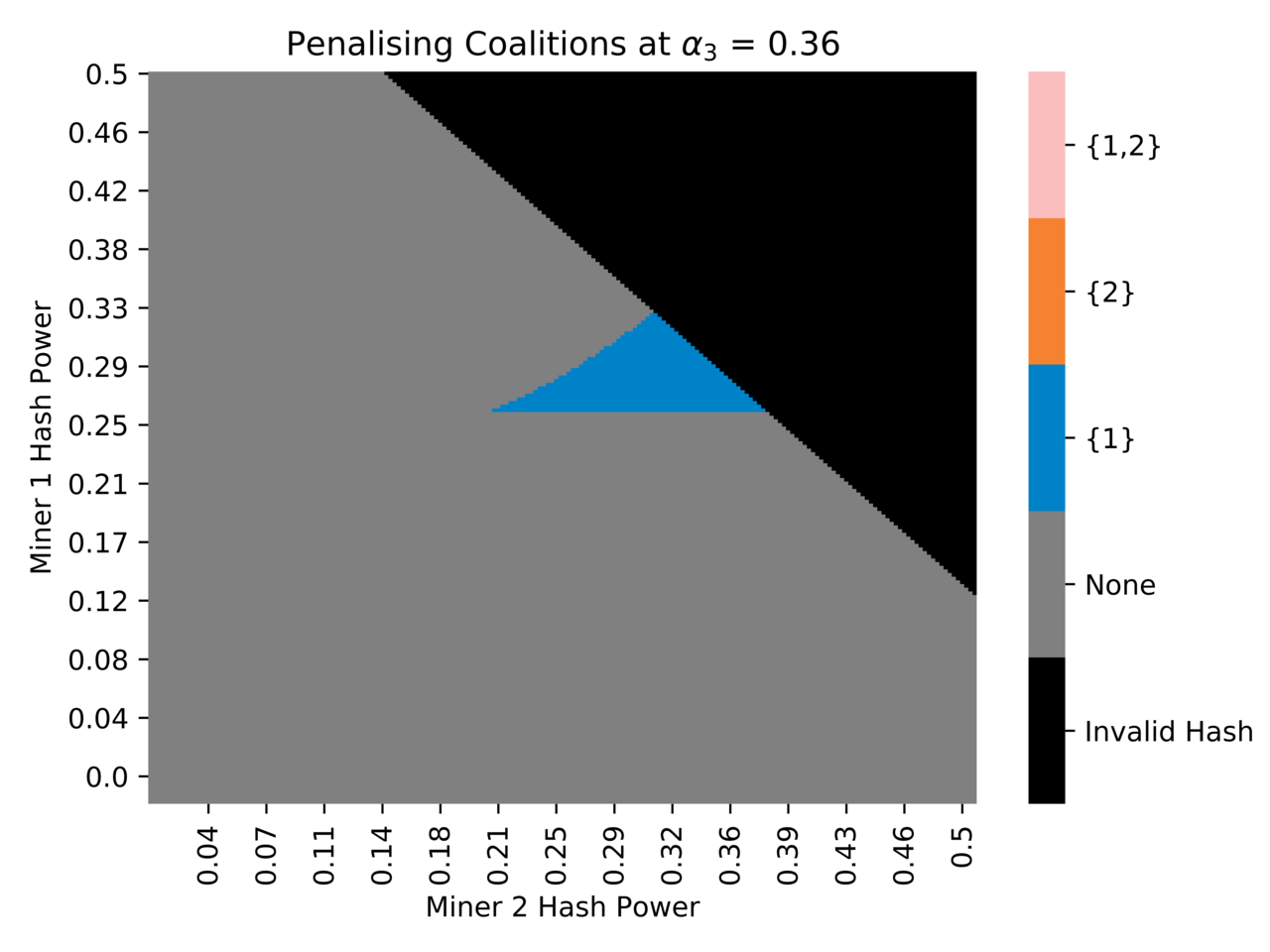}
\includegraphics[width = 0.24\textwidth]{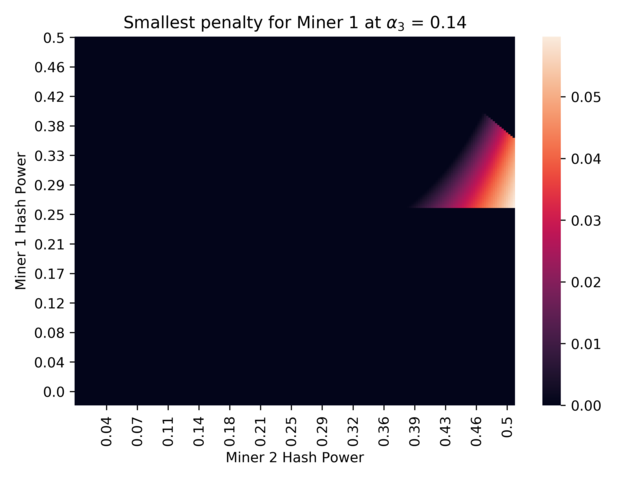}
\includegraphics[width = 0.24\textwidth]{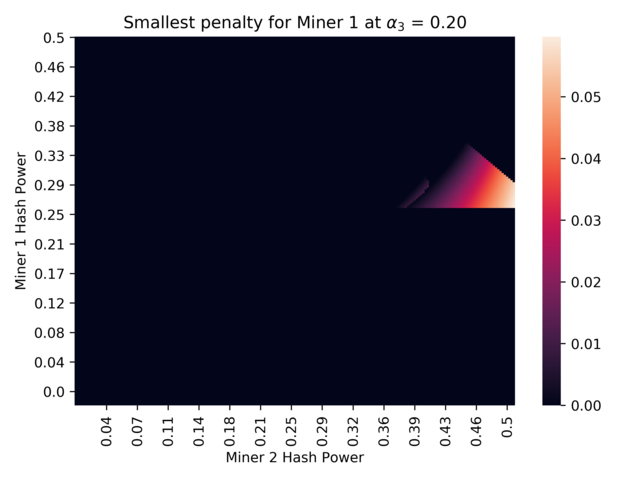}
\includegraphics[width = 0.24\textwidth]{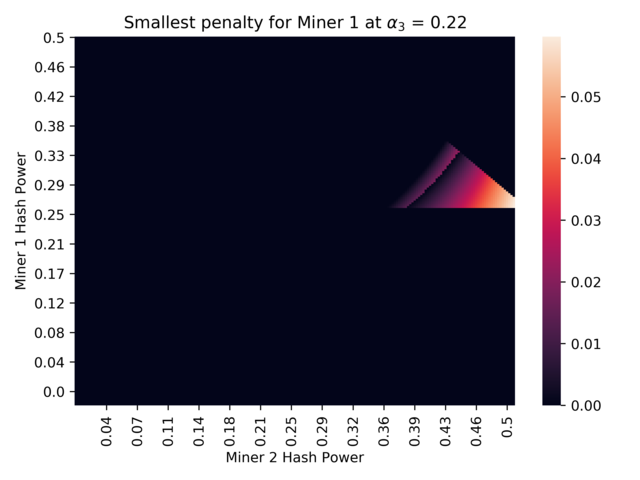}
\includegraphics[width = 0.24\textwidth]{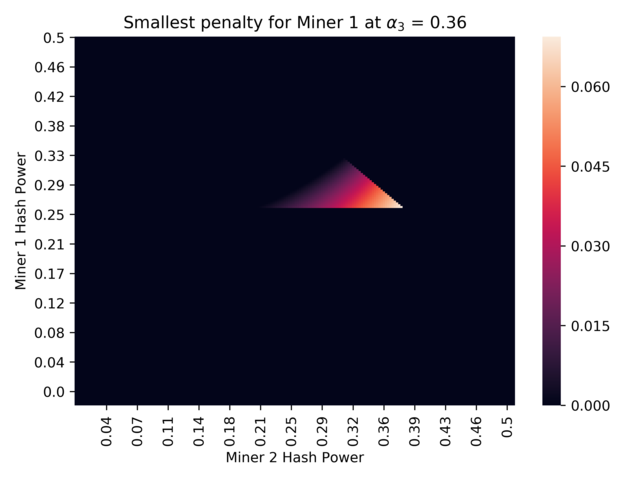}
\caption{All Penalising Coalitions for miner 1 when $\alpha_3 \in \{0.14, 0.2, 0.22, 0.36\}$, and the smallest penalty they incur.}
\label{fig:3SSM-valid_coalition}
\end{figure}

In Figure \ref{fig:3SSM-valid_coalition} we plot hash rates where there exist penalising coalitions against miner 1 along with the smallest given penalty they can incur on miner 1. Furthermore, in the top row of the plot, we specify precisely which coalitions $C \subset [2,3]$ satisfy the conditions of Definition \ref{def:multiplayer-penalising}. The plots show that for $\alpha_3 \in \{0.14,0.36\}$ there is only one kind of penalising coalition ($C = \{2\}$ or $C = \{3\}$ respectively), but for $\alpha_3 \in \{0.2,0.22\}$, $C = \{2\}, \{3\}$ and $\{2,3\}$ are all penalising coalitions at different hash rates and for some values of $\alpha$.

\end{document}